\documentclass[10pt]{article} 
\usepackage[preprint]{tmlr}

\usepackage{amsmath,amsfonts,bm}

\def\eqref#1{equation~\ref{#1}}

\def\1{\bm{1}}

\DeclareMathAlphabet{\mathsfit}{\encodingdefault}{\sfdefault}{m}{sl}
\SetMathAlphabet{\mathsfit}{bold}{\encodingdefault}{\sfdefault}{bx}{n}

\DeclareMathOperator*{\argmin}{arg\,min}

\usepackage[utf8]{inputenc} 
\usepackage[T1]{fontenc}    
\usepackage{hyperref}       
\usepackage{url}            
\usepackage{booktabs}       
\usepackage{amsfonts}       
\usepackage{nicefrac}       
\usepackage{microtype}      
\usepackage{xcolor}         

\usepackage{microtype}
\usepackage{graphicx}
\usepackage{subcaption}
\usepackage{booktabs} 
\usepackage{hyperref}
\usepackage{xr-hyper}
\usepackage{times,enumitem}
\usepackage{ulem}
\usepackage{url}
\usepackage[f]{esvect}
\usepackage{multirow}
\usepackage{bigints}
\usepackage{amsfonts,amsmath,amssymb,amsthm, bbm, bm, mathtools}
\usepackage{wrapfig}
\usepackage{verbatim,float,url}
\usepackage[english]{babel}
\usepackage{cancel}
\usepackage{color}
\usepackage[dvipsnames]{xcolor}
\usepackage{multirow}
\usepackage{booktabs}
\usepackage{pifont}
\usepackage[flushleft]{threeparttable}
\usepackage{graphics}
\usepackage{tikz}
\usetikzlibrary{automata, arrows.meta, positioning, fit, shapes}
\usepackage{makecell}
\usepackage{algorithm}
\usepackage{algpseudocode} 

\theoremstyle{plain}
\newtheorem{theorem}{Theorem}[section]

\theoremstyle{definition}

\newtheorem{assumption}[theorem]{Assumption}
\theoremstyle{remark}
\newtheorem{remark}[theorem]{Remark}

\title{Latent Space Graph-fused LASSO for\\ Clustering in Attributed Graphs}

\author{\name Yik Lun Kei \email ykei@ucsc.edu \\
      \addr Department of Statistics\\ The University of California, Santa Cruz
      \AND
      \name Oscar Hernan Madrid Padilla \email oscar.madrid@stat.ucla.edu \\
      \addr Department of Statistics and Data Science\\ The University of California, Los Angeles
      \AND
      \name Rebecca Killick \email r.killick@lancaster.ac.uk \\
      \addr School of Mathematical Sciences\\ Lancaster University
      \AND
      \name James D. Wilson \email jdwilson4@usfca.edu\\
      \addr Department of Mathematics and Statistics\\ University of San Francisco
      \AND
      \name Xi Chen \email xchen376@ucsc.edu\\
      \addr Department of Statistics\\ The University of California, Santa Cruz
      \AND
      \name Robert B. Lund \email rolund@ucsc.edu\\
      \addr Department of Statistics\\ The University of California, Santa Cruz
}

\def\month{MM}  
\def\year{YYYY} 
\def\openreview{\url{https://openreview.net/forum?id=XXXX}} 

\begin{document}

\maketitle

\begin{abstract}
This manuscript studies latent space graph-fused LASSO methods for clustering graphs with multivariate nodal attributes. The framework includes node-specific priors of low-dimensional representations coupled with a neural decoder that bridges the observed attributes with the latent variables. Both structural and attribute information are incorporated for the clustering through a graph-fused LASSO regularization on prior nodal means. The optimization problem is solved via the alternating direction method of multipliers with Langevin dynamics for posterior inference. Simulations on grid and block graphs, together with applications to California county temperatures and Los Angeles traffic data, demonstrate the effectiveness of the proposed method.
\end{abstract}

\section{Introduction}

Graphs are often used to describe relational phenomena, where entities are denoted by nodes and their relations are delineated by edges. Graphical structures naturally arise in various domains, including anthropology \citep{sandel2026lethal}, biology \citep{han2023spatial}, and neurology \citep{yang2022functional}. A fundamental task in network analysis involves partitioning graph nodes into disjoint sets, where nodes within a set are more associated than those in different sets. Identifying nodal clusters can provide valuable insights (see \cite{shai2017case} for case studies). Most networks possess nodal attributes, while many existing network clustering methods do not incorporate node-specific information \citep{malliaros2013clustering, abbe2018community, macgregor2023fast, tsitsulin2023graph}. The development of clustering attributed networks is still in its infancy. This manuscript demonstrates how to leverage multivariate nodal attributes, promoting clustering methods that account for both the network structure and nodal attributes. 

Recent advances in vision–language models highlight the importance of integrating heterogeneous sources of information in a coherent manner. Similarly, clustering nodes in attributed networks requires an effective combination of structural and nodal information. Previous research often treated these two components separately. For example, a rich literature has studied edge-based clustering, where communities are identified through patterns of edge connectivity \citep{mucha2010community, amini2013pseudo, wilson2017community}. Beyond clustering, statistical modeling of network connectivity has also been investigated \citep{durante2014nonparametric, leifeld2018temporal, lee2020varying}. In contrast, nodal-based approaches have focused on identifying group structures through similarities and shared dependence patterns in nodal observations \citep{chen2023community, roy2024relational, zhu2025simultaneous}. 

Some techniques have considered the joint analyses of node and edge-based characteristics. \cite{binkiewicz2017casc} performed covariate-assisted spectral clustering by augmenting the graph's Laplacian with pairwise similarities of nodal covariates. \cite{shen2024bayesian} propose a Bayesian stochastic block model, where nodal information is incorporated via a covariate-dependent random partition prior. \cite{hu2024nac} leverage individual and neighboring covariates with a node-specific weight matrix that improves clustering. Recently, the graph-fused LASSO regularizations in \cite{wang2016trend}, \cite{chen2023more}, and \cite{Padilla2023graphon} have produced promising results, capably smoothing signals between neighboring nodes while preserving boundaries between regions having distinct signals. Following \cite{hallac2015network} and \cite{yu2025graph}, our framework adapts the graph-fused LASSO regularization to achieve the filtering behavior \citep{tibshirani2005sparsity} for clustering nodes.

In many analyses, nodal attributes provide valuable clustering information, especially in cases with a non-informative graph. An intuitive solution introduces a latent variable at each node. By compressing nodal features into a low-dimensional representation, complex patterns can be integrated with the network structure to facilitate analysis. In our approach, latent representations are modeled by node-specific priors, and the latent space is regularized by a graph-fused LASSO penalty that incorporates network information. The resulting low-dimensional representations preserve graphical structure and attribute information, promoting effective nodal clustering.

Latent space models naturally bridge observed attributes with low-dimensional representations. \cite{handcock2007model} and \cite{sewell2015latent} embed nodes in a Euclidean space, where edge probabilities increase with latent space proximity. Extending this idea, \cite{wang2023joint} govern both nodes and their attributes in a shared latent space and develop a variational Bayesian EM algorithm for estimation. Similarly, \cite{zhu2023disentangling} construct a latent space model for both attractive and repulsive nodes, thereby disentangling network polarization. In contrast, our framework models nodal attributes, conditioning on their low-dimensional representations. Different from a variational auto-encoder \citep{kingma2013auto, kipf2016variational}, which encourages the approximate posterior to align with a zero-mean Gaussian prior, we focus on estimating Gaussian prior means for our clustering. With a graph-fused LASSO regularization imposed on prior parameter differences across edges, nodes exhibiting similar patterns are encouraged to have similar prior parameters. Consequently, $k$-means methods that cluster via centroids work well in the low-dimensional latent space.

Extending graph-fused LASSO to latent spaces and multivariate nodal attributes, this manuscript makes the following contributions:

\begin{itemize}

\item A decoder-only latent space model that bridges nodal attributes with latent variables is introduced. The latent variables are regarded as low-dimensional representations of the nodal features and capture network structural patterns. Clustering is performed on the latent space.

\item An approach is developed to estimate the model's Gaussian prior means. The graph-fused LASSO regularization enforces piecewise constant means across graph edges, producing similar prior parameters in neighboring nodes while preserving regions having disparate signals.

\item The optimization problem encountered is solved via the alternating direction method of multipliers (ADMM). Simulation studies on grid and block graphs and applications to temperature and traffic data demonstrate the effectiveness of our framework.

\end{itemize}

The rest of the manuscript is organized as follows. Section \ref{sec_model} describes our proposed framework. Section \ref{sec_learning} presents an objective function and develops the ADMM procedure that solves our associated optimization problem. Section \ref{sec_analysis} provides a theoretical analysis of the results. Section \ref{sec_clustering} discusses model selection and nodal clustering methods. Section \ref{sec_experiments} illustrates the proposed methods through simulations and data applications. Section \ref{sec_discussion} discusses limitations and ideas for future work.

\section{Decoder-only Latent Space Models}
\label{sec_model}

\subsection{Model Specification}

We consider an undirected graph $\mathcal{G} = (\mathcal{V}, \mathcal{E})$, where $\mathcal{V}$ and $\mathcal{E}$ are the respective node and edge sets. Each node $i \in \mathcal{V}$ is associated with a multivariate attribute vector $\bm{Y}_i \coloneqq (Y_{1,i}, Y_{2,i}, \dots, Y_{n,i}) \in \mathbb{R}^n$. For each $\bm{Y}_i$, we assume existence of a latent variable $\bm{Z}_i \in \mathbb{R}^d$ such that the conditional distribution of $\bm{Y}_i$ given $\bm{Z}_i$ is specified via the decoder
\[
\bm{Y}_i | \bm{Z}_i \sim \mathcal{N} \big(\bm{h}_{\bm{\phi}}(\bm{Z}_i), \bm{I}_n \big),
\]
where $\bm{h}_{\bm{\phi}}: \mathbb{R}^d \rightarrow \mathbb{R}^n$ is a neural network parameterized by $\bm{\phi}$ and elaborated upon in Section \ref{sec_decoder}. Moreover, we impose a prior distribution on the latent $\bm{Z}_i$ via
\[
\bm{Z}_i \sim \mathcal{N} (\bm{\mu}_i, \bm{I}_d).
\] 
Denote the conditional probability density function of $\bm{Y}_i$ given $\bm{Z}_i$ as $p_{\bm{\phi}}(\bm{y}_i | \bm{z}_i)$, and the marginal  probability density function of $\bm{Z}_i$ by $p_{\bm{\mu}_i}( \bm{z}_i)$. Anchored on the decoder $p_{\bm{\phi}}(\bm{y}_i | \bm{z}_i)$, the latent variable $\bm{Z}_i \in \mathbb{R}^d$ is a low-dimensional representation of $\bm{Y}_i \in \mathbb{R}^{n}$ at node $i$. The prior parameter, $\bm{\mu}_i \in \mathbb{R}^d$, is learned for each node $i \in \mathcal{V}$. 

Figure \ref{overview_GFL} illustrates the framework. The shaded circles in the top layer depict the observed $\{\bm{Y}_i\}_{i \in \mathcal{V}}$, and the dashed circles in the bottom layer depict the latent $\{ \bm{Z}_i \}_{ i \in \mathcal{V} }$. The decoder $p_{\bm{\phi}}(\bm{y}_i|\bm{z}_i)$ with neural network parameter $\bm{\phi}$ is shared across all nodes, while the means $\bm{\mu}_i$ vary across nodes. The decoder $p_{\bm{\phi}}(\bm{y}_i|\bm{z}_i)$ bridges observed attributes with latent variables in a bottom-up manner. Intuitively, the decoder helps infer the low-dimensional representation $\bm{Z}_i$ in a top-down manner. The simplicity of the proposed framework is noted, without the need for an encoder.

\begin{figure}[!htb]
\centering
\resizebox{0.5\columnwidth}{!}{%
\begin{tikzpicture} [on grid, state/.style={draw, minimum size=1.1cm}]

\node (z1) [circle, state, dashed] {$\bm{Z}_{1}$};
\node (y1) [circle, state, above = 2.5cm of z1, fill=lightgray] {$\bm{Y}_{1}$};
\node (label1) [draw, dashed, draw = none, below = 1cm of z1] {\small $\mathcal{N}(\bm{\mu}_1,\bm{I}_d)$};

\node (z2) [circle, state, dashed, below right = 2cm and 2cm of z1] {$\bm{Z}_{2}$};
\node (y2) [circle, state, above = 2.5cm of z2, fill=lightgray] {$\bm{Y}_{2}$};
\node (label2) [draw, dashed, draw = none, below = 1cm of z2] {\small $\mathcal{N}(\bm{\mu}_2,\bm{I}_d)$};

\node (z3) [circle, state, dashed, above right = 3.5cm and 2cm of z2] {$\bm{Z}_{3}$};
\node (y3) [circle, state, above = 2.5cm of z3, fill=lightgray] {$\bm{Y}_{3}$};
\node (label3) [draw, dashed, draw = none, below = 1cm of z3] {\small $\mathcal{N}(\bm{\mu}_3,\bm{I}_d)$};

\node (z4) [circle, state, dashed, below right = 1cm and 3.5cm of z3] {$\bm{Z}_{4}$};
\node (y4) [circle, state, above = 2.5cm of z4, fill=lightgray] {$\bm{Y}_{4}$};
\node (label4) [draw, dashed, draw = none, below = 1cm of z4] {\small $\mathcal{N}(\bm{\mu}_4,\bm{I}_d)$};

\node (z5) [circle, state, dashed, above right = 1cm and 3 cm of z4] {$\bm{Z}_{5}$};
\node (y5) [circle, state, above = 2.5cm of z5, fill=lightgray] {$\bm{Y}_{5}$};
\node (label5) [draw, dashed, draw = none, below = 1cm of z5] {\small $\mathcal{N}(\bm{\mu}_5,\bm{I}_d)$};

\node[draw, dotted, fit={(y1) (y2) (y3) (z1) (z2) (z3) (label2)}, minimum width=7.7cm, minimum height=7cm, rounded corners, label=above: Cluster 1] (c1) {};

\node[draw, dotted, fit={(y4) (y5) (z4) (z5) (label4)}, minimum width=6cm, minimum height=5.5cm, rounded corners, label=above: Cluster 2, xshift=6mm] (c2) {};

\path [-stealth, thick]

(z1) edge [dashed] node[left]{$p_{\bm{\phi}}(\bm{y}_1|\bm{z}_1)$} (y1)

(z2) edge [dashed] node[right]{$p_{\bm{\phi}}(\bm{y}_2|\bm{z}_2)$} (y2)

(z3) edge [dashed] node[right]{$p_{\bm{\phi}}(\bm{y}_3|\bm{z}_3)$} (y3)

(z4) edge [dashed] node[right]{$p_{\bm{\phi}}(\bm{y}_4|\bm{z}_4)$} (y4)

(z5) edge [dashed] node[right]{$p_{\bm{\phi}}(\bm{y}_5|\bm{z}_5)$} (y5)

(y1) edge[-] (y2)
(y2) edge[-] (y3)
(y1) edge[-] (y3)
(y3) edge[-] (y4)
(y4) edge[-] (y5)

; 

\end{tikzpicture}
} 

\caption{Illustration of prior distributions and a decoder on a graph having five nodes and two clusters.}
\label{overview_GFL}
\end{figure}

\subsection{Neural Decoder}
\label{sec_decoder}

The distribution of interest is the marginal $p(\bm{y}_i)$, which is often complicated and unknown a priori. Instead of imposing a parametric form on $p(\bm{y}_i)$, our framework introduces an auxiliary latent variable $\bm{Z}_i \in \mathbb{R}^d$ to model $p(\bm{y}_i)$ from a flexible neural mixture:
\begin{equation}
\label{marginal_likelihood}
p_{\bm{\phi}, \bm{\mu}_i}(\bm{y}_i) = \int p_{\bm{\phi}}(\bm{y}_i | \bm{z}_i) p_{\bm{\mu}_i}(\bm{z}_i) d\bm{z}_i.
\end{equation}

\begin{remark}
The Gaussian assumption is placed on $p(\bm{y}_i | \bm{z}_i)$ instead of the marginal $p(\bm{y}_i)$. The induced marginal distribution of $\bm{Y}_i \in \mathbb{R}^n$ can be highly non-Gaussian and depends on $\bm{h}_{\bm{\phi}}(\bm{\cdot})$.

\end{remark}

\begin{remark}
Our framework assumes that the nodal features $Y_{j,i}$ for $j = 1, \dots, n$ are conditionally independent given $\bm{Z}_i$. The auxiliary latent variable $\bm{Z}_i \in \mathbb{R}^d$ serves as a low-dimensional summary of the relationships in $\bm{Y}_i \in \mathbb{R}^n$ while retaining a flexible marginal distribution $p(\bm{y}_i)$. In particular, the covariance of $\bm{Y}_i \in \mathbb{R}^n$ is
\begin{align*}
\mathrm{Cov}(\bm{Y}_i) & = \mathbb{E}\big[\mathrm{Cov}(\bm{Y}_i|\bm{Z}_i)\big] + \mathrm{Cov}\big[\mathbb{E}(\bm{Y}_i|\bm{Z}_i)\big]\\
& = \bm{I}_n + \mathrm{Cov}\big[\bm{h}_{\bm{\phi}}(\bm{Z}_i)\big],
\end{align*}
which depends on $\bm{h}_{\bm{\phi}}(\bm{\cdot})$, thereby allowing for marginal dependence between $Y_{j,i}$ and $Y_{j',i}$.
\end{remark}

\begin{remark}
In contrast to unbounded latent space models \citep{mattei2018leveraging}, the negative log-likelihood of (\ref{marginal_likelihood}) with our model specification is lower bounded:
\[
- \ln \left( \int p_{\bm{\phi}}(\bm{y}_i | \bm{z}_i) p_{\bm{\mu}_i}(\bm{z}_i) d\bm{z}_i \right) \geq \frac{n}{2} \ln (2\pi),
\]
rendering associated minimization problems well posed.
\end{remark}

For flexibility of the marginal $p(\bm{y}_i)$, we parameterize $\bm{h}_{\bm{\phi}}(\bm{\cdot})$ in $p_{\bm{\phi}}(\bm{y}_i|\bm{z}_i)$ using a feed-forward ReLU neural network, which possesses universal approximation properties \citep{hornik1989multilayer, hornik1991approximation, yarotsky2017error}. This parameterization is chosen for representation learning, capturing the complex patterns in nodal attributes via $\mathrm{Cov}(\bm{Y}_i) = \bm{I}_n + \mathrm{Cov}[\bm{h}_{\bm{\phi}}(\bm{Z}_i)]$.  This yields node-level representations tailored for graph-based regularization. Moreover, the feed-forward ReLU neural networks can be expressive and parsimonious; it also permits the theoretical analysis in Section \ref{sec_analysis}. While other architectures can be used \citep{lecun2015deep}, ReLU neural networks have been extensively studied and are well suited for clustering attributed networks with our framework. 

\section{Learning and Inference}
\label{sec_learning}

This section develops parameter estimation methods. A penalized likelihood is formulated first; the resulting constrained optimization is subsequently solved with an ADMM-type procedure. Denote the joint log-likelihood of $\{ \bm{Y}_i  \}_{i \in \mathcal{V}}$ as $L(\bm{\phi}, \bm{\mu}) = \sum_{i \in \mathcal{V}} \ell(\bm{\phi}, \bm{\mu}_i)$, where the node $i$ marginal log-likelihood is $\ell(\bm{\phi},\bm{\mu}_i) = \ln( \int p_{\bm{\phi}}(\bm{y}_i | \bm{z}_i) p_{\bm{\mu}_i}(\bm{z}_i) d\bm{z}_i)$. To cluster the prior parameters, the penalized optimization
\begin{equation}
\hat{\bm{\phi}},\hat{\bm{\mu}} = \argmin_{ \bm{\phi},\bm{\mu} } \bigg[-L(\bm{\phi},\bm{\mu}) + \lambda \sum_{(i,j) \in \mathcal{E}} \|\bm{\mu}_i - \bm{\mu}_j\|_2\bigg]
\label{eqn:original}
\end{equation}
is considered, where $\lambda > 0$ is a tuning parameter for the penalty term. The $i$th row of the $\mathbb{R}^{N \times d}$ matrix $\bm{\mu}$ is denoted by $\bm{\mu}_i \in \mathbb{R}^d$, where $N \coloneqq |\mathcal{V}|$.

The graph-fused LASSO regularization governs associations between nodes, incorporating structure and attribute information into our latent representations. The regularization here is a total variation sum of the multivariate pairwise differences $\bm{\mu}_i - \bm{\mu}_j \in \mathbb{R}^d$ over edges $(i,j) \in \mathcal{E}$. The regularization induces sparsity in the differences while permitting simultaneous shifts across the $d$ dimensions. By penalizing the total variation of the prior parameters across edges, our framework can identify boundaries between clusters having disparate signals.

To optimize (\ref{eqn:original}), which involves the latent variables, we first manipulate the objective function. Introduce the slack variables $\bm{\nu}_{i,j} \in \mathbb{R}^{d}$ and recast the problem as the constrained optimization 
\begin{equation}
\begin{split}
\label{eqn:original_sub_to}
\hat{\bm{\phi}},\hat{\bm{\mu}} =& \argmin_{ \bm{\phi},\bm{\mu} } \bigg[-L(\bm{\phi},\bm{\mu}) + \lambda \sum_{(i,j) \in \mathcal{E}} \|\bm{\nu}_{i,j}\|_2\bigg], \qquad \text{subject to} \quad 
\bm{\mu}_i - \bm{\mu}_j = \bm{\nu}_{i,j}.
\end{split}
\end{equation}
The augmented Lagrangian is formulated as
\begin{align*}
\mathcal{L}(\bm{\phi}, \bm{\mu}, \bm{\nu}, \bm{\rho}) = -L(\bm{\phi},\bm{\mu}) + \lambda \sum_{(i,j) \in \mathcal{E}} \|\bm{\nu}_{i,j}\|_2 + \sum_{(i,j) \in \mathcal{E}} \bm{\rho}_{i,j}^\top (\bm{\mu}_i - \bm{\mu}_j - \bm{\nu}_{i,j}) + \frac{\gamma}{2} \sum_{(i,j) \in \mathcal{E}} \lVert \bm{\mu}_i - \bm{\mu}_j - \bm{\nu}_{i,j} \rVert_2^2,   
\end{align*}
where $\bm{\rho}_{i,j} \in \mathbb{R}^d$ is the Lagrange multiplier for each $(i,j) \in \mathcal{E}$ and $\gamma>0$ is an additional penalty parameter for the augmentation term. Let $\bm{w}_{i,j} = \gamma^{-1}\bm{\rho}_{i,j} \in \mathbb{R}^{d}$ denote the scaled dual variable. The augmented Lagrangian is equivalently formulated as
\begin{align*}
\mathcal{L}&(\bm{\phi}, \bm{\mu}, \bm{\nu}, \bm{w}) =  -L(\bm{\phi}, \bm{\mu}) + \lambda \sum_{(i,j) \in \mathcal{E}} \|\bm{\nu}_{i,j}\|_2 + \frac{\gamma}{2} \sum_{(i,j) \in \mathcal{E}} \left( \lVert \bm{\mu}_i - \bm{\mu}_j - \bm{\nu}_{i,j}  + \bm{w}_{i,j} \rVert_2^2 - \lVert \bm{w}_{i,j} \rVert_2^2 \right).
\end{align*}

An ADMM-type procedure \citep{boyd2011distributed, deng2017parallel, wang2019global} is employed to recursively optimize our objective function:
\begin{align}
\bm{\phi}^{(a+1)} & = \argmin_{\bm{\phi}} \bigg[-L(\bm{\phi},\bm{\mu}^{(a)})\bigg],
\nonumber
\\
\label{loss_mu}
\bm{\mu}_i^{(a+1)} & = \argmin_{\bm{\mu}_i} \bigg[-L(\bm{\phi}^{(a+1)},\bm{\mu}_i) + \frac{\gamma}{2} \sum_{j \in \mathcal{B}(i)} \lVert \bm{\mu}_i - \bm{\mu}_j - \bm{\nu}_{i,j}^{(a)} + \bm{w}_{i,j}^{(a)} \rVert_2^2 \bigg],\qquad\forall\ i \in \mathcal{V},
\\
\label{loss_nu}
\bm{\nu}_{i,j}^{(a+1)} & = \argmin_{\bm{\nu}_{i,j}} \bigg[\lambda \|\bm{\nu}_{i,j}\|_2 + \frac{\gamma}{2} \lVert \bm{\mu}_i^{(a+1)} - \bm{\mu}_j^{(a+1)} - \bm{\nu}_{i,j} + \bm{w}_{i,j}^{(a)} \rVert_2^2\bigg],\qquad\forall\ (i,j) \in \mathcal{E}, 
\\
\bm{w}_{i,j}^{(a+1)} & = \bm{\mu}_i^{(a+1)} - \bm{\mu}_j^{(a+1)} - \bm{\nu}_{i,j}^{(a+1)} + \bm{w}_{i,j}^{(a)},\qquad\forall\ (i,j) \in \mathcal{E}.
\nonumber
\end{align}
Here, $a$ indexes the ADMM iterations, and $\mathcal{B}(i) \coloneqq \{ j \in \mathcal{V} : (i,j) \in \mathcal{E} \}$ is the set of neighbors of node $i$.

The update for $\bm{\mu}_i$ at an iteration, with the neighboring parameters fixed at their preceding iteration values, is
\begin{equation}
\bm{\mu}_i = \frac{1}{1+\gamma |\mathcal{B}(i)|} \bigg[ \mathbb{E} ( \bm{Z}_i | \bm{Y}_i ) + \gamma \sum_{j \in \mathcal{B}(i)} ( \bm{\mu}_j + \bm{\nu}_{i,j} - \bm{w}_{i,j}) \bigg].
\label{sol_mu}
\end{equation}
Moreover, the gradient with respect to the decoder parameter is
\begin{equation}
- \nabla_{\bm{\phi}}\ L(\bm{\phi}, \bm{\mu}) = - \sum_{i \in \mathcal{V}} \mathbb{E} \Big( \nabla_{\bm{\phi}} \ln \big( p_{\bm{\phi}}(\bm{Y}_i|\bm{Z}_i) \big) \Big| \bm{Y}_i \Big).
\label{grad_decoder}
\end{equation}
The parameter $\bm{\phi}$ is updated via back-propagation. The update in (\ref{loss_nu}) is equivalent to solving a group LASSO problem \citep{yuan2006model, vert2010fast, kei2025separable}. One can update $\bm{\nu}_{i,j} \in \mathbb{R}^d$ for each edge $(i,j) \in \mathcal{E}$ via
\begin{equation}
\label{update_nu}
\bm{\nu}_{i,j}= \left( 1 - \frac{\lambda}{\gamma \|\bm{s}_{i,j}\|_2}\right)_+ \bm{s}_{i,j},
\end{equation}
where $\bm{s}_{i,j} = \bm{\mu}_i - \bm{\mu}_j + \bm{w}_{i,j} \in \mathbb{R}^d$ and $(x)_+ = \max(0,x)$.

Calculating (\ref{sol_mu}) and the gradient in (\ref{grad_decoder}) requires evaluating a conditional expectation $\mathbb{E} ( \bm{\cdot} | \bm{Y}_i )$ under the posterior density $p(\bm{z}_i|\bm{y}_i) \propto p(\bm{y}_i|\bm{z}_i) \times p(\bm{z}_i)$. Here, Langevin dynamics are used to generate approximate posterior samples for estimating conditional expectations \citep{du2019implicit, nijkamp2020anatomy, purohit2024posterior, kei2025change}. In particular, let $k$ be the time step in the Langevin dynamics and let $\delta > 0$ be a small step size. The Langevin dynamics to generate approximate posterior samples iterate
\begin{equation}
\bm{z}_i^{k+1} = \bm{z}_i^{k} + \delta \big[ \nabla_{\bm{z}_i} \ln\big( p_{\bm{\phi}}(\bm{y}_i|\bm{z}_i^{k})\big) - (\bm{z}_i^{k} - \bm{\mu}_i) \big] + \sqrt{2 \delta} \bm{\epsilon}
\label{Langevin_sampling}
\end{equation}
in $k$, where $\bm{\epsilon} \sim \mathcal{N}(\bm{0},\bm{I}_d)$ is a random perturbation. The gradient of the log-likelihood $\ln(p_{\bm{\phi}}(\bm{y}_i|\bm{z}_i))$ with respect to $\bm{z}_i$ is calculated via back-propagation. The derivations of the updates and the computational complexity of the ADMM procedure are provided in Appendix \ref{appendix_par_learning}.

\section{Theoretical Analysis}
\label{sec_analysis}

Let $\mathcal{P}$ be the parameter space for the neural-network weights $\bm{\phi}$. Suppose that $\bm{h}_{\bm{\phi}}$ has $L$ layers, $B$ neurons in total, and $W$ parameters. We use the notation
\[
\| \bm{h}_{ \bm{\phi}} (\bm{z}) \|_{\infty}\,\coloneqq\,  \underset{j =1,\ldots,n}{\max}\, \| h_{ \bm{\phi}, j}(\bm{z}) \|_{\infty},
\]
with $h_{ \bm{\phi}, j}(\bm{z})$ denoting the $j$th coordinate of $\bm{h}_{ \bm{\phi} }(\bm{z}) \in \mathbb{R}^n$, and
\begin{align*}
\| h_{\bm{\phi}, j}(\bm{z}) \|_{\infty}\, \coloneqq\,\underset{\bm{z} \in \mathbb{R}^d }{\sup}\, \vert h_{\bm{\phi}, j}(\bm{z}) \vert,
\qquad \text{and} \qquad
\| \bm{\mu}\|_{\infty}\, \coloneqq\,  \underset{k=1, \ldots, d, \, i \in \mathcal{V}  }{\max} \,\,\vert \mu_{k,i}\vert,    
\end{align*}
where $\mu_{k,i}$ is the $k$th coordinate of $\bm{\mu}_i \in \mathbb{R}^d$. To facilitate our analysis, the following assumption is imposed.

\begin{assumption}
\label{as1}
The vectors $\{ \bm{Y}_i\}_{i\in\mathcal{V}}$ are independent across nodes, although the coordinates within each $\bm{Y}_i \in \mathbb{R}^n$ may be dependent. We assume that $\max\{d,n\}=O(1)$ and that the coordinates of the observations are uniformly sub-Gaussian; that is, for some constant $\sigma > 0$,
\begin{align*}
\max_{i\in\mathcal{V}}\max_{1\leq j\leq n} \|Y_{j,i}\|_{\psi_2}\leq \sigma,
\qquad \text{and} \qquad
\|X\|_{\psi_2} \coloneqq \inf\left\{t>0:\mathbb{E} \left[ \exp\left(\frac{X^2}{t^2}\right) \right] \leq 2  \right\}.
\end{align*}
\end{assumption}

Consider the following constrained optimization version of our estimator:
\begin{equation}
\label{eqn:ideal_estimator}
\begin{array}{ccc}
\hat{\bm\phi},\hat{\bm\mu}\,\coloneqq & \underset{\bm{\phi}  \in \mathcal{P}, \,\,\bm\mu \in \mathbb{R}^{\vert \mathcal{V}\vert \times d} }{\arg \min} & -L(\bm \phi,\bm\mu)\\
\end{array}
\end{equation}
subject to the constraints
\begin{equation}
\label{constraints}
\max \big\{ \| \bm{h}_{\bm \phi}(\bm{z}) \|_{\infty}, \|\bm\mu\|_{\infty} \big\} \leq C, 
\quad
\sum_{(i,j) \in \mathcal{E}} \|\bm{\mu}_i - \bm{\mu}_j\|_2 \leq U.
\end{equation}
The main excess-risk bound is stated below.

\begin{theorem}
\label{thm1}
Suppose that $\mathcal{G}$ is connected and that Assumption \ref{as1} holds. Let $p_i^*$ be the true marginal density of $\bm{Y}_i$, and let $\bm{\phi}^*$ and $\bm{\mu}^*$ be any minimizer of
\[
\begin{array}{cc}
\underset{\bm{\phi}\in\mathcal{P},\,\bm{\mu}\in\mathbb{R}^{|\mathcal{V}|\times d}}{\operatorname{minimize}} & \displaystyle \frac{1}{|\mathcal{V}|}\sum_{i\in\mathcal{V}} \mathrm{KL}\bigl(p_i^*\,\big|\,p_{\bm{\phi},\bm{\mu}_i}(\bm{y}_i)\bigr)
\end{array}
\]
subject to the constraints in (\ref{constraints}), where $p_{\bm{\phi},\bm{\mu}_i}(\bm{y}_i)$ is the marginal density of $\bm{Y}_i$ given by
\[
p_{\bm{\phi},\bm{\mu}_i}(\bm{y}_i)= \frac{1}{(2\pi)^{(n+d)/2}} \int \exp\left\{-\frac{1}{2}\|\bm{y}_i-\bm{h}_{\bm{\phi}}(\bm{z}_i)\|_2^2 -\frac{1}{2}\|\bm{z}_i-\bm{\mu}_i\|_2^2\right\}d\bm{z}_i.
\]
Assume that the oracle objective displayed above has a finite minimum value. Then there exists a constant $K>0$, depending only on $C$, $\sigma$, $n$, and $d$, such that the estimator in (\ref{eqn:ideal_estimator}) satisfies
\begin{align}
\label{eqn:rate}
\frac{1}{|\mathcal{V}|}\sum_{i\in\mathcal{V}} \mathbb{E}&\left[\mathrm{KL}\bigl(p_i^*\,\big|\,p_{\hat{\bm{\phi}},\hat{\bm{\mu}}_i}(\bm{y}_i)\bigr)\right]
\leq
K\frac{\sqrt{LW\ln(eB)}+\sqrt{1+U}}{\sqrt{|\mathcal{V}|}} + \frac{1}{|\mathcal{V}|}\sum_{i\in\mathcal{V}} \mathrm{KL}\bigl(p_i^*\,\big|\,p_{\bm{\phi}^*,\bm{\mu}_i^*}(\bm{y}_i)\bigr).
\end{align}
The expectation on the left-hand side above is taken over the sample used to construct $\{\hat{\bm{\phi}}, \hat{\bm{\mu}}\}$ and the Kullback-Leibler divergence (KL) is evaluated with respect to an independent draw from $p_i^*$.
\end{theorem}

 
Theorem \ref{thm1} is general: the true data generating mechanism is not assumed to lie in the model class induced by the constraints in (\ref{constraints}). The assumed conditions are relatively mild. Specifically, Assumption \ref{as1} imposes a sub-Gaussian tail condition on the observations. This assumption accommodates a broad class of distributions and does not impose independence between the coordinates within a node. The requirement $\max\{d,n\} = O(1)$ keeps the latent and observed dimensions fixed while allowing the number of nodes $|\mathcal{V}|$ to grow. The constraint $\sum_{(i,j) \in \mathcal{E}} \| \bm{\mu}_i - \bm{\mu}_j \|_2 \leq U$ in (\ref{constraints}) controls the total variation of node-specific priors across edges, ensuring that learned priors are piecewise smooth over the graph.  This is essential when clustering nodes that share similar prior parameters. The result's proof is presented in Appendix \ref{appendix_excess_risk}.

Interpreting the bound in (\ref{eqn:rate}), the first term measures the expressive complexity of the neural network family and depends on the number of layers $L$, the number of neurons $B$, and the total number of parameters $W$. The second term measures the best possible approximation error in terms of KL divergence, achievable within our constrained model class. Since the true densities $p_i^*$ are not assumed to lie in our model class, this term accounts for model mis-specification. Even with a highly parameterized neural network, the statistical error can vanish once a sufficiently large number of nodes are observed (the approximation error goes to zero).

\section{Model Selection and Nodal Clustering}
\label{sec_clustering}

\subsection{Model Selection}

The optimization problem in (\ref{eqn:original_sub_to}) involves the tuning parameter $\lambda$. Given a list of candidate $\lambda$ values, selection of it is performed through cross validation. Specifically, the observed data is split into training and testing sets via nodes. Let $\mathcal{V}_{\text{test}}$ be the node set of the testing data. During training, the node $i$ data $\bm{Y}_i$ is replaced with the element-wise average $\bar{\bm{Y}}_i$ from all neighbors. The node $i$ data is held out for testing when $i \in \mathcal{V}_{\text{test}}$. 

After parameters are learned from the training data, $\hat{\bm{\mu}}$ is used for parameters at removed nodes and the decoder calculates the log-likelihood of the non-training data. The $\lambda$ with the highest log-likelihood is selected. In particular, the log-likelihoods for the testing data are calculated by
\begin{align*}
\sum_{i \in \mathcal{V}_{\text{test}}} \ln \big(p(\bm{y}_i)\big) \approx \sum_{i \in \mathcal{V}_{\text{test}}} \ln \left[ \frac{1}{s} \sum_{u=1}^s \big[ p_{\hat{\bm{\phi}}}(\bm{y}_i | \bm{z}_{u,i}) \big] \right],
\end{align*}
where $\{\bm{z}_{u,i}\}_{u=1}^s$ are samples drawn from $p_{\hat{\bm{\mu}}_i}(\bm{z}_i)$ with the learned parameter $\hat{\bm{\mu}}_i$ for node $i$. Eventually, with the selected $\lambda$, we learn the parameters again via the full data, resulting in the final estimated parameters for clustering.

\subsection{Nodal Clusterings}

For attributed graphs, features observed at neighboring nodes often exhibit similar patterns. To capture this behavior, we jointly model nodal attributes and graphical structure, allowing learned priors to integrate both sources of information. Specifically, the graph-fused LASSO regularization encourages adjacent nodes with similar patterns to have similar prior means, effectively promoting a piecewise-constant structure over the graph. The estimated prior parameters form well-separated regions in a low-dimensional Euclidean space. Therefore, $k$-means is applied to $\{ \hat{\bm{\mu}}_i \}_{i \in \mathcal{V}}$:  a centroid-based clustering method well aligns with the geometry induced by the regularization. The resulting partitions are more informative, reflecting both attribute similarity and structural proximity (rather than solely relying on the features $\{\bm{Y}_i\}_{i \in \mathcal{V}}$ or the graph structure $\mathcal{G} = (\mathcal{V}, \mathcal{E})$ alone).

Since the $k$-means algorithm requires the number of clusters $k$, we estimate this parameter via a silhouette score \citep{rousseeuw1987silhouettes}. A range of $k$ in $\{ 2, \dots, K \}$ is considered; the $k$ having the highest silhouette score $S \in [-1,1]$, defined by
\[
S = \frac{1}{|\mathcal{V}|} \sum_{i \in \mathcal{V}} \frac{d_i^{\text{out}} - d_i^{\text{in}}}{\max(d_i^{\text{in}}, d_i^{\text{out}})},
\]
is selected. Here, $d_i^{\text{in}}$ is the mean distance between point $i$ and all points in the same cluster, and $d_i^{\text{out}}$ is the minimum mean distance between point $i$ and any other cluster. A higher $S$ suggests more cohesive clusters, with distinct separation between different clusters. 

\section{Experiments}
\label{sec_experiments} 

\subsection{Simulations}

To generate realistic networks, grid and block graphs are now simulated. Grid graphs imitate spatial aspects such as urban layouts or power systems, where entities reside on a lattice and clusters reflect spatial regions. Block graphs mimic social relations, where individuals with similar attributes form cohesive clusters. To simulate nodal attributes, autoregressive (AR) and vector autoregressive (VAR) series are used for the multivariate data. Both models inject temporal dependence by linearly regressing on past series values.

We simulate 25 grid graphs at the node sizes $N\coloneqq|\mathcal{V}|$ described below, with $n=90$ observed features per node. To evaluate performance, the following clustering metrics are considered: normalized mutual information (NMI), adjusted Rand index (ARI), accuracy (ACC), homogeneity (HOM), completeness (COM), and cluster purity (PUR). These evaluation metrics require a known truth to measure from; see \cite{emmons2016analysis} and \cite{su2022comprehensive} for additional details. Nine competitors, $k$-means \citep{hartigan1979algorithm}, DTW, \citep{sakoe2003dtw}, CASC \citep{binkiewicz2017casc}, SDP \citep{yan2021sdp}, NAC \citep{hu2024nac}, SCORE \citep{Jin2015SCORE}, VGAE \citep{kipf2016variational}, DGI \citep{velickovic2018deep}, GraphMAE \citep{hou2022graphmae}, are provided for comparison.

The $k$-means and DTW methods only use the nodal attributes for clustering, while the SCORE method only uses the graphical structure. The remaining methods are hybrid approaches that cluster via both node attributes and graph structure. The simulations assess whether integrating both sources of information leads to better clustering than approaches that use the graph or attributes alone, as well as compared with other hybrid approaches. The true number of clusters is supplied to competitors while our method selects the number of clusters for evaluation.

Throughout, our method is dubbed GFL. The latent dimension is taken as $d=3$, and the two hidden layers in the neural network each contain $32$ neurons. The tuning $\lambda$'s are $\{ 0.1, 0.25, 0.5, 0.75, 1.0\}$ and $10\%$ of the nodes were held out for model selection. For each $\lambda$, $50$ iterations of the ADMM procedure are run and $\gamma=\lambda$ is taken. In each ADMM iteration, the neural network is trained using $20$ steps of an Adam optimizer with a learning rate of $10^{-4}$. For Langevin dynamic sampling, we take $\delta=0.4$ and $500$ samples are drawn for each node using $50$ MCMC iterations. Additional details and results are provided in Appendix \ref{appendix_additional_result}.

\subsection*{Scenario 1}

\begin{figure}[!ht]
\begin{center}
\includegraphics[width=0.55\columnwidth]{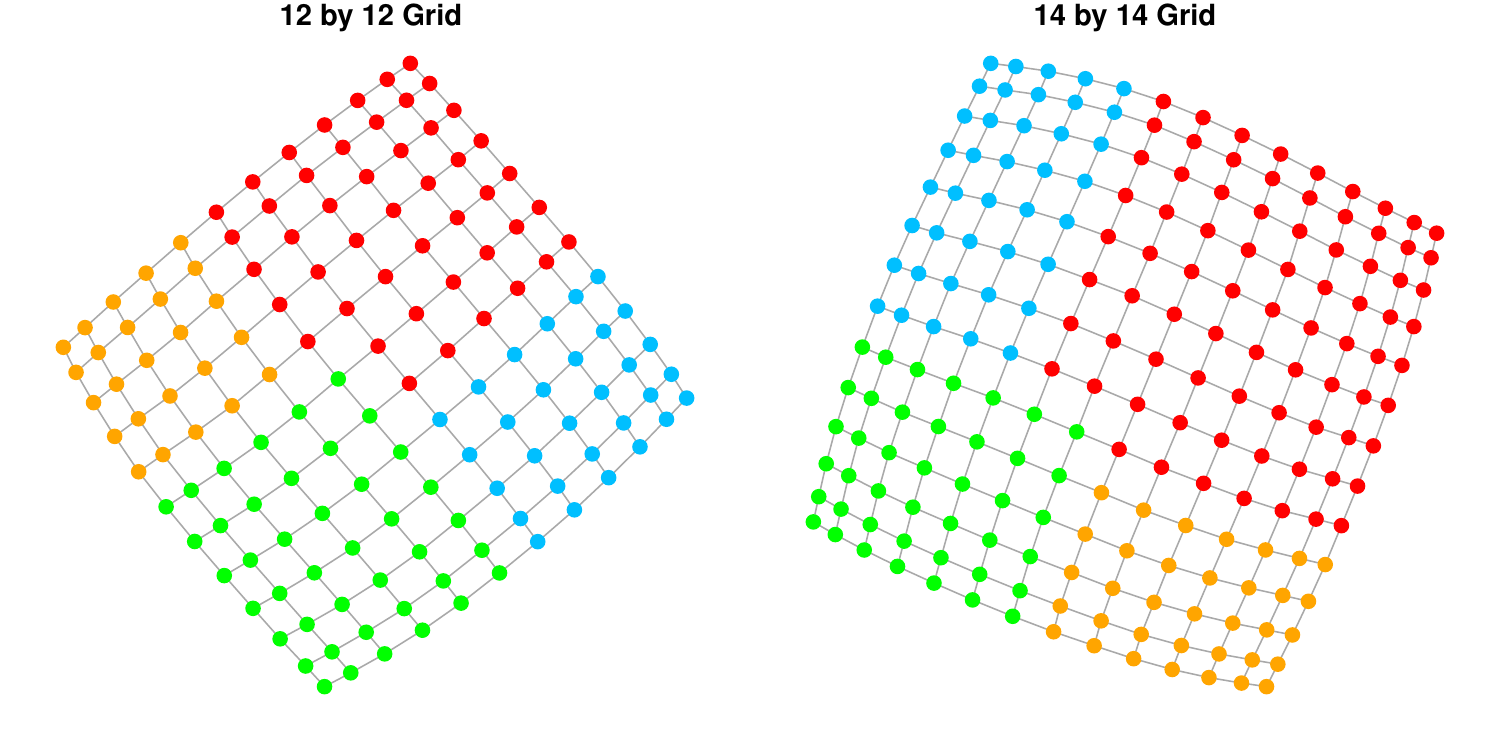}
\end{center}
\caption{Grid graphs with four unbalanced clusters. The nodal colors are ground truth cluster labels. No clusters of nodes are distinguishable from the graph topology alone, necessitating the use of nodal attributes for clustering.}
\label{fig_grid}
\end{figure}

The grid graphs have four unbalanced regional clusters: top left $\mathcal{C}_1$, top right $\mathcal{C}_2$, bottom left $\mathcal{C}_3$, and bottom right $\mathcal{C}_4$. The edge density for the $12 \times 12$ grid is about $2|\mathcal{E}||\mathcal{V}|^{-2} \approx 0.026$, and that for the $14 \times 14$ grid decreases to about $2|\mathcal{E}||\mathcal{V}|^{-2} \approx 0.019$ --- these graphs are sparse. The unbalanced cluster sizes for $N=144$ are $25$, $30$, $42$, and $47$, and those for $N=196$ are $35$, $40$, $42$, and $79$. Figure \ref{fig_grid} illustrates the simulated grids.

For node $i$ in cluster $\mathcal{C}_k$, its time series $\bm{Y}_{i} = \{ Y_{t,i} \}_{t=1}^{n}$ is generated via a cluster-specific AR model of order one:
\begin{equation*}
Y_{t,i} = \mu_{k} + \psi_{k} ( Y_{t-1,i} - \mu_{k} ) + \varepsilon_{t,i}, \qquad t = 2, \ldots, n, \qquad i \in \mathcal{C}_k,
\end{equation*}
where $\mu_{k}$ is the mean of cluster $\mathcal{C}_k$, $\psi_{k} \in (-1,1)$ is the AR coefficient, and $\varepsilon_{t,i} \sim \mathcal{N}(0, \sigma_k^2)$ is Gaussian white noise. We choose $\psi_{k} = 0.5$ and $\sigma_{k}^2 = 1$ for all four clusters. The process is rendered stationary by taking the initial state
\begin{equation*}
Y_{1,i} \sim \mathcal{N} \left( \mu_{k}, \frac{\sigma_k^2}{1-\psi_k^2} \right), \qquad i \in \mathcal{C}_k.
\label{initial_state}
\end{equation*}
To incorporate regional heterogeneity, the four cluster means are taken as $\mu_1 = -0.75$, $\mu_2 = 0.0$, $\mu_3 = 0.75$, and $\mu_4 = 1.5$. Table \ref{tab:sim1} reports the sample means and standard deviations of the evaluation metrics.

The grid structure does not reveal cluster boundaries, making clustering based on the graph topology alone challenging. In this setting, cluster information resides in the nodal attributes. The hybrid methods, CASC, SDP, and NAC, which rely on both the node adjacency matrix and nodal attributes perform poorly, suggesting that an uninformative spatial structure hinders clustering even when the nodal features exhibit cluster differences. Increasing the graph size $N$ from $144$ to $196$ does not improve their performances, and can even degrade matters. Enlarging the grid jeopardizes graph-based approaches that emphasize structural cues. The SCORE method, which does not utilize nodal features for clustering, performs better than most hybrid methods. The error here stems from the method's tendency to produce four equal-sized squared regions, making some boundary nodes incorrectly assigned. The two graph neural network methods, DGI and GraphMAE, perform well due to their flexible ability to learn meaningful nodal representations. Overall, the proposed GFL method outperforms all competitor methods.

\begin{table*}[!ht]
\caption{Sample means (one standard deviation) of the evaluation metrics. The best metric's score is bolded.}
\label{tab:sim1}
\centering
\resizebox{0.85\textwidth}{!}{%
\begin{tabular}{cccccccc}
\toprule
$N$ & Method & NMI $\uparrow$ & ARI $\uparrow$ & ACC $\uparrow$ & HOM $\uparrow$ & COM $\uparrow$ & PUR $\uparrow$\\
\midrule
\multirow{10}{2em}{$144$} 
& GFL & $\bm{96.99}\% (0.05)$ & $\bm{96.62}\% (0.06)$ & $\bm{98.33}\% (0.04)$ & $\bm{96.52}\% (0.06)$ & $\bm{97.61}\% (0.04)$ & $\bm{99.17}\% (0.01)$ \\
& $k$-means & $71.62\% (0.11)$ & $64.26\% (0.20)$ & $76.56\% (0.21)$ & $73.09\% (0.11)$ & $71.92\% (0.12)$ & $83.25\% (0.18)$ \\
& DTW & $52.44\% (0.04)$ & $39.74\% (0.09)$ & $59.69\% (0.15)$ & $53.47\% (0.10)$ & $53.29\% (0.07)$ & $71.86\% (0.15)$ \\
& CASC & $34.06\% (0.03)$ & $23.36\% (0.05)$ & $28.69\% (0.05)$ & $36.67\% (0.13)$ & $33.13\% (0.02)$ & $45.28\% (0.09)$ \\
& SDP & $29.17\% (0.05)$ & $17.74\% (0.05)$ & $37.08\% (0.12)$ & $31.70\% (0.15)$ & $28.30\% (0.05)$ & $49.36\% (0.10)$ \\
& NAC & $32.12\% (0.09)$ & $21.62\% (0.08)$ & $44.75\% (0.13)$ & $34.55\% (0.16)$ & $31.35\% (0.09)$ & $53.78\% (0.12)$ \\
& SCORE & $74.29\% (0.06)$ & $68.16\% (0.14)$ & $84.78\% (0.17)$ & $77.40\% (0.05)$ & $72.86\% (0.09)$ & $84.78\% (0.17)$ \\
& VGAE	& $55.26\% (0.10)$	&$46.07\% (0.13)$&	$71.28\% (0.10)$	&$55.13\% (0.10)$&	$55.40\% (0.10)$	&$73.00\% (0.09)$ \\
& DGI	&$83.08\%(0.05)$	&$83.08\%(0.06)$	&$93.56\%(0.03)$	&$82.89\%(0.05)$	&$83.27\%(0.05)$	&$93.56\%(0.03)$ \\
& GraphMAE	&$80.28\%(0.04)$	&$79.30\%(0.05)$	&$92.11\%(0.02)$	&$79.82\%(0.04)$	&$80.76\%(0.04)$	&$92.11\%(0.02)$\\
\midrule
\multirow{10}{2em}{$196$} 
& GFL & $\bm{98.40}\% (0.04)$ & $\bm{97.95}\% (0.05)$ & $\bm{98.20}\% (0.06)$ & $\bm{97.65}\% (0.06)$ & $\bm{99.37}\% (0.01)$ & $\bm{99.84}\% (0.00)$ \\
& $k$-means & $80.40\% (0.09)$ & $79.92\% (0.17)$ & $89.84\% (0.18)$ & $83.22\% (0.06)$ & $79.38\% (0.12)$ & $89.84\% (0.18)$ \\
& DTW & $53.30\% (0.06)$ & $42.29\% (0.12)$ & $62.51\% (0.19)$ & $55.24\% (0.11)$ & $53.29\% (0.08)$ & $71.84\% (0.16)$ \\
& CASC & $29.27\% (0.03)$ & $20.43\% (0.04)$ & $28.96\% (0.06)$ & $32.21\% (0.14)$ & $28.14\% (0.02)$ & $44.51\% (0.09)$ \\
& SDP & $29.73\% (0.04)$ & $19.10\% (0.05)$ & $28.92\% (0.07)$ & $32.67\% (0.14)$ & $28.60\% (0.03)$ & $48.45\% (0.10)$ \\
& NAC & $26.83\% (0.09)$ & $17.98\% (0.08)$ & $36.10\% (0.15)$ & $29.63\% (0.17)$ & $25.83\% (0.08)$ & $50.04\% (0.12)$ \\
& SCORE & $65.64\% (0.05)$ & $54.75\% (0.11)$ & $78.94\% (0.16)$ & $69.43\% (0.06)$ & $63.70\% (0.08)$ & $78.94\% (0.16)$ \\
& VGAE	& $53.12\%(0.09)$	& $39.79\%(0.10)$	& $67.88\%(0.08)$	& $53.20\%(0.09)$	& $53.06\%(0.08)$	& $70.02\%(0.07)$\\
& DGI	& $84.46\%(0.04)$	& $85.65\%(0.03)$	& $94.35\%(0.02)$	& $84.31\%(0.04)$	& $84.62\%(0.04)$	& $94.35\%(0.02)$\\
& GraphMAE	& $82.72\%(0.04)$	& $83.05\%(0.04)$	& $93.59\%(0.02)$	& $82.62\%(0.04)$	& $82.81\%(0.04)$	& $93.59\%(0.02)$\\
\bottomrule
\end{tabular}
}
\end{table*}

\subsection*{Scenario 2}

Our next scenario simulates graphs with block structures. Specifically, for the size $N \coloneqq |\mathcal{V}|$, a probability matrix $\bm{P} \in [0,1]^{N \times N}$ is first defined as
\begin{align*}
\bm{P}_{i,j} =
\begin{cases}
0.30, &\ i,j \in \mathcal{C}_k,\ k \in \{1,2,3\},\\
0.15, &\ \text{otherwise,}
\end{cases}  
\end{align*}
where $\mathcal{C}_1, \mathcal{C}_2, \mathcal{C}_3$ are three clusters that partition $\mathcal{V}$. Then edges are sampled via an adjacency matrix format as $\bm{A}_{i,j} \sim \text{Bernoulli}({\bm P}_{i,j})$. The edge density is roughly $2 |\mathcal{E}||\mathcal{V}|^{-2} \approx 0.4$, larger than that of a grid graph. For $N=120$ nodes, the three unbalanced cluster sizes are taken as $30$, $40$ and $50$; for $N=210$, the cluster sizes are $60$, $70$, and $80$. The nodal attributes are generated as in Scenario 1. We choose $\psi_{k} = 0.5$ and $\sigma_{k}^2 = 1$ for all three clusters. Cluster means are $\mu_1 = -0.9$, $\mu_2 = 0.0$, and $\mu_3 = 0.9$. Table \ref{tab:sim2} reports sample means and standard deviations of our clustering evaluation metrics.

Scenario 2 provides cluster information through both graph connectivity and nodal attributes. With $120$ nodes, $k$-means applied directly to the series performs reasonably well, since the clusters are well separated via their series means. The graph neural network methods DGI and GraphMAE perform worse, suggesting that their learned nodal representations may not fully capture the available cluster information. Graph-based methods such as CASC, SDP, NAC, and SCORE also perform poorly, reflecting the challenges posed by the limited graph size and unbalanced cluster sizes. As the graph size increases from $N=120$ to $N = 210$, these methods generally improve, with CASC, SCORE, and VGAE showing substantial gains. This improvement is expected since more data is available for distinguishing clusters. Nevertheless, the proposed GFL method consistently outperforms all competing methods, demonstrating its ability to effectively integrate structural and attribute information, even when the graph size is limited.

\begin{table}[!ht]
\caption{Sample means (one standard deviation) of the evaluation metrics. The best metric's score is bolded.}
\label{tab:sim2}
\centering
\resizebox{0.85\columnwidth}{!}{%
\begin{tabular}{cccccccc}
\toprule
$N$ & Method & NMI $\uparrow$ & ARI $\uparrow$ & ACC $\uparrow$ & HOM $\uparrow$ & COM $\uparrow$ & PUR $\uparrow$\\
\midrule
\multirow{10}{2em}{$120$} 
& GFL & $\bm{92.67}\% (0.07)$ & $\bm{92.77}\% (0.10)$ & $\bm{95.77}\% (0.08)$ & $\bm{91.16}\% (0.12)$ & $\bm{95.15}\% (0.03)$ & $\bm{98.77}\% (0.01)$ \\
& $k$-means & $87.54\% (0.11)$ & $88.24\% (0.18)$ & $93.47\% (0.19)$ & $90.11\% (0.05)$ & $86.92\% (0.14)$ & $93.47\% (0.19)$ \\
& DTW & $53.49\% (0.09)$ & $47.61\% (0.16)$ & $65.90\% (0.19)$ & $52.04\% (0.15)$ & $58.26\% (0.10)$ & $81.30\% (0.17)$ \\
& CASC & $27.10\% (0.08)$ & $24.39\% (0.09)$ & $45.07\% (0.13)$ & $29.78\% (0.16)$ & $26.38\% (0.07)$ & $59.03\% (0.13)$ \\
& SDP & $31.05\% (0.14)$ & $25.18\% (0.13)$ & $45.90\% (0.15)$ & $33.76\% (0.19)$ & $30.29\% (0.14)$ & $57.97\% (0.14)$ \\
& NAC & $33.67\% (0.08)$ & $29.02\% (0.10)$ & $52.80\% (0.18)$ & $36.45\% (0.15)$ & $32.85\% (0.08)$ & $62.20\% (0.14)$ \\
& SCORE & $34.63\% (0.10)$ & $35.13\% (0.15)$ & $61.73\% (0.20)$ & $37.21\% (0.16)$ & $34.03\% (0.10)$ & $67.77\% (0.16)$ \\
& VGAE	& $ 50.80\%(0.15)$	& $ 55.36\%(0.17)$	& $ 80.67\%(0.10)$	& $ 50.66\%(0.15)$	& $ 50.95\%(0.15)$	& $ 81.03\%(0.10)$\\
& DGI	& $ 45.05\%(0.05)$	& $ 38.63\%(0.07)$	& $ 69.37\%(0.07)$	& $ 45.16\%(0.05)$	& $ 44.94\%(0.05)$	& $ 70.37\%(0.05)$\\
& GraphMAE	& $ 52.81\%(0.09)$	& $ 50.60\%(0.14)$	& $ 74.13\%(0.12)$	& $ 52.19\%(0.09)$	& $ 53.51\%(0.08)$	& $ 76.23\%(0.09)$\\
\midrule
\multirow{10}{2em}{$210$}
& GFL & $\bm{96.11}\% (0.06)$ & $\bm{95.74}\% (0.10)$ & $\bm{97.30}\% (0.08)$ & $\bm{94.91}\% (0.10)$ & $\bm{98.07}\% (0.02)$ & $\bm{99.58}\% (0.01)$ \\
& $k$-means & $88.49\% (0.12)$ & $89.76\% (0.19)$ & $93.92\% (0.19)$ & $91.14\% (0.05)$ & $87.95\% (0.14)$ & $93.92\% (0.19)$ \\
& DTW & $68.96\% (0.10)$ & $69.69\% (0.17)$ & $85.58\% (0.18)$ & $71.00\% (0.10)$ & $69.20\% (0.11)$ & $86.25\% (0.18)$ \\
& CASC & $80.07\% (0.12)$ & $83.52\% (0.18)$ & $91.64\% (0.19)$ & $82.75\% (0.08)$ & $79.49\% (0.14)$ & $91.64\% (0.19)$ \\
& SDP & $40.16\% (0.05)$ & $36.32\% (0.10)$ & $59.30\% (0.16)$ & $42.80\% (0.13)$ & $39.62\% (0.06)$ & $64.32\% (0.13)$ \\
& NAC & $42.56\% (0.08)$ & $41.38\% (0.13)$ & $69.05\% (0.17)$ & $45.22\% (0.14)$ & $41.99\% (0.09)$ & $70.21\% (0.16)$ \\
& SCORE & $73.30\% (0.10)$ & $77.83\% (0.17)$ & $89.56\% (0.18)$ & $75.88\% (0.08)$ & $72.84\% (0.12)$ & $89.56\% (0.18)$ \\
& VGAE	& $83.64\%(0.05)$	& $88.22\%(0.04)$	& $95.85\%(0.01)$	& $83.57\%(0.05)$	& $83.71\%(0.05)$	& $95.85\%(0.01)$\\
& DGI	& $56.07\%(0.05)$	& $54.95\%(0.08)$	& $80.86\%(0.05)$	& $56.01\%(0.05)$	& $56.13\%(0.05)$	& $80.86\%(0.05)$\\
& GraphMAE	& $71.01\%(0.08)$	& $72.37\%(0.10)$	& $88.44\%(0.08)$	& $70.49\%(0.08)$	& $71.56\%(0.08)$	& $88.84\%(0.06)$\\
\bottomrule
\end{tabular}
}
\vskip -0.1in
\end{table}

\subsection*{Scenario 3}

Scenario 3 extends the block graphs in Scenario 2 by adding correlation across series residing in the same cluster. Here, a VAR model of order one is applied to incorporate cluster-specific correlations. The $N$-dimensional time series $\{ {{\bm Y}_{t}} \}_{t = 1}^{n}$ are generated simultaneously over all $N$ nodes via
\[
{\bm Y}_{t} = {\bm \beta} + {\Phi}({\bm Y}_{t-1} - {\bm \beta}) + {\bm \xi}_{t}, \qquad t = 2, \ldots, n.
\]
Here, $\bm \beta = (\beta_1, \beta_2, \ldots, \beta_N)^\top$ is the nodal mean vector with $\beta_i = \mu_k$ for $i \in \mathcal{C}_k$. The error $\bm{\xi}_t$ is sampled as $\mathcal{N}({\bm 0}_N, \bm{\Sigma})$ for each $t$, where the covariance matrix $\bm \Sigma=\{\Sigma_{i,j} \}_{i,j=1, \ldots, N}$ with cluster-specific correlations is defined by
\[
\Sigma_{i,j} = 
\begin{cases}
    1 & i = j, \\
    \rho & i \neq j \text{ and }  i,j\in \mathcal{C}_k, \ k \in [3], \\
    0 & \text{otherwise}.
\end{cases}
\]
We take $\rho=0.15$, which induces moderate correlation. The VAR coefficient is taken as $\Phi = 0.5 \bm{I}_N$ and the means for the three clusters are repeated from Scenario 2. Our simulations start with ${\bm Y}_1 = {\bm 0}$ and burn-in is used for $100$ iterations to ensure stationary nodal series. Compared to Scenario 2, the series means are not as well separated, making clustering more difficult. Table \ref{tab:sim3} reports the means and standard deviations of our evaluation metrics. 

With $120$ nodes, $k$-means, CASC, SDP, and NAC exhibit good performance. In particular, CASC, which leverages nodal covariates via pair-wise similarity, benefits from the positive intra-cluster correlations. In contrast, the graph neural network methods DGI and GraphMAE show moderate performance, suggesting that learning informative nodal representations may be challenging with a relatively small graph. When the number of nodes increases to $210$, competitors improve significantly (more than in Scenario 2), exploiting the correlation between nodes to enhance clustering. Overall, the proposed GFL method achieves strong clustering performance in the presence of positive inter-nodal correlations. Although its performance declines slightly at $N=210$, GFL remains competitive, demonstrating its effectiveness in leveraging graph connectivity and nodal attributes under inter-nodal dependence.

\begin{table}[!ht]
\caption{Sample means (one standard deviation) of the evaluation metrics. The best metric's score is bolded.}
\label{tab:sim3}
\centering
\resizebox{0.85\columnwidth}{!}{%
\begin{tabular}{cccccccc}
\toprule
$N$ & Method & NMI $\uparrow$ & ARI $\uparrow$ & ACC $\uparrow$ & HOM $\uparrow$ & COM $\uparrow$ & PUR $\uparrow$\\
\midrule
\multirow{10}{2em}{$120$} 
& GFL & $\bm{94.41}\% (0.08)$ & $\bm{93.44}\% (0.13)$ & $\bm{95.83}\% (0.09)$ & $92.89\% (0.13)$ & $\bm{97.24}\% (0.04)$ & $\bm{99.17}\% (0.01)$ \\
& $k$-means & $89.47\% (0.19)$ & $85.75\% (0.28)$ & $88.07\% (0.25)$ & $90.67\% (0.19)$ & $90.81\% (0.18)$ & $94.03\% (0.19)$ \\
& DTW & $57.71\% (0.12)$ & $50.26\% (0.20)$ & $66.33\% (0.20)$ & $55.94\% (0.17)$ & $63.06\% (0.13)$ & $82.67\% (0.17)$ \\
& CASC & $91.70\% (0.12)$ & $91.75\% (0.19)$ & $94.63\% (0.19)$ & $\bm{94.29}\% (0.05)$ & $91.07\% (0.15)$ & $94.63\% (0.19)$ \\
& SDP & $84.20\% (0.17)$ & $83.33\% (0.24)$ & $88.20\% (0.23)$ & $86.72\% (0.15)$ & $83.65\% (0.19)$ & $91.07\% (0.20)$ \\
& NAC & $76.78\% (0.11)$ & $79.23\% (0.18)$ & $90.10\% (0.18)$ & $79.59\% (0.09)$ & $75.94\% (0.13)$ & $90.10\% (0.18)$ \\
& SCORE & $35.66\% (0.11)$ & $35.59\% (0.16)$ & $60.97\% (0.21)$ & $38.30\% (0.17)$ & $34.97\% (0.12)$ & $68.20\% (0.16)$ \\
& VGAE	& $65.81\%(0.09)$	& $71.99\%(0.09)$	& $89.57\%(0.04)$	& $65.71\%(0.09)$	& $65.92\%(0.09)$	& $89.57\%(0.04)$\\
& DGI	& $54.83\%(0.09)$	& $54.62\%(0.13)$	& $80.03\%(0.09)$	& $54.54\%(0.09)$	& $55.14\%(0.09)$	& $80.63\%(0.08)$\\
& GraphMAE	& $68.19\%(0.08)$	& $68.24\%(0.10)$	& $85.37\%(0.07)$	& $67.44\%(0.09)$	& $69.07\%(0.07)$	& $85.47\%(0.07)$\\
\midrule
\multirow{10}{2em}{$210$} 
& GFL & $92.42\% (0.11)$ & $88.99\% (0.17)$ & $91.83\% (0.13)$ & $88.28\% (0.17)$ & $\bm{99.23}\% (0.02)$ & $\bm{99.83}\% (0.00)$ \\
& $k$-means & $\bm{96.31}\% (0.13)$ & $\bm{95.33}\% (0.19)$ & $95.83\% (0.19)$ & $\bm{98.94}\% (0.02)$ & $95.78\% (0.15)$ & $95.83\% (0.19)$ \\
& DTW & $72.38\% (0.12)$ & $72.59\% (0.18)$ & $86.51\% (0.18)$ & $74.61\% (0.11)$ & $72.28\% (0.14)$ & $87.09\% (0.18)$ \\
& CASC & $91.36\% (0.13)$ & $91.87\% (0.19)$ & $94.63\% (0.19)$ & $94.02\% (0.05)$ & $90.81\% (0.15)$ & $94.63\% (0.19)$ \\
& SDP & $89.90\% (0.12)$ & $90.83\% (0.19)$ & $94.29\% (0.19)$ & $92.56\% (0.04)$ & $89.35\% (0.15)$ & $94.29\% (0.19)$ \\
& NAC & $89.80\% (0.12)$ & $90.94\% (0.19)$ & $94.32\% (0.19)$ & $92.50\% (0.04)$ & $89.21\% (0.15)$ & $94.32\% (0.19)$ \\
& SCORE & $76.32\% (0.11)$ & $80.27\% (0.18)$ & $90.42\% (0.18)$ & $78.96\% (0.08)$ & $75.79\% (0.13)$ & $90.42\% (0.18)$ \\
& VGAE	& $92.17\%(0.05)$	& $94.84\%(0.03)$	& $\bm{98.23}\%(0.01)$	& $92.18\%(0.05)$	& $92.16\%(0.05)$	& $98.23\%(0.01)$\\
& DGI	& $78.46\%(0.05)$	& $82.48\%(0.05)$	& $93.79\%(0.02)$	& $78.41\%(0.05)$	& $78.52\%(0.05)$	& $93.79\%(0.02)$\\
& GraphMAE	& $81.75\%(0.09)$	& $83.40\%(0.10)$	& $93.35\%(0.05)$	& $81.46\%(0.09)$	& $82.04\%(0.09)$	& $93.35\%(0.05)$\\
\bottomrule
\end{tabular}
}
\vskip -0.1in
\end{table}

\subsection{A California County Climate Network}
\label{sec_temperature}




California contains both the highest and lowest elevations in the contiguous United States. Its coastal proximity to the Pacific Ocean, neighboring desert environments, and broad latitudinal span lead to significant temperature heterogeneity, making clustering its counties a challenging problem. There are $N = 58$ counties in the state, all of which report data. Monthly average temperatures over the $14$-year period January 2011 - December 2024 are studied here with $n=168$. The data were extracted from the National Centers for Environmental Information.

A network with $58$ nodes representing all counties was generated; two counties are declared connected by an edge if they share common border.  The goal here is to utilize both spatial proximity of the counties and their temperatures to identify useful climate regions within the state. Recent clusterings (by station, not county) of California's climate appear in \cite{abatzoglou2009classification}. While we examine the recent $14$-year record to mitigate the impact of non-stationarity on results, California climate trend studies appear in \cite{ladochy2007recent} and \cite{walton2018assessment}. 

\begin{figure*}[!ht]
\centering
\includegraphics[width=0.45\linewidth]{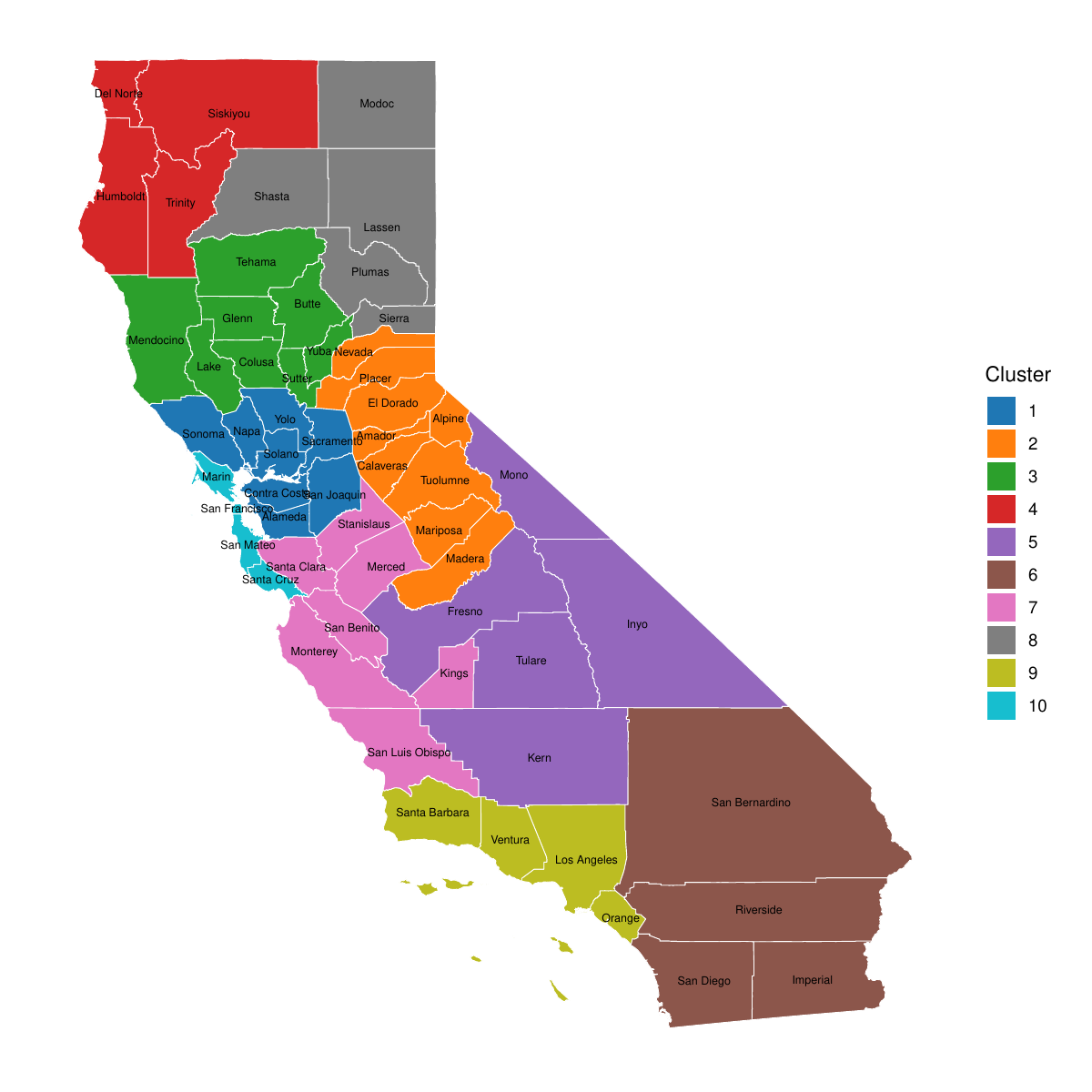}
\caption{California county clustering from GFL.}
\label{fig_experiment_CA}
\end{figure*}

Some pre-processing was implemented to obtain stationary series. Specifically, with $\{ Y_t \}_{t=1}^n$ denoting a county's monthly temperature, we seasonally convert to a standardized measurement by examining 
\[
S_{kT+\nu} \coloneqq \frac{Y_{kT+\nu}-\hat{\mu}_\nu}{ \hat{\sigma}_\nu},
\]
where $k \in \{ 0, 1, \ldots , n_{\rm yr}-1 \}$ and $\nu \in \{ 1, 2, \ldots , T \}$ is the calendar month of the year. Here, $n_{\rm yr}=14$ and $T=12$ is the number of calendar months. Estimates of the month $\nu$ mean $\mu_\nu$ and standard deviation $\sigma_\nu$ are
\[
\hat{\mu}_\nu = \frac{1}{n_{\rm yr}} \sum_{k=0}^{n_{\rm yr}-1} Y_{kT+\nu}, 
\qquad
\hat{\sigma}_\nu^2 = \frac{1}{n_{\rm yr}-1} \sum_{k=0}^{n_{\rm yr}-1} (Y_{kT+\nu}-\hat{\mu}_\nu)^2.
\]
This standardization removes the monthly mean and variance seasonality common to temperate zone temperature records, but preserves county-level variability. Figure \ref{fig_experiment_CA} displays results from our methods and and its competitor. Ten clusters were selected by silhouette scores.

The separation between coastal and inland regions is apparent, and the latitudinal transition from north to south is noted. In particular, clusters 2 and 8 are dominated by mountainous areas, whereas Clusters 1, 3, 7, 9 are characterized by lower elevations. The Bay Area is further divided into a Delta region (Cluster 1) and coastal region (Cluster 10), suggesting that the clustering method has captured the intricate heterogeneity within the area. Moreover, Fresno, Tulare, and Kern Counties are leading agricultural counties in the United States, and they are grouped in Cluster 5 at the southern area of the San Joaquin Valley, a basin that experiences high temperatures and has irrigation. The overall clustering is appealing, placing the Desert Southeast, Mid-Coastal, Lower Coastal, Bay Area, San Joaquin Valley, Sierra Nevada Mountains, Coastal Northwest, and Mountainous Northeast counties into distinct regions. While our clustering utilizes county proximities and temperature dynamics, the clustered regions reveal both geographic and agricultural characteristics, with potential implications for resource allocation and policy planning.

\begin{figure}[!ht]
    \centering
    \begin{subfigure}{0.3\columnwidth}
        \centering
        \includegraphics[width=\linewidth]{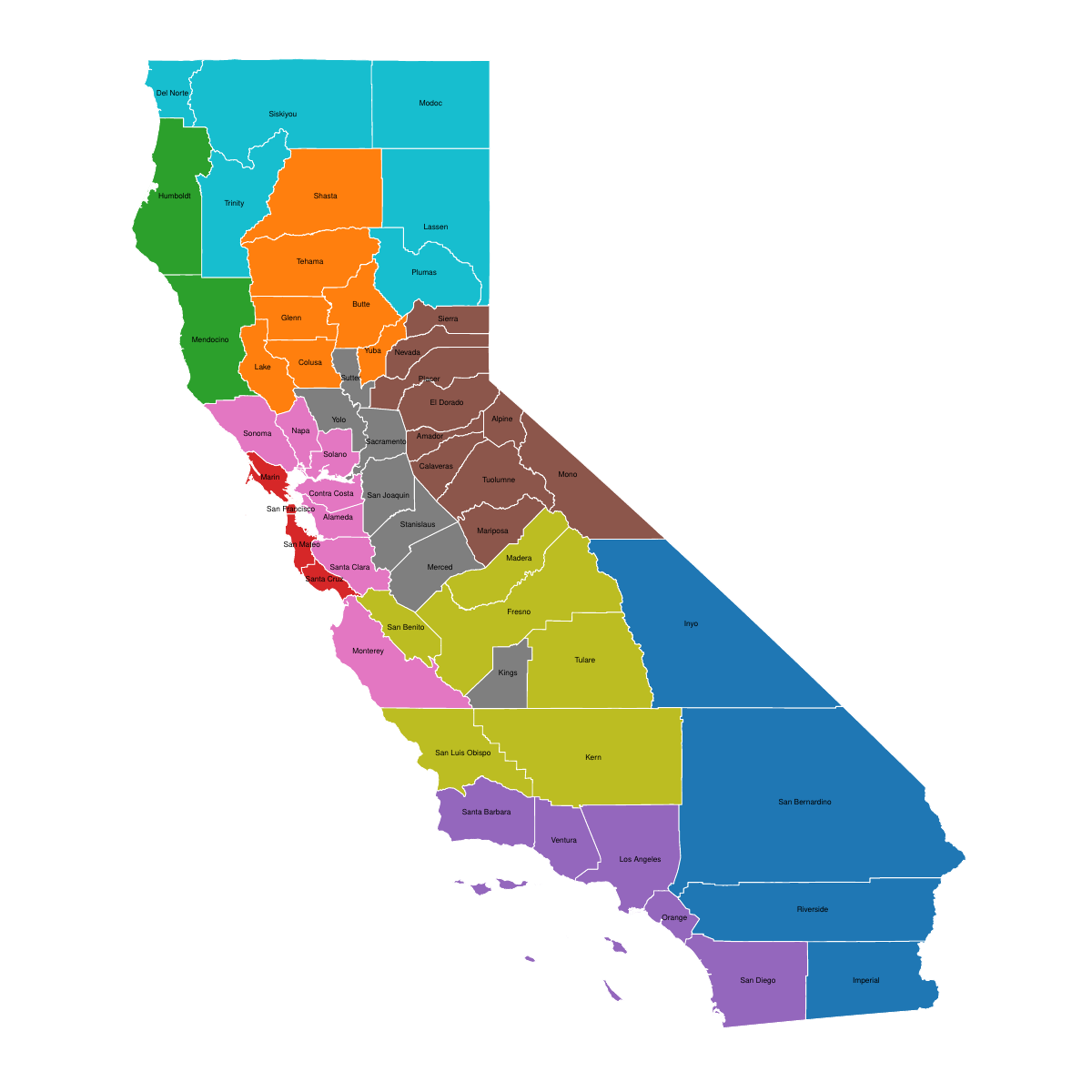}
        \caption{$k$-means}
    \end{subfigure}
    \begin{subfigure}{0.3\columnwidth}
        \centering
        \includegraphics[width=\linewidth]{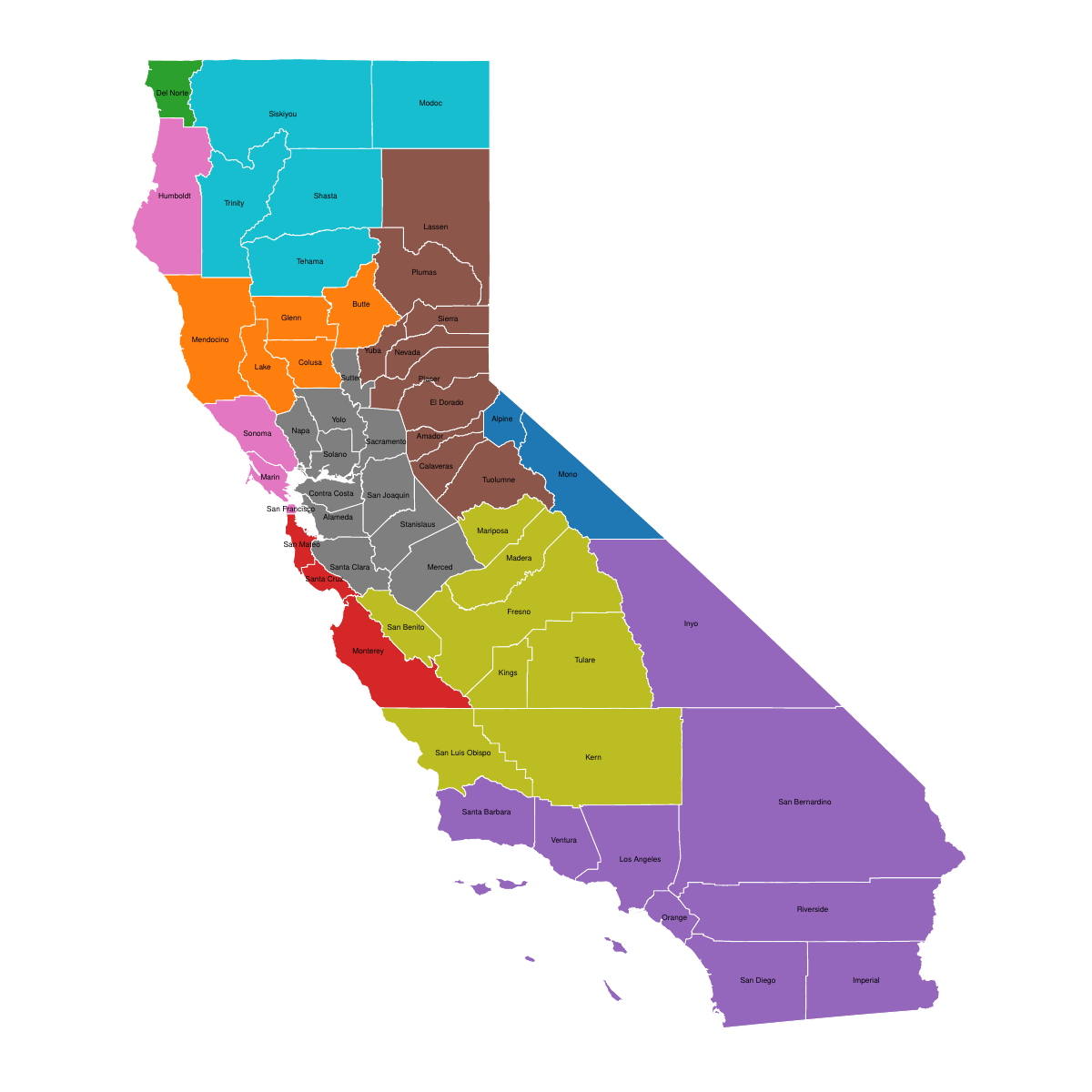}
        \caption{DTW}
    \end{subfigure}
    \begin{subfigure}{0.3\columnwidth}
        \centering
        \includegraphics[width=\linewidth]{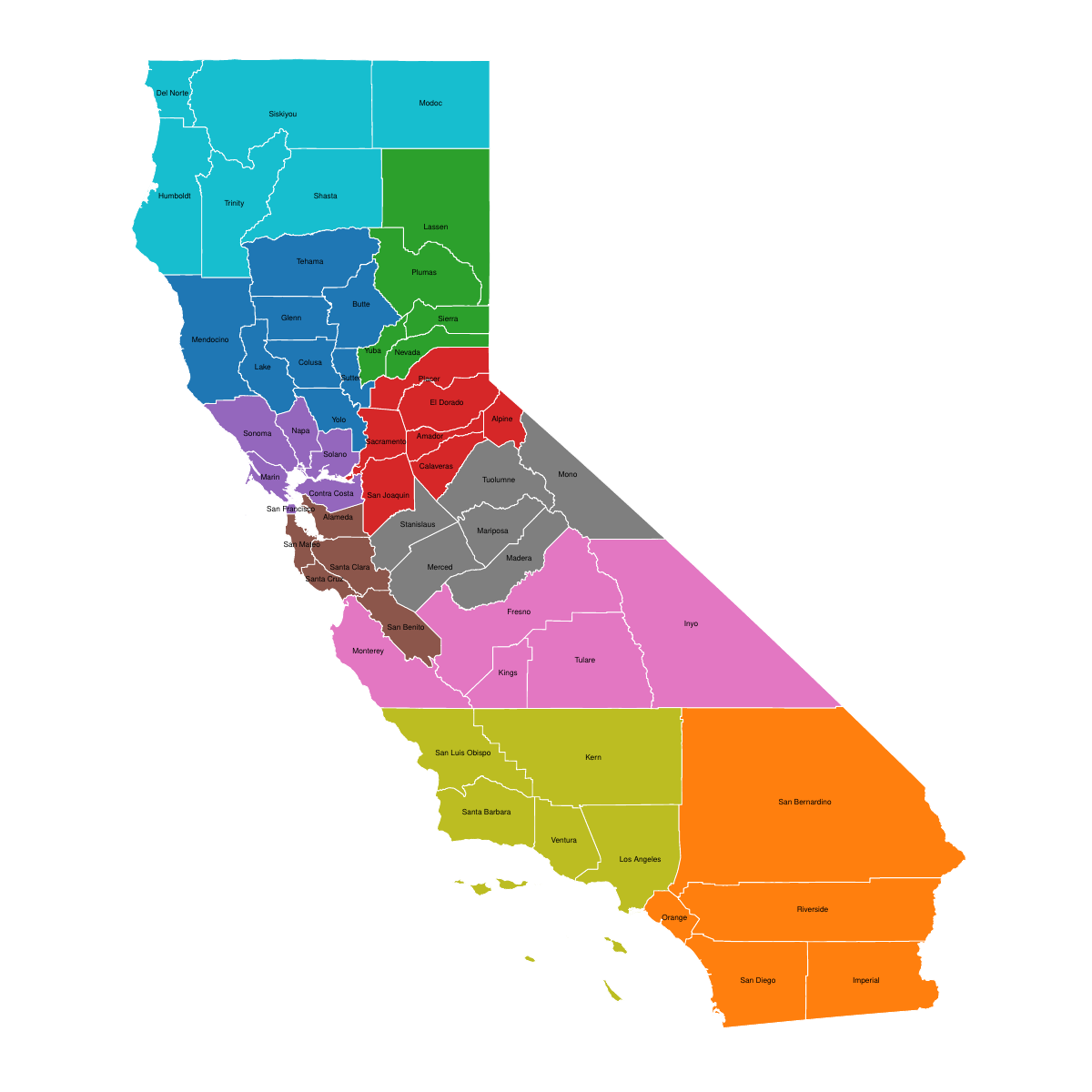}
        \caption{CASC}
    \end{subfigure}
    \begin{subfigure}{0.3\columnwidth}
        \centering
        \includegraphics[width=\linewidth]{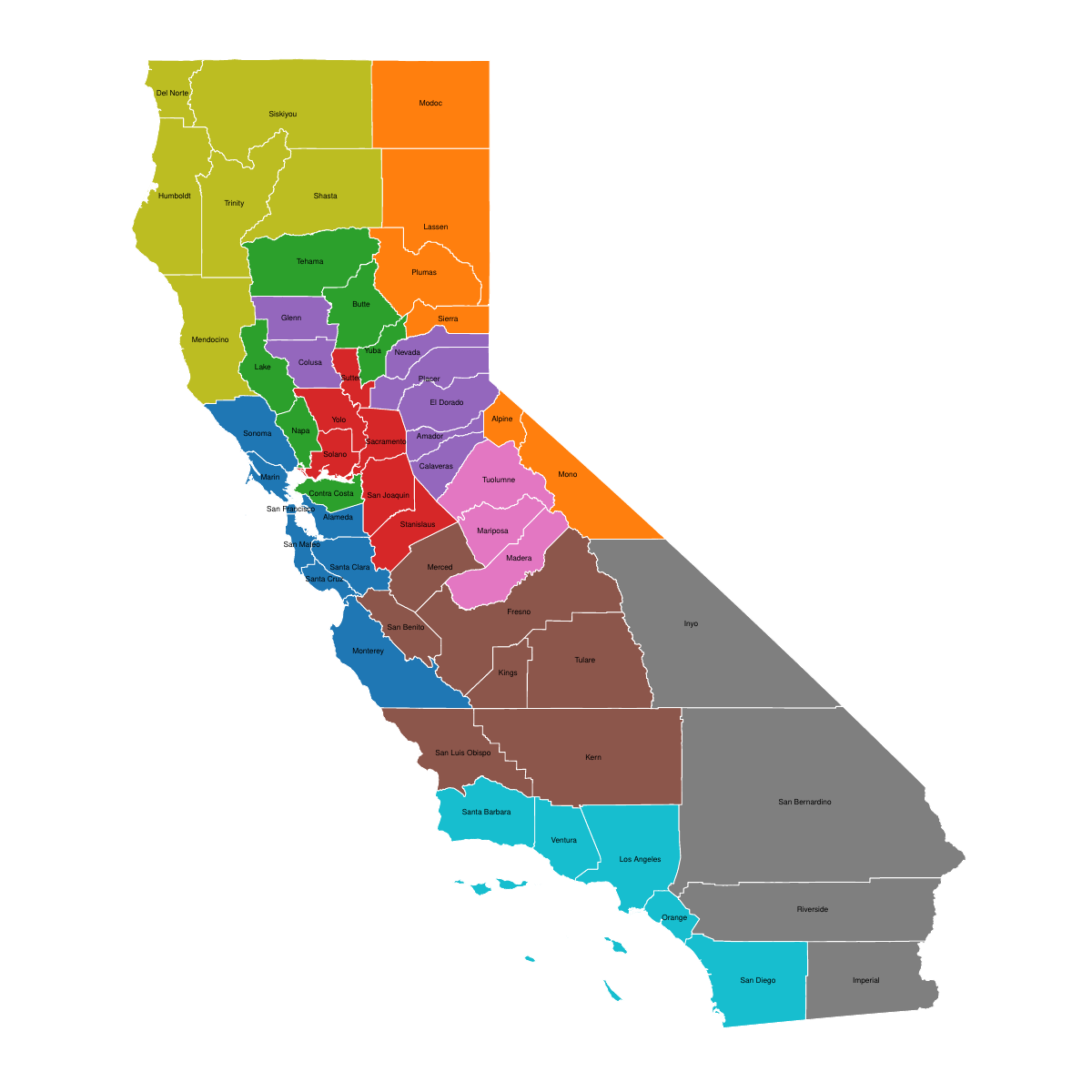}
        \caption{SDP}
    \end{subfigure}
    \begin{subfigure}{0.3\columnwidth}
        \centering
        \includegraphics[width=\linewidth]{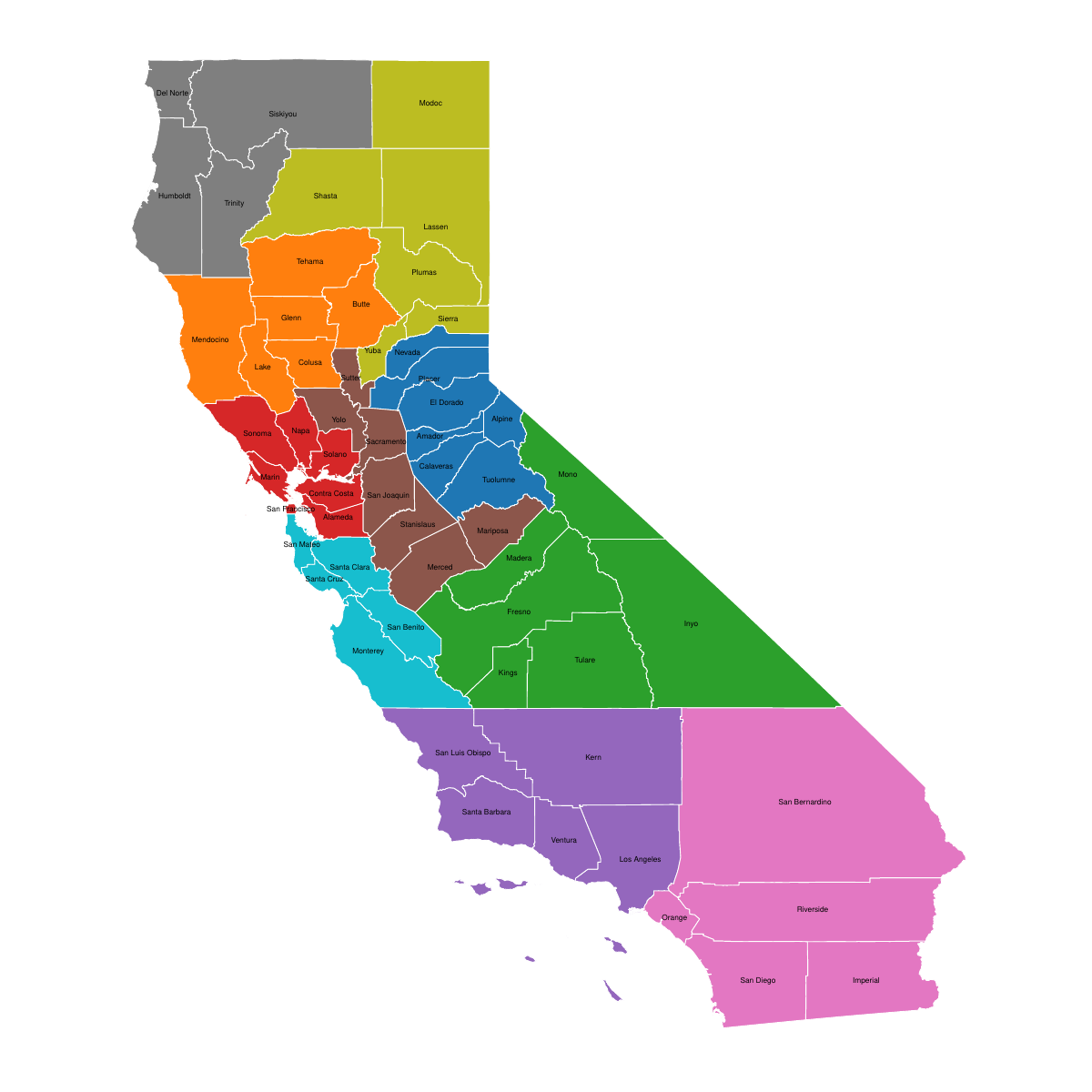}
        \caption{NAC}
    \end{subfigure}
    \begin{subfigure}{0.3\columnwidth}
        \centering
        \includegraphics[width=\linewidth]{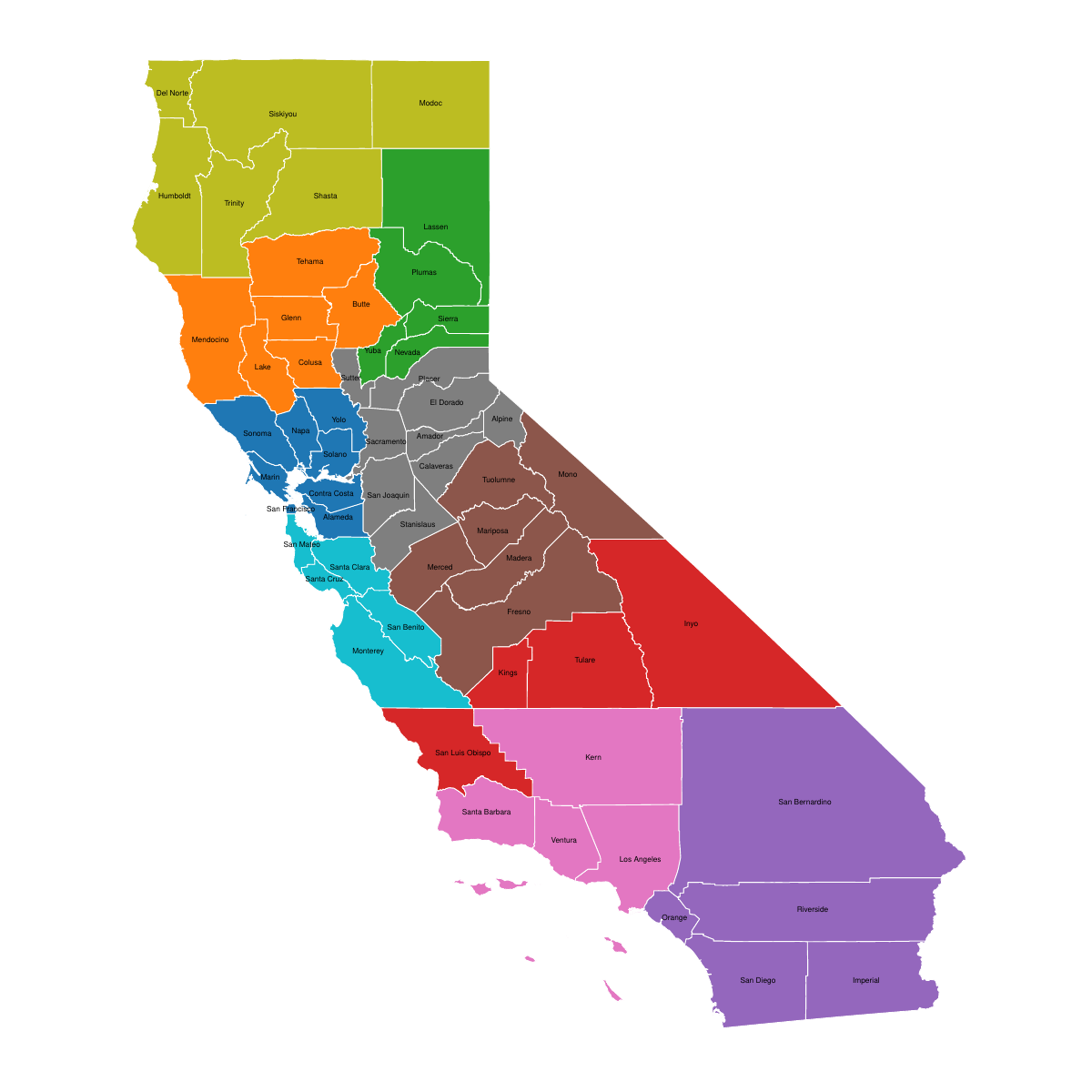}
        \caption{SCORE}
    \end{subfigure}
    \begin{subfigure}{0.3\columnwidth}
        \centering
        \includegraphics[width=\linewidth]{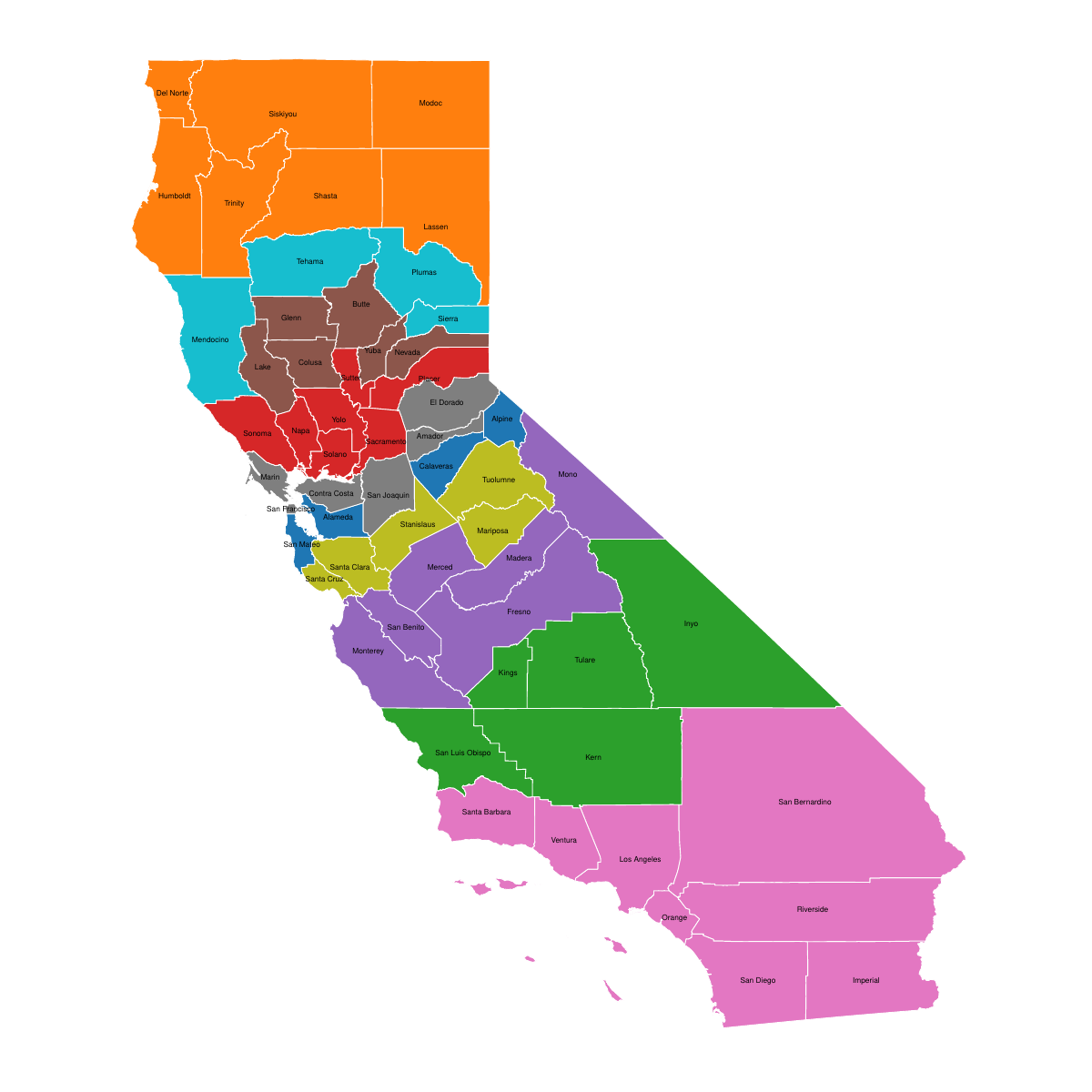}
        \caption{VGAE}
    \end{subfigure}
    \begin{subfigure}{0.3\columnwidth}
        \centering
        \includegraphics[width=\linewidth]{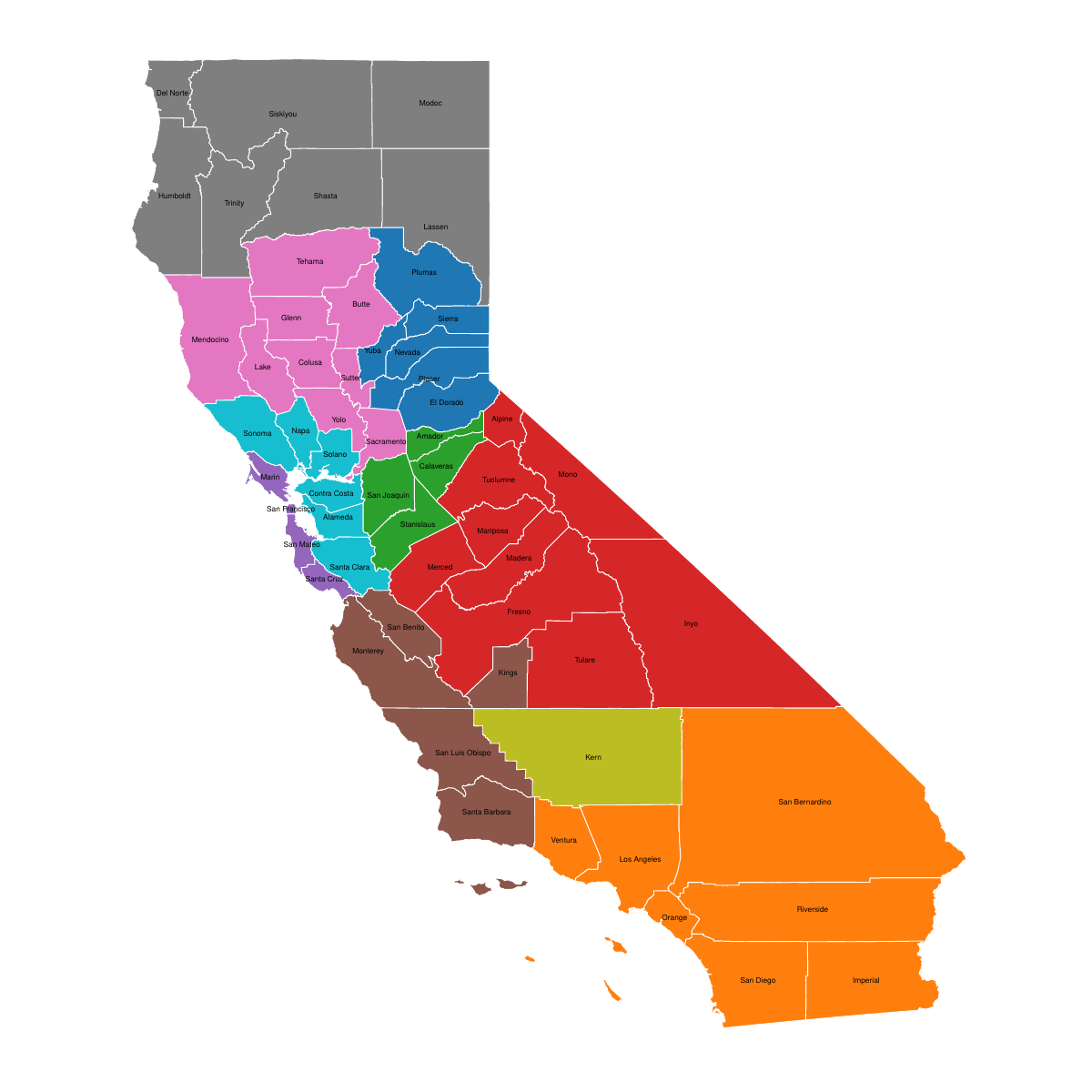}
        \caption{DGI}
    \end{subfigure}
    \begin{subfigure}{0.3\columnwidth}
        \centering
        \includegraphics[width=\linewidth]{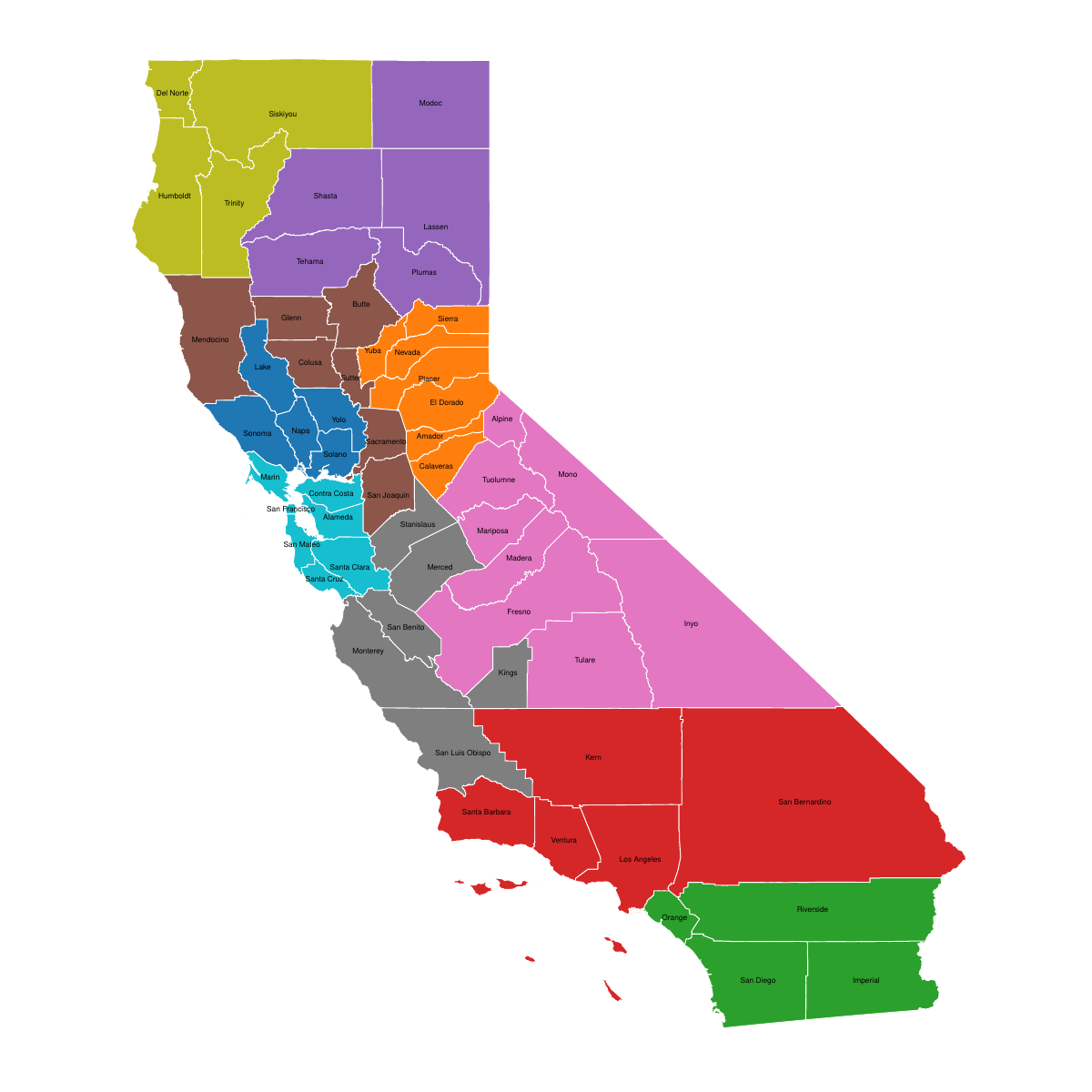}
        \caption{GraphMAE}
    \end{subfigure}
    \caption{California county clustering from competitor methods with ten clusters.}
    \label{fig_CA_map_competitor}
\end{figure}

Figure \ref{fig_CA_map_competitor} displays clusterings from competitor methods. Ten clusters are selected to be comparable with our result; if the numbers of clusters for competitors are selected by silhouette scores, fewer clusters and unrealistic results arise. For the $k$-means, DTW, and SDP methods, Inyo County in the Owens Valley is grouped with the Desert South cluster. The CASC, SCORE, and three graph neural network methods produce clusters that mix coastal and inland counties. In most methods, the coastal San Diego and Orange Counties are placed in the Desert South cluster. The Kern County in the lower San Joaquin Valley is grouped with the Lower Coastal region by the CASC, SDP, NAC, SCORE, and GraphMAE methods. For the $k$-means, CASC, SCORE, VGAE, and DGI methods, the Northeast Mountain counties mix with the Coastal Northwest counties. With understanding of California's geography, these clusterings are difficult to justify.

\subsection{Los Angeles Traffic Sensor Network}
\label{sec_traffic}

Traffic conditions in urban highway systems exhibit spatially interconnected patterns of traffic flow and congestion. Identifying groups of sensors having similar patterns can help route traffic, facilitating traffic management and urban planning. Such a clustering can be informed by observed traffic speeds and spatial relationships among sensors (nearby sensors may experience similar conditions while different highway segments may exhibit distinct patterns). To illustrate our method, we investigate the METR-LA traffic dataset \citep{jagadish2014big}, which contains traffic speed measurements collected at 5-minute intervals from sensors along the Los Angeles highway system over approximately four months.

We adopt the road network constructed in \citet{li2018dcrnn_traffic}; this  network contains $N=207$ nodes, each representing a traffic sensor. The graph is constructed based on pairwise distances between sensors, with edge weights set via a Gaussian kernel. Specifically, sensors separated by distances exceeding a threshold are not connected, whereas edge weights between nearby sensors decrease with the intrasensor distance. The weighted adjacency matrix is then binarized by setting non-zero entries to unity.

\begin{figure*}[!ht]
\centering
\includegraphics[width=0.55\linewidth]{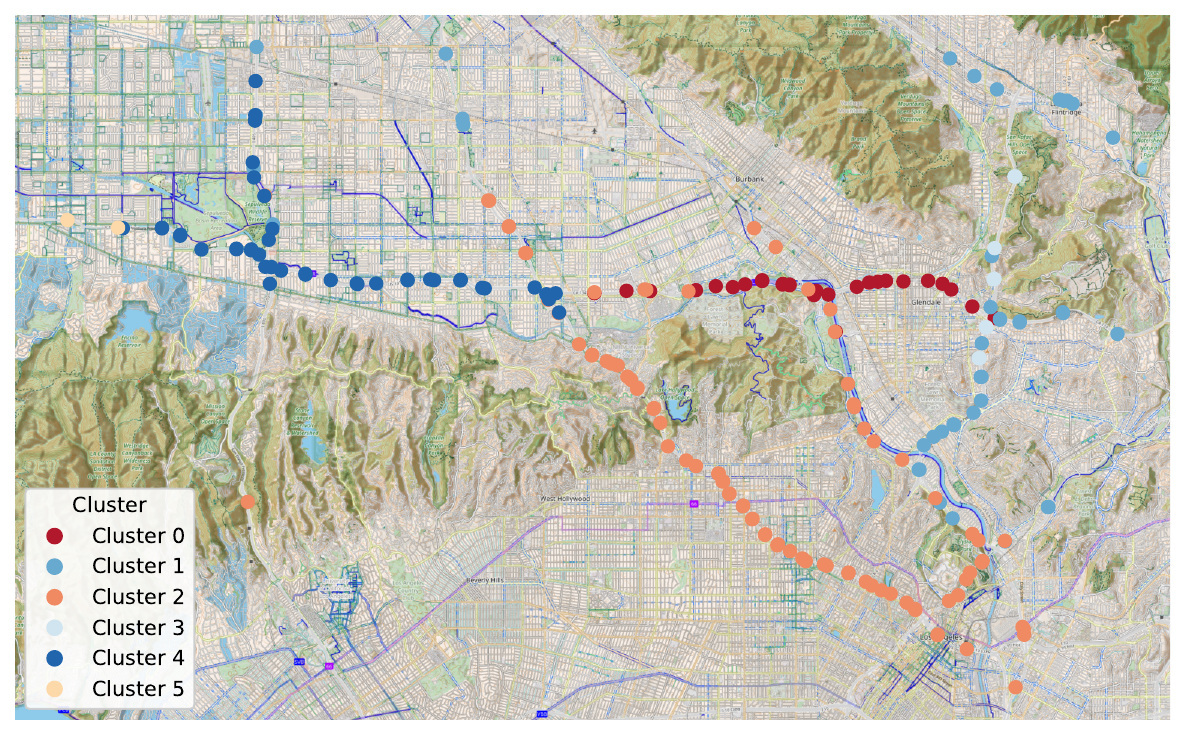}
\caption{Los Angeles traffic sensor clustering from GFL.}
\label{fig_experiment_LA}
\end{figure*}

To characterize daytime traffic patterns, we focus on a 9:00 AM - 3:00 PM period and compute the daily average speed at each sensor. The data hence contains $n=119$ daily observations for $N=207$ sensors, where $Y_{t,i}$ denoting the average speed at sensor $i$ on day $t$. To account for differences in mean traffic speeds and variability across sensors, each nodal series is standardized via $S_{t,i}=\hat{\sigma}_i^{-1}(Y_{t,i}-\hat{\mu}_i)$ for $t=1,\ldots,n$, where $\hat{\mu}_i$ and $\hat{\sigma}_i$ denote the sample mean and standard deviation of sensor $i$'s speed measurements over all days. The standardization removes sensor-specific differences in mean speed and scale, allowing our clustering to focus on relative variations in speed measurements. Figure~\ref{fig_experiment_LA} presents our clustering results for the traffic sensor network. Six clusters were selected via silhouette scores.

Four major clusters are identified. In the vicinity of Glendale, the highway network forms a loop bounded by State Route SR-134 (Ventura Freeway) to the north, SR-2 (Glendale Freeway) to the east, and Interstate I-5 (Golden State Freeway) to the west. The clustering results reveal distinct spatial patterns along these corridors. Near Hollywood, three highway corridors branch in different directions: United States Highway US-101 extends westward (Cluster 4), SR-134 continues eastward (Cluster 0), and US-101 extends southeastward toward downtown Los Angeles (Cluster 2). Notably, sensors along US-101 and I-5, which lead to downtown Los Angeles, are placed in Cluster 2, suggesting similarities in their traffic patterns toward the central business district, despite being located along different highways. Conversely, the intersection marks a division in the traffic patterns along US-101: sensors along the western segment north of the mountains are assigned to Cluster 4, and those along the southeastern segment toward downtown Los Angeles are assigned to Cluster 2. Clusters 3 and 5 contain relatively few sensors and are spatially interspersed in the four clusters. Overall, these results suggest that our method capably identifies sensor groups with similar variations in traffic speeds, while incorporating their spatial relationships.

\begin{figure}[!ht]
    \centering
    \begin{subfigure}{0.3\columnwidth}
        \centering
        \includegraphics[width=\linewidth]{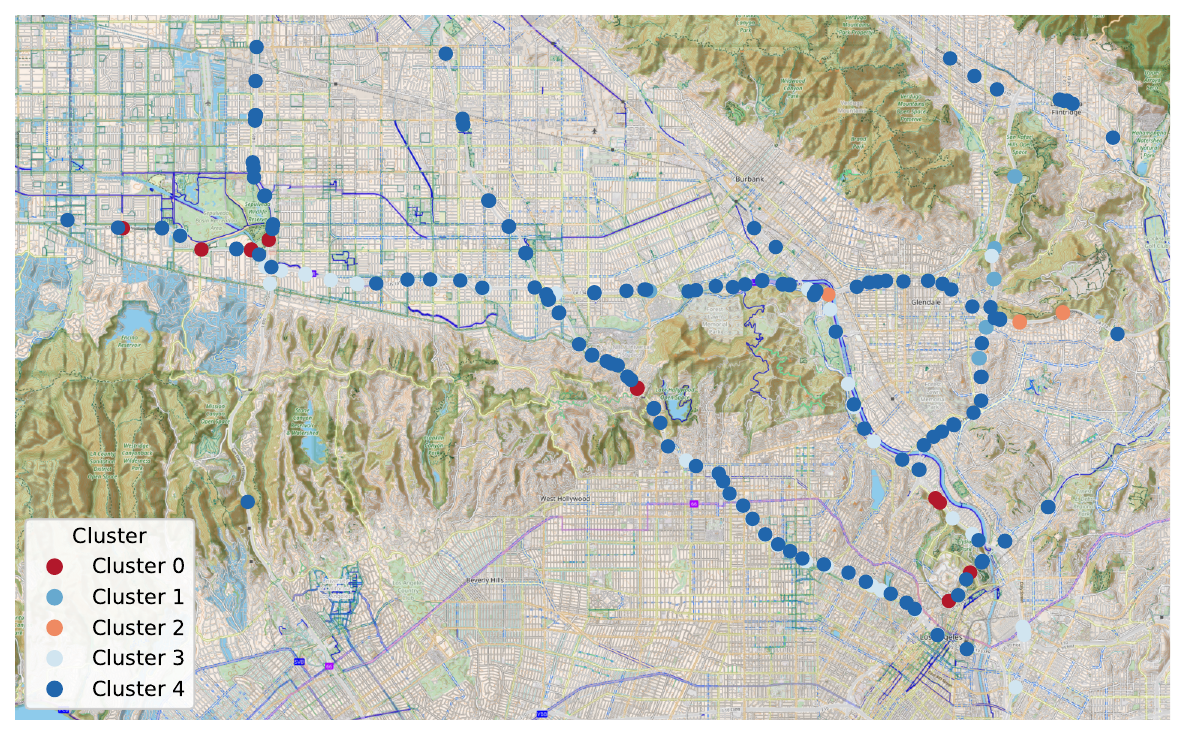}
        \caption{$k$-means}
    \end{subfigure}
    \begin{subfigure}{0.3\columnwidth}
        \centering
        \includegraphics[width=\linewidth]{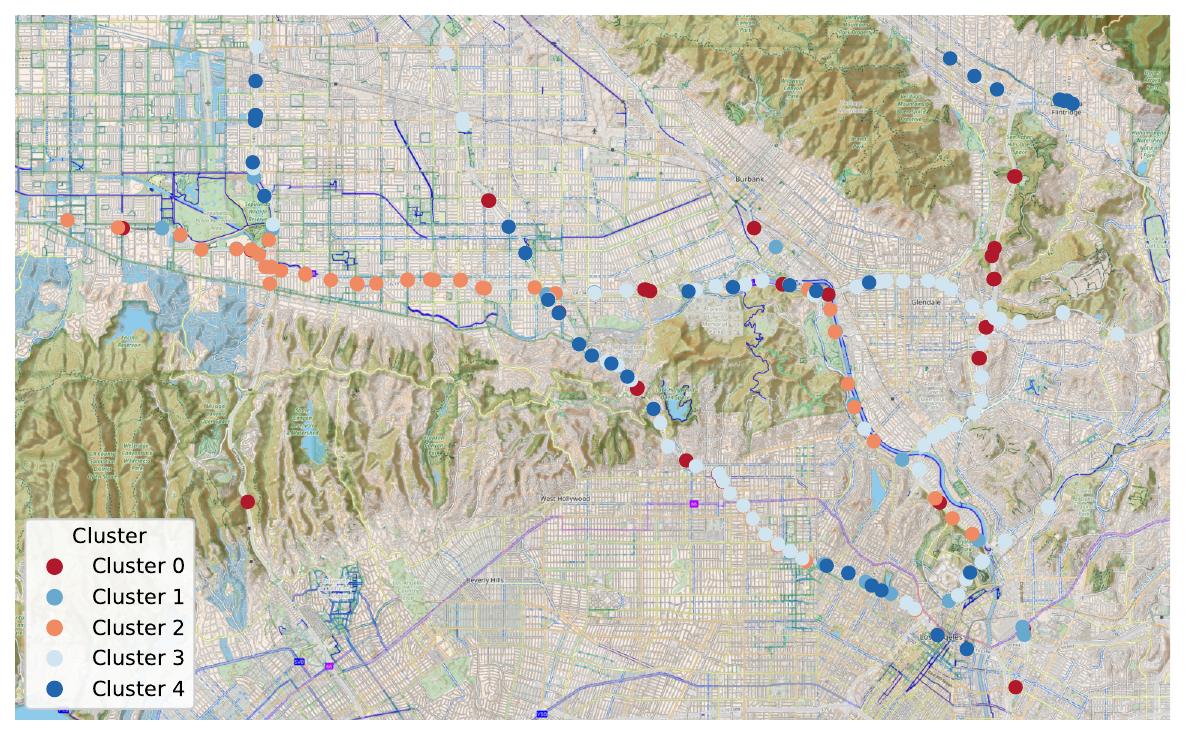}
        \caption{DTW}
    \end{subfigure}
    \begin{subfigure}{0.3\columnwidth}
        \centering
        \includegraphics[width=\linewidth]{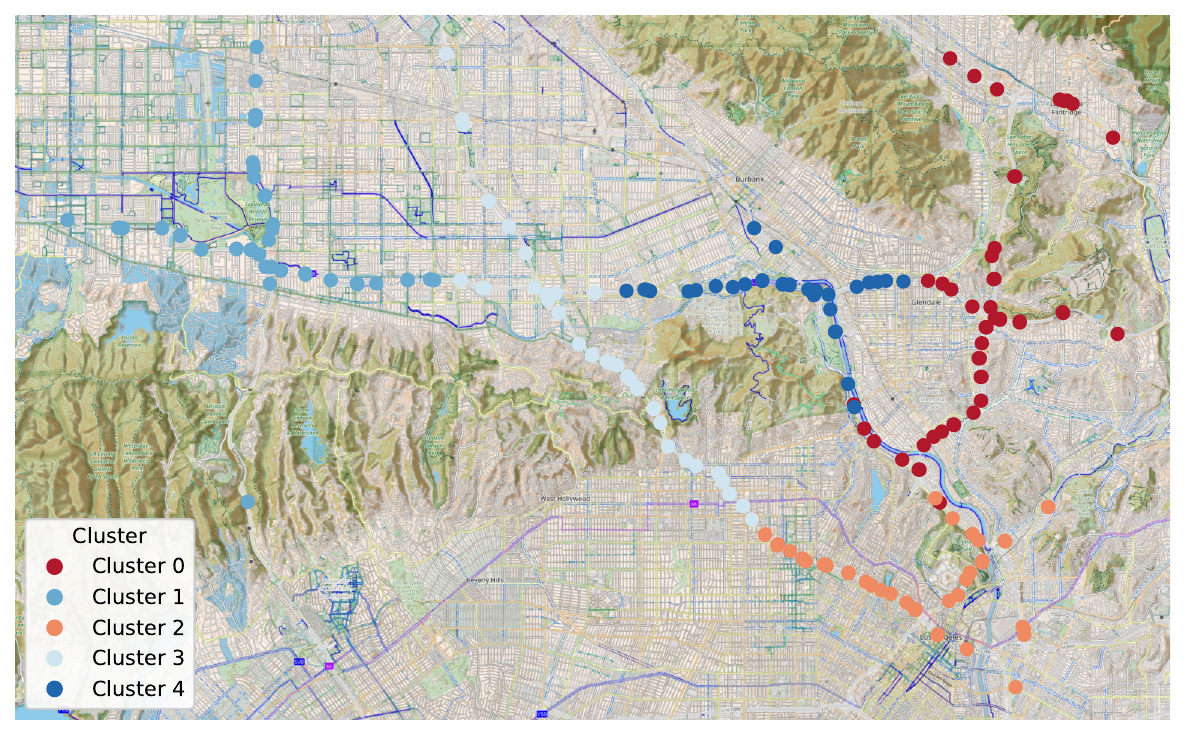}
        \caption{CASC}
    \end{subfigure}
    \begin{subfigure}{0.3\columnwidth}
        \centering
        \includegraphics[width=\linewidth]{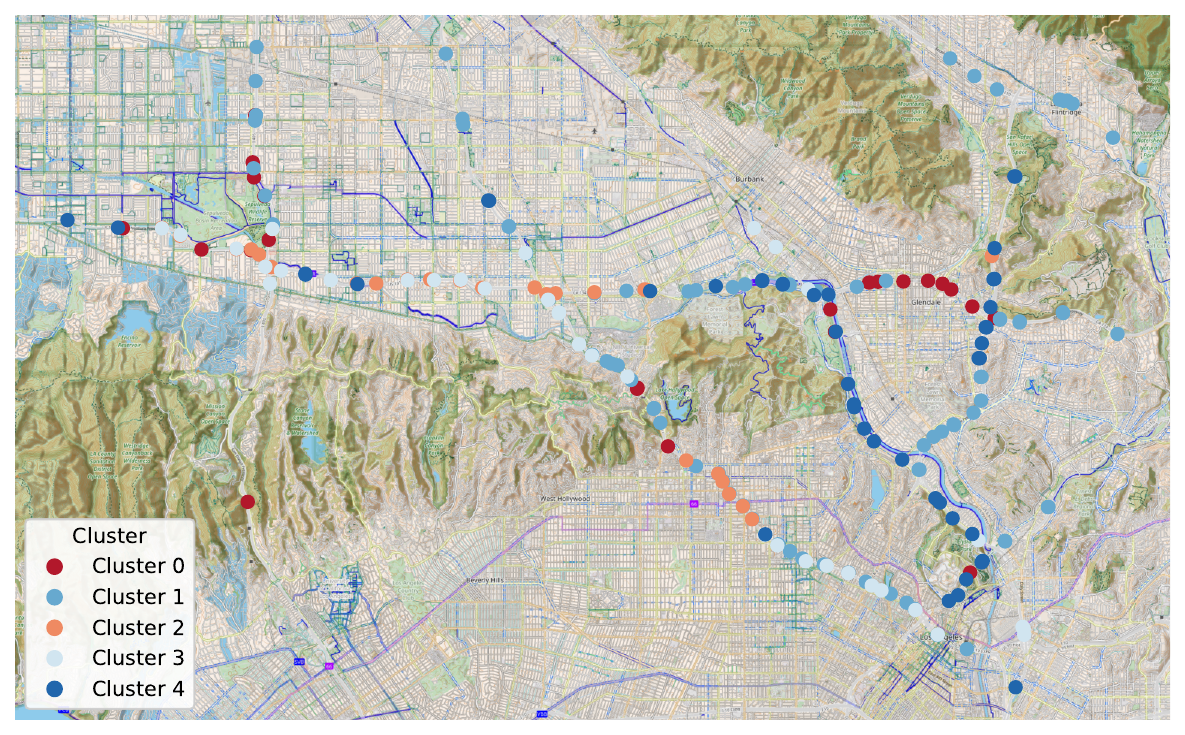}
        \caption{SDP}
    \end{subfigure}
    \begin{subfigure}{0.3\columnwidth}
        \centering
        \includegraphics[width=\linewidth]{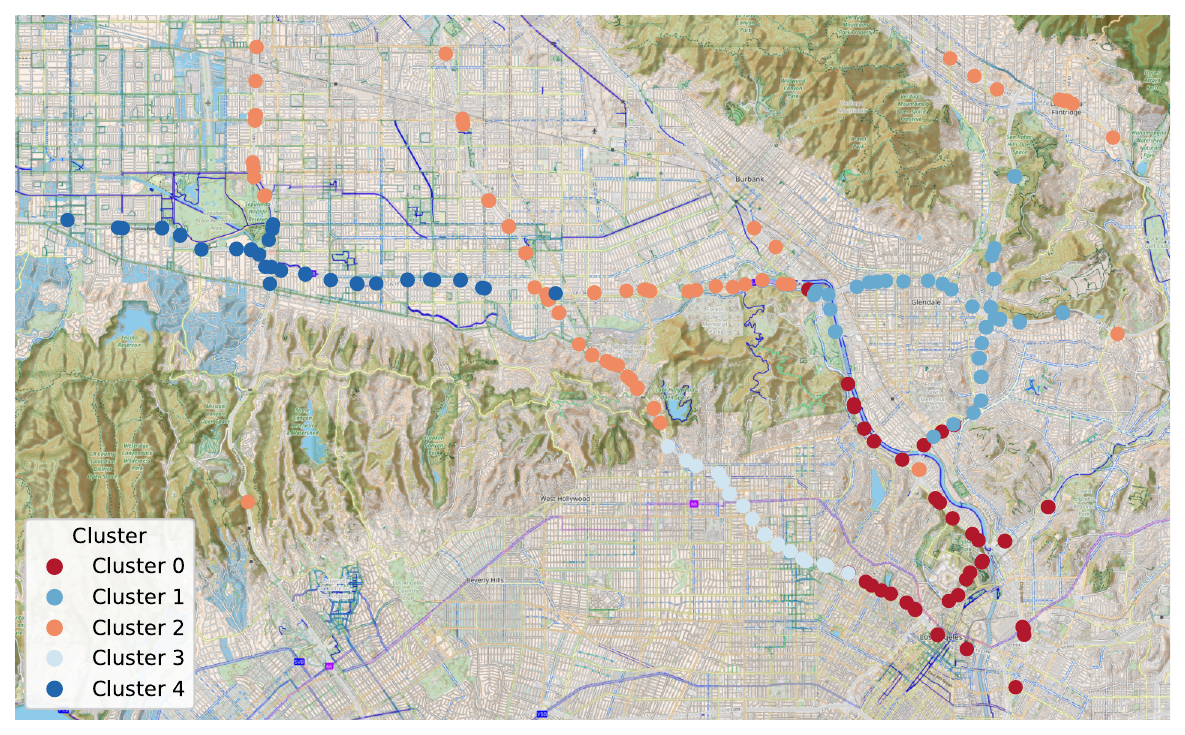}
        \caption{NAC}
    \end{subfigure}
    \begin{subfigure}{0.3\columnwidth}
        \centering
        \includegraphics[width=\linewidth]{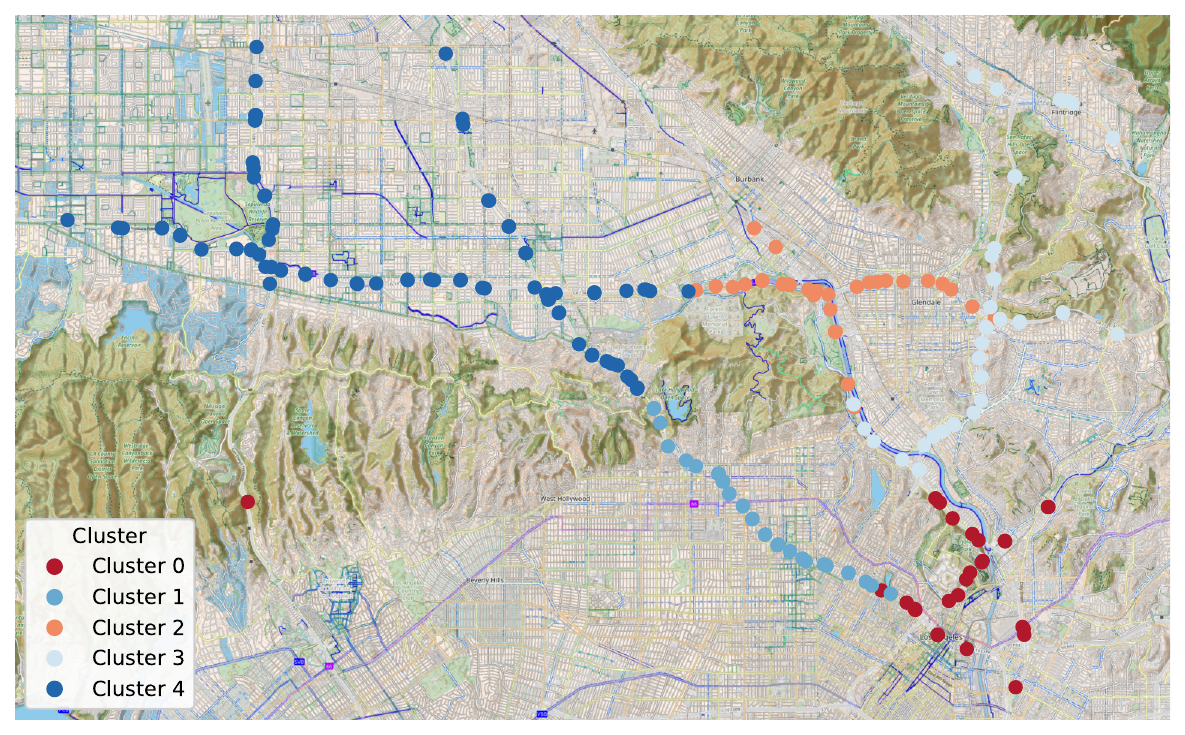}
        \caption{SCORE}
    \end{subfigure}
    \begin{subfigure}{0.3\columnwidth}
        \centering
        \includegraphics[width=\linewidth]{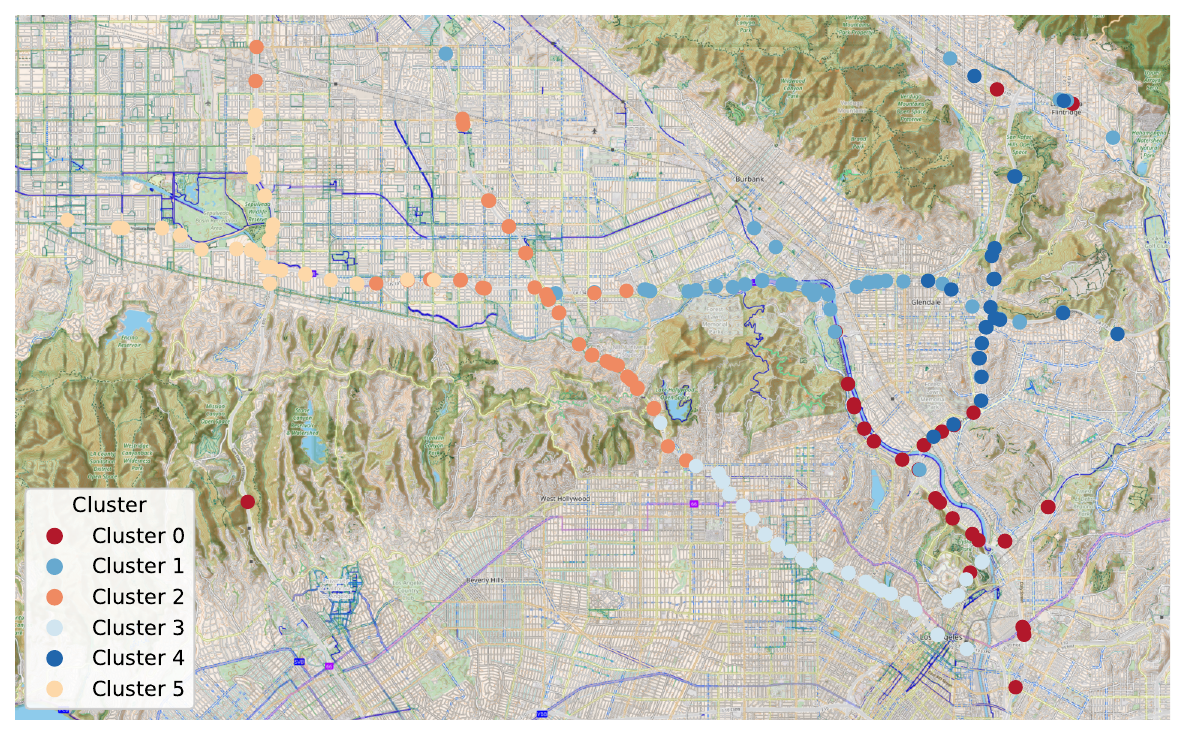}
        \caption{VGAE}
    \end{subfigure}
    \begin{subfigure}{0.3\columnwidth}
        \centering
        \includegraphics[width=\linewidth]{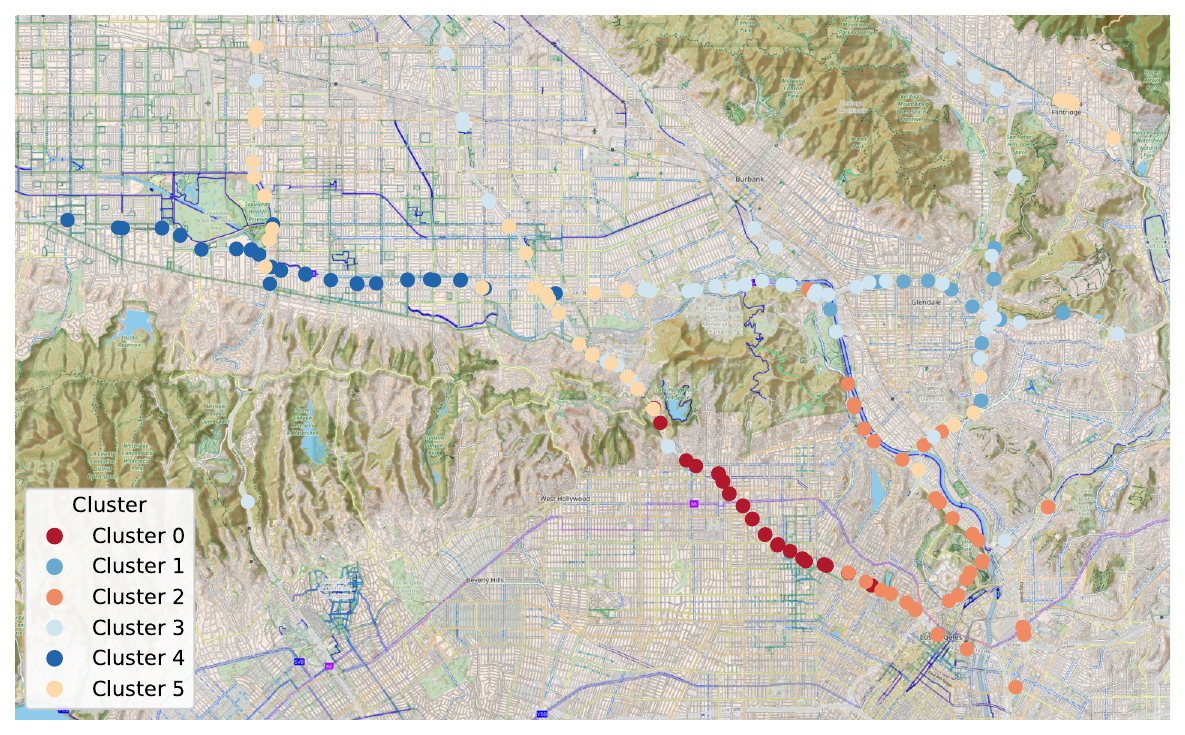}
        \caption{DGI}
    \end{subfigure}
    \begin{subfigure}{0.3\columnwidth}
        \centering
        \includegraphics[width=\linewidth]{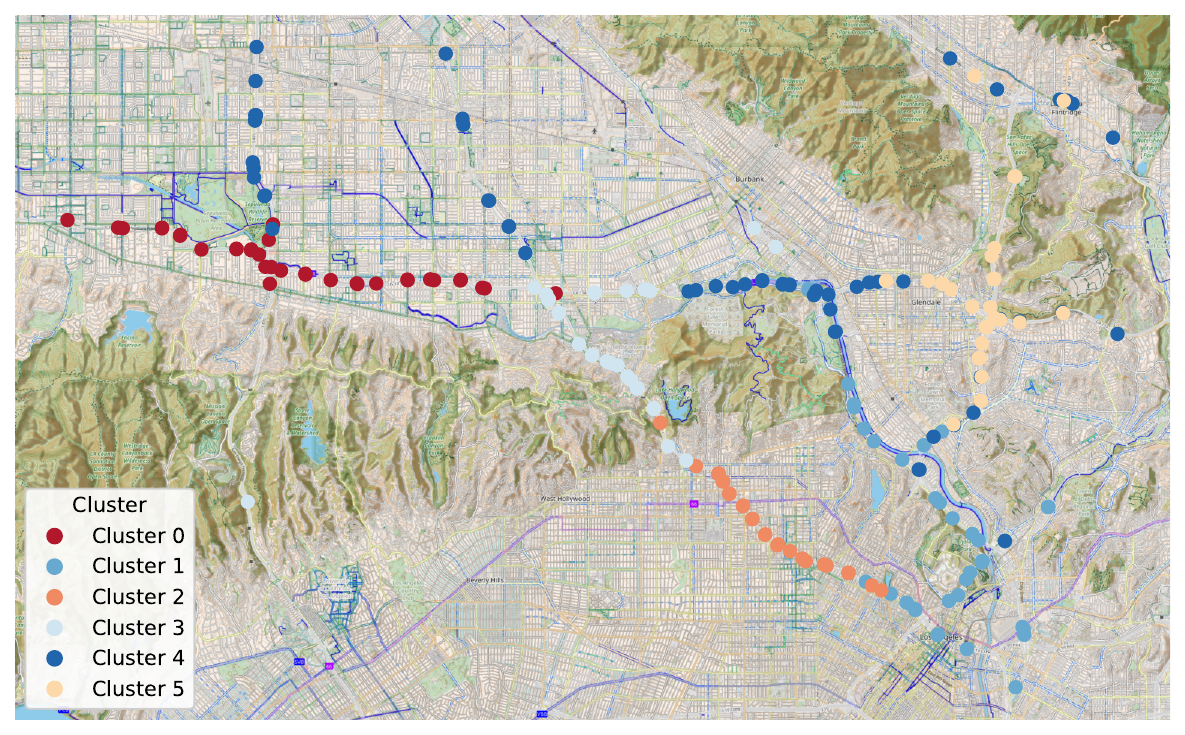}
        \caption{GraphMAE}
    \end{subfigure}
    \caption{Los Angeles traffic sensor clustering from competitor methods with five clusters.}
    \label{fig_Traffic_competitor}
\end{figure}

Figure~\ref{fig_Traffic_competitor} presents clusterings for the traffic sensor network from the competing methods. Five clusters are selected to be comparable to our result. The $k$-means method assigns most sensors to a single cluster, failing to distinguish heterogeneous traffic patterns across the highway system. DTW identifies one cluster along the westward segment of US-101, but the remaining clusters are spatially dispersed and lack any geographical continuity. Similarly, SDP produces clusters that are spatially disconnected. In contrast, NAC, VGAE, DGI, and GraphMAE have more spatially contiguous clusters, but tend to partition individual highways into multiple segments. For example, US-101 is divided into three or four clusters by these methods, resulting in frequent changes in cluster assignments along the same highway. Such fragmentation suggests that these methods may place greater emphasis on spatial differences than on traffic patterns. Overall, the competing methods show limitations in balancing spatial connectivity with traffic dynamics.

\section{Discussion}
\label{sec_discussion}

This manuscript developed a decoder-only latent space model for clustering nodes in networks possessing node-level attributes. By combining node-specific Gaussian priors, graph-fused LASSO regularization, and a neural network decoder, graph structures and attributes are integrated into low-dimensional representations to facilitate clustering. Simulations and applications produced good clustering performance on various graph and attribute types.

Several limitations of this work are worth discussing. Foremost, the type of nodal data could change. For this, the likelihood would need to be modified for the situation. The assumption of independent attributes across nodes could also be relaxed, as it was in our Scenario 3 simulation. On the network front, dyadic relations often have a degree of strength, perhaps continuous values in lieu of zero-one connections \citep{krivitsky2012exponential, wilson2017stochastic} used here. Adapting our LASSO regularization to handle such weights would improve the resulting clustering. Varying network edges and dynamic networks can also arise \citep{malinovskaya2023statistical, kei2023partially}. Methods that accommodate both nodal and dyadic attributes would extend applicability of our methods.

\newpage
\bibliographystyle{tmlr}
\bibliography{references}

\begin{thebibliography}{68}
\providecommand{\natexlab}[1]{#1}
\providecommand{\url}[1]{\texttt{#1}}
\expandafter\ifx\csname urlstyle\endcsname\relax
  \providecommand{\doi}[1]{doi: #1}\else
  \providecommand{\doi}{doi: \begingroup \urlstyle{rm}\Url}\fi

\bibitem[Abatzoglou et~al.(2009)Abatzoglou, Redmond, and Edwards]{abatzoglou2009classification}
John~T Abatzoglou, Kelly~T Redmond, and Laura~M Edwards.
\newblock Classification of regional climate variability in the state of {C}alifornia.
\newblock \emph{Journal of Applied Meteorology and Climatology}, 48\penalty0 (8):\penalty0 1527--1541, 2009.

\bibitem[Abbe(2018)]{abbe2018community}
Emmanuel Abbe.
\newblock Community detection and stochastic block models: recent developments.
\newblock \emph{Journal of Machine Learning Research}, 18\penalty0 (177):\penalty0 1--86, 2018.

\bibitem[Amini et~al.(2013)Amini, Chen, Bickel, and Levina]{amini2013pseudo}
Arash~A. Amini, Aiyou Chen, Peter~J. Bickel, and Elizaveta Levina.
\newblock {Pseudo-likelihood methods for community detection in large sparse networks}.
\newblock \emph{The Annals of Statistics}, 41\penalty0 (4):\penalty0 2097 -- 2122, 2013.
\newblock \doi{10.1214/13-AOS1138}.

\bibitem[Bartlett et~al.(2019)Bartlett, Harvey, Liaw, and Mehrabian]{bartlett2019nearly}
Peter~L Bartlett, Nick Harvey, Christopher Liaw, and Abbas Mehrabian.
\newblock Nearly-tight vc-dimension and pseudodimension bounds for piecewise linear neural networks.
\newblock \emph{Journal of Machine Learning Research}, 20\penalty0 (63):\penalty0 1--17, 2019.

\bibitem[Binkiewicz et~al.(2017)Binkiewicz, Vogelstein, and Rohe]{binkiewicz2017casc}
Norbert Binkiewicz, Joshua~T Vogelstein, and Karl Rohe.
\newblock Covariate-assisted spectral clustering.
\newblock \emph{Biometrika}, 104\penalty0 (2):\penalty0 361--377, 2017.

\bibitem[Boyd et~al.(2011)Boyd, Parikh, Chu, Peleato, Eckstein, et~al.]{boyd2011distributed}
Stephen Boyd, Neal Parikh, Eric Chu, Borja Peleato, Jonathan Eckstein, et~al.
\newblock Distributed optimization and statistical learning via the alternating direction method of multipliers.
\newblock \emph{Foundations and Trends{\textregistered} in Machine Learning}, 3\penalty0 (1):\penalty0 1--122, 2011.

\bibitem[Chen et~al.(2023{\natexlab{a}})Chen, Fan, and Zhu]{chen2023community}
Elynn~Y Chen, Jianqing Fan, and Xuening Zhu.
\newblock Community network auto-regression for high-dimensional time series.
\newblock \emph{Journal of Econometrics}, 235\penalty0 (2):\penalty0 1239--1256, 2023{\natexlab{a}}.

\bibitem[Chen et~al.(2023{\natexlab{b}})Chen, Jewell, and Witten]{chen2023more}
Yiqun Chen, Sean Jewell, and Daniela Witten.
\newblock More powerful selective inference for the graph fused {LASSO}.
\newblock \emph{Journal of Computational and Graphical Statistics}, 32\penalty0 (2):\penalty0 577--587, 2023{\natexlab{b}}.

\bibitem[Deng et~al.(2017)Deng, Lai, Peng, and Yin]{deng2017parallel}
Wei Deng, Ming-Jun Lai, Zhimin Peng, and Wotao Yin.
\newblock Parallel multi-block admm with o (1/k) convergence.
\newblock \emph{Journal of Scientific Computing}, 71\penalty0 (2):\penalty0 712--736, 2017.

\bibitem[Du \& Mordatch(2019)Du and Mordatch]{du2019implicit}
Yilun Du and Igor Mordatch.
\newblock Implicit generation and modeling with energy based models.
\newblock \emph{Advances in Neural Information Processing Systems}, 32:\penalty0 3608 -- 3618, 2019.

\bibitem[Durante \& Dunson(2014)Durante and Dunson]{durante2014nonparametric}
Daniele Durante and David~B Dunson.
\newblock Nonparametric {B}ayes dynamic modelling of relational data.
\newblock \emph{Biometrika}, 101\penalty0 (4):\penalty0 883--898, 2014.

\bibitem[Emmons et~al.(2016)Emmons, Kobourov, Gallant, and B{\"o}rner]{emmons2016analysis}
Scott Emmons, Stephen Kobourov, Mike Gallant, and Katy B{\"o}rner.
\newblock Analysis of network clustering algorithms and cluster quality metrics at scale.
\newblock \emph{PLoS One}, 11\penalty0 (7):\penalty0 e0159161, 2016.

\bibitem[Hallac et~al.(2015)Hallac, Leskovec, and Boyd]{hallac2015network}
David Hallac, Jure Leskovec, and Stephen Boyd.
\newblock Network {LASSO}: Clustering and optimization in large graphs.
\newblock In \emph{Proceedings of the 21st ACM SIGKDD International Conference on Knowledge Discovery and Data Mining}, pp.\  387--396, 2015.

\bibitem[Han et~al.(2023)Han, Bushong, Segawa, Tiard, Wong, Brady, Momcilovic, Wolf, Zhang, Petcherski, Madany, Xu, Lee, Poyurovsky, Olszewski, Holloway, Gomez, John, Dubinett, Koehler, Shirihai, Stiles, Lisberg, Soatto, Sadeghi, Ellisman, and Shackelford]{han2023spatial}
Mingqi Han, Eric~A. Bushong, Mayuko Segawa, Alexandre Tiard, Alex Wong, Morgan~R. Brady, Milica Momcilovic, Dane~M. Wolf, Ralph Zhang, Anton Petcherski, Matthew Madany, Shili Xu, Jason~T. Lee, Masha~V. Poyurovsky, Kellen Olszewski, Travis Holloway, Adrian Gomez, Maie~St. John, Steven~M. Dubinett, Carla~M. Koehler, Orian~S. Shirihai, Linsey Stiles, Aaron Lisberg, Stefano Soatto, Saman Sadeghi, Mark~H. Ellisman, and David~B. Shackelford.
\newblock Spatial mapping of mitochondrial networks and bioenergetics in lung cancer.
\newblock \emph{Nature}, 615\penalty0 (7953):\penalty0 712--719, 2023.

\bibitem[Handcock et~al.(2007)Handcock, Raftery, and Tantrum]{handcock2007model}
Mark~S Handcock, Adrian~E Raftery, and Jeremy~M Tantrum.
\newblock Model-based clustering for social networks.
\newblock \emph{Journal of the Royal Statistical Society Series A: Statistics in Society}, 170\penalty0 (2):\penalty0 301--354, 2007.

\bibitem[Hartigan \& Wong(1979)Hartigan and Wong]{hartigan1979algorithm}
John~A Hartigan and Manchek~A Wong.
\newblock Algorithm as 136: A k-means clustering algorithm.
\newblock \emph{Journal of the Royal Statistical Society. Series C (Applied Statistics)}, 28\penalty0 (1):\penalty0 100--108, 1979.

\bibitem[Hornik(1991)]{hornik1991approximation}
Kurt Hornik.
\newblock Approximation capabilities of multilayer feedforward networks.
\newblock \emph{Neural Networks}, 4\penalty0 (2):\penalty0 251--257, 1991.

\bibitem[Hornik et~al.(1989)Hornik, Stinchcombe, and White]{hornik1989multilayer}
Kurt Hornik, Maxwell Stinchcombe, and Halbert White.
\newblock Multilayer feedforward networks are universal approximators.
\newblock \emph{Neural Networks}, 2\penalty0 (5):\penalty0 359--366, 1989.

\bibitem[Hou et~al.(2022)Hou, Liu, Cen, Dong, Yang, Wang, and Tang]{hou2022graphmae}
Zhenyu Hou, Xiao Liu, Yukuo Cen, Yuxiao Dong, Hongxia Yang, Chunjie Wang, and Jie Tang.
\newblock Graphmae: Self-supervised masked graph autoencoders.
\newblock In \emph{Proceedings of the 28th ACM SIGKDD Conference on Knowledge Discovery and Data Mining}, pp.\  594--604, 2022.

\bibitem[Hu \& Wang(2024)Hu and Wang]{hu2024nac}
Yaofang Hu and Wanjie Wang.
\newblock Network-adjusted covariates for community detection.
\newblock \emph{Biometrika}, 111\penalty0 (4):\penalty0 1221--1240, 2024.

\bibitem[Jagadish et~al.(2014)Jagadish, Gehrke, Labrinidis, Papakonstantinou, Patel, Ramakrishnan, and Shahabi]{jagadish2014big}
Hosagrahar~V Jagadish, Johannes Gehrke, Alexandros Labrinidis, Yannis Papakonstantinou, Jignesh~M Patel, Raghu Ramakrishnan, and Cyrus Shahabi.
\newblock Big data and its technical challenges.
\newblock \emph{Communications of the ACM}, 57\penalty0 (7):\penalty0 86--94, 2014.

\bibitem[Jin(2015)]{Jin2015SCORE}
Jiashun Jin.
\newblock Fast community detection by {SCORE}.
\newblock \emph{Annals of Statistics}, 43\penalty0 (1):\penalty0 57--89, 2015.
\newblock \doi{10.1214/14-AOS1265}.

\bibitem[Kei et~al.(2023)Kei, Chen, and Padilla]{kei2023partially}
Yik~Lun Kei, Yanzhen Chen, and Oscar Hernan~Madrid Padilla.
\newblock A partially separable model for dynamic valued networks.
\newblock \emph{Computational Statistics \& Data Analysis}, 187:\penalty0 107811, 2023.

\bibitem[Kei et~al.(2025{\natexlab{a}})Kei, Li, Chen, and Padilla]{kei2025separable}
Yik~Lun Kei, Hangjian Li, Yanzhen Chen, and Oscar Hernan~Madrid Padilla.
\newblock Change point detection on a separable model for dynamic networks.
\newblock \emph{Transactions on Machine Learning Research}, 2025{\natexlab{a}}.
\newblock ISSN 2835-8856.

\bibitem[Kei et~al.(2025{\natexlab{b}})Kei, Li, Li, Chen, and Padilla]{kei2025change}
Yik~Lun Kei, Jialiang Li, Hangjian Li, Yanzhen Chen, and Oscar Hernan~Madrid Padilla.
\newblock Change point detection in dynamic graphs with decoder-only latent space model.
\newblock \emph{Transactions on Machine Learning Research}, 2025{\natexlab{b}}.
\newblock ISSN 2835-8856.

\bibitem[Kingma \& Welling(2014)Kingma and Welling]{kingma2013auto}
Diederik~P Kingma and Max Welling.
\newblock Auto-encoding variational {B}ayes.
\newblock \emph{International Conference on Learning Representations}, 2014.

\bibitem[Kipf \& Welling(2016)Kipf and Welling]{kipf2016variational}
Thomas~N Kipf and Max Welling.
\newblock Variational graph auto-encoders.
\newblock \emph{arXiv preprint arXiv:1611.07308}, 2016.

\bibitem[Krivitsky(2012)]{krivitsky2012exponential}
Pavel~N Krivitsky.
\newblock Exponential-family random graph models for valued networks.
\newblock \emph{Electronic Journal of Statistics}, 6:\penalty0 1100--1128, 2012.

\bibitem[LaDochy et~al.(2007)LaDochy, Medina, and Patzert]{ladochy2007recent}
Steve LaDochy, Richard Medina, and William Patzert.
\newblock Recent {C}alifornia climate variability: spatial and temporal patterns in temperature trends.
\newblock \emph{Climate Research}, 33\penalty0 (2):\penalty0 159--169, 2007.

\bibitem[LeCun et~al.(2015)LeCun, Bengio, and Hinton]{lecun2015deep}
Yann LeCun, Yoshua Bengio, and Geoffrey Hinton.
\newblock Deep learning.
\newblock \emph{Nature}, 521\penalty0 (7553):\penalty0 436--444, 2015.

\bibitem[Lee et~al.(2020)Lee, Li, and Wilson]{lee2020varying}
Jihui Lee, Gen Li, and James~D Wilson.
\newblock Varying-coefficient models for dynamic networks.
\newblock \emph{Computational Statistics \& Data Analysis}, 152:\penalty0 107052, 2020.

\bibitem[Leifeld et~al.(2018)Leifeld, Cranmer, and Desmarais]{leifeld2018temporal}
Philip Leifeld, Skyler~J Cranmer, and Bruce~A Desmarais.
\newblock Temporal exponential random graph models with btergm: Estimation and bootstrap confidence intervals.
\newblock \emph{Journal of Statistical Software}, 83:\penalty0 1--36, 2018.

\bibitem[Li et~al.(2018)Li, Yu, Shahabi, and Liu]{li2018dcrnn_traffic}
Yaguang Li, Rose Yu, Cyrus Shahabi, and Yan Liu.
\newblock Diffusion convolutional recurrent neural network: Data-driven traffic forecasting.
\newblock In \emph{International Conference on Learning Representations}, 2018.

\bibitem[Macgregor(2023)]{macgregor2023fast}
Peter Macgregor.
\newblock Fast and simple spectral clustering in theory and practice.
\newblock \emph{Advances in Neural Information Processing Systems}, 36:\penalty0 34410--34425, 2023.

\bibitem[Madrid~Padilla \& Chen(2023)Madrid~Padilla and Chen]{Padilla2023graphon}
Oscar~Hernan Madrid~Padilla and Yanzhen Chen.
\newblock Graphon estimation via nearest-neighbour algorithm and two-dimensional fused-{LASSO} denoising.
\newblock \emph{Canadian Journal of Statistics}, 51\penalty0 (1):\penalty0 95--110, 2023.

\bibitem[Madrid~Padilla et~al.(2018)Madrid~Padilla, Sharpnack, Scott, and Tibshirani]{padilla2018dfs}
Oscar~Hernan Madrid~Padilla, James Sharpnack, James~G Scott, and Ryan~J Tibshirani.
\newblock The dfs fused {LASSO}: Linear-time denoising over general graphs.
\newblock \emph{Journal of Machine Learning Research}, 18\penalty0 (176):\penalty0 1--36, 2018.

\bibitem[Malinovskaya et~al.(2023)Malinovskaya, Killick, Leeming, and Otto]{malinovskaya2023statistical}
Anna Malinovskaya, Rebecca Killick, Kathryn Leeming, and Philipp Otto.
\newblock Statistical monitoring of european cross-border physical electricity flows using novel temporal edge network processes.
\newblock \emph{arXiv preprint arXiv:2312.16357}, 2023.

\bibitem[Malliaros \& Vazirgiannis(2013)Malliaros and Vazirgiannis]{malliaros2013clustering}
Fragkiskos~D Malliaros and Michalis Vazirgiannis.
\newblock Clustering and community detection in directed networks: A survey.
\newblock \emph{Physics Reports}, 533\penalty0 (4):\penalty0 95--142, 2013.

\bibitem[Mattei \& Frellsen(2018)Mattei and Frellsen]{mattei2018leveraging}
Pierre-Alexandre Mattei and Jes Frellsen.
\newblock Leveraging the exact likelihood of deep latent variable models.
\newblock \emph{Advances in Neural Information Processing Systems}, 31, 2018.

\bibitem[Mucha et~al.(2010)Mucha, Richardson, Macon, Porter, and Onnela]{mucha2010community}
Peter~J Mucha, Thomas Richardson, Kevin Macon, Mason~A Porter, and Jukka-Pekka Onnela.
\newblock Community structure in time-dependent, multiscale, and multiplex networks.
\newblock \emph{Science}, 328\penalty0 (5980):\penalty0 876--878, 2010.

\bibitem[Nijkamp et~al.(2020)Nijkamp, Hill, Han, Zhu, and Wu]{nijkamp2020anatomy}
Erik Nijkamp, Mitch Hill, Tian Han, Song-Chun Zhu, and Ying~Nian Wu.
\newblock On the anatomy of {MCMC}-based maximum likelihood learning of energy-based models.
\newblock In \emph{Proceedings of the AAAI Conference on Artificial Intelligence}, volume~34, pp.\  5272--5280, 2020.

\bibitem[Purohit et~al.(2024)Purohit, Repasky, Lu, Qiu, Xie, and Cheng]{purohit2024posterior}
Vishal Purohit, Matthew Repasky, Jianfeng Lu, Qiang Qiu, Yao Xie, and Xiuyuan Cheng.
\newblock Posterior sampling via {L}angevin dynamics based on generative priors.
\newblock \emph{arXiv preprint arXiv:2410.02078}, 2024.

\bibitem[Rousseeuw(1987)]{rousseeuw1987silhouettes}
Peter~J Rousseeuw.
\newblock Silhouettes: a graphical aid to the interpretation and validation of cluster analysis.
\newblock \emph{Journal of Computational and Applied Mathematics}, 20:\penalty0 53--65, 1987.

\bibitem[Roy et~al.(2024)Roy, Roy, and Ghosal]{roy2024relational}
Arkaprava Roy, Anindya Roy, and Subhashis Ghosal.
\newblock Relational graph in vector autoregression: A case study on the effect of the great recession on connectivity of economic indicators.
\newblock \emph{arXiv preprint arXiv:2410.22617}, 2024.

\bibitem[Sadhanala \& Tibshirani(2019)Sadhanala and Tibshirani]{sadhanala2019additive}
Veeranjaneyulu Sadhanala and Ryan~J Tibshirani.
\newblock Additive models with trend filtering.
\newblock \emph{The Annals of Statistics}, 47\penalty0 (6):\penalty0 3032--3068, 2019.

\bibitem[Sakoe \& Chiba(2003)Sakoe and Chiba]{sakoe2003dtw}
Hiroaki Sakoe and Seibi Chiba.
\newblock Dynamic programming algorithm optimization for spoken word recognition.
\newblock \emph{IEEE Transactions on Acoustics, Speech, and Signal Processing}, 26\penalty0 (1):\penalty0 43--49, 2003.

\bibitem[Sandel et~al.(2026)Sandel, He, Ren, Kei, Lee, Clark, Reddy, Negrey, Birungi, Apamaku, Kanweri, Kalunga, Aliganyira, Ramírez-Amaya, Nakayima, Katumba, Kamugyisha, Acosta-Florez, van Boekholt, Mbabazi, Akamumpa, Namaganda, Tumusiime, Angedakin, Reinert, Madrid-Padilla, Cucuringu, Wipf, Langergraber, Watts, and Mitani]{sandel2026lethal}
Aaron~A. Sandel, Yixuan He, Junpeng Ren, Yik~Lun Kei, Kevin~C. Lee, Isabelle~R. Clark, Rachna~B. Reddy, Jacob~D. Negrey, Charles Birungi, Blessing~A. Apamaku, Diana Kanweri, Davis Kalunga, Christopher Aliganyira, Sebastián Ramírez-Amaya, Phionah Nakayima, Raymond Katumba, Brian Kamugyisha, Daniela Acosta-Florez, Bas van Boekholt, Godfrey Mbabazi, Erone Akamumpa, Sharifah Namaganda, Alfred Tumusiime, Samuel Angedakin, Gesine Reinert, Oscar Madrid-Padilla, Mihai Cucuringu, David Wipf, Kevin~E. Langergraber, David~P. Watts, and John~C. Mitani.
\newblock Lethal conflict after group fission in wild chimpanzees.
\newblock \emph{Science}, 392\penalty0 (6794):\penalty0 216--220, 2026.

\bibitem[Sewell \& Chen(2015)Sewell and Chen]{sewell2015latent}
Daniel~K Sewell and Yuguo Chen.
\newblock Latent space models for dynamic networks.
\newblock \emph{Journal of the American Statistical Association}, 110\penalty0 (512):\penalty0 1646--1657, 2015.

\bibitem[Shai et~al.(2021)Shai, Stanley, Granell, Taylor, and Mucha]{shai2017case}
Saray Shai, Natalie Stanley, Clara Granell, Dane Taylor, and Peter~J. Mucha.
\newblock Case studies in network community detection.
\newblock In \emph{The Oxford Handbook of Social Networks}. Oxford University Press, 2021.
\newblock ISBN 9780190251765.
\newblock \doi{10.1093/oxfordhb/9780190251765.013.16}.

\bibitem[Shen et~al.(2024)Shen, Amini, Josephs, and Lin]{shen2024bayesian}
Luyi Shen, Arash Amini, Nathaniel Josephs, and Lizhen Lin.
\newblock Bayesian community detection for networks with covariates.
\newblock \emph{Bayesian Analysis}, 1\penalty0 (1):\penalty0 1--28, 2024.

\bibitem[Su et~al.(2022)Su, Xue, Liu, Wu, Yang, Zhou, Hu, Paris, Nepal, Jin, et~al.]{su2022comprehensive}
Xing Su, Shan Xue, Fanzhen Liu, Jia Wu, Jian Yang, Chuan Zhou, Wenbin Hu, Cecile Paris, Surya Nepal, Di~Jin, et~al.
\newblock A comprehensive survey on community detection with deep learning.
\newblock \emph{IEEE Transactions on Neural Networks and Learning Systems}, 35\penalty0 (4):\penalty0 4682--4702, 2022.

\bibitem[Tibshirani et~al.(2005)Tibshirani, Saunders, Rosset, Zhu, and Knight]{tibshirani2005sparsity}
Robert Tibshirani, Michael Saunders, Saharon Rosset, Ji~Zhu, and Keith Knight.
\newblock Sparsity and smoothness via the fused lasso.
\newblock \emph{Journal of the Royal Statistical Society Series B: Statistical Methodology}, 67\penalty0 (1):\penalty0 91--108, 2005.

\bibitem[Tsitsulin et~al.(2023)Tsitsulin, Palowitch, Perozzi, and M{\"u}ller]{tsitsulin2023graph}
Anton Tsitsulin, John Palowitch, Bryan Perozzi, and Emmanuel M{\"u}ller.
\newblock Graph clustering with graph neural networks.
\newblock \emph{Journal of Machine Learning Research}, 24\penalty0 (127):\penalty0 1--21, 2023.

\bibitem[Veli{\v{c}}kovi{\'{c}} et~al.(2019)Veli{\v{c}}kovi{\'{c}}, Fedus, Hamilton, Li{\`{o}}, Bengio, and Hjelm]{velickovic2018deep}
Petar Veli{\v{c}}kovi{\'{c}}, William Fedus, William~L. Hamilton, Pietro Li{\`{o}}, Yoshua Bengio, and R~Devon Hjelm.
\newblock {Deep Graph Infomax}.
\newblock In \emph{International Conference on Learning Representations}, 2019.

\bibitem[Vert \& Bleakley(2010)Vert and Bleakley]{vert2010fast}
Jean-Philippe Vert and Kevin Bleakley.
\newblock Fast detection of multiple change-points shared by many signals using group lars.
\newblock \emph{Advances in Neural Information Processing Systems}, 23, 2010.

\bibitem[Walton \& Hall(2018)Walton and Hall]{walton2018assessment}
Daniel Walton and Alex Hall.
\newblock An assessment of high-resolution gridded temperature datasets over {C}alifornia.
\newblock \emph{Journal of Climate}, 31\penalty0 (10):\penalty0 3789--3810, 2018.

\bibitem[Wang et~al.(2023)Wang, Paul, and De~Boeck]{wang2023joint}
Selena Wang, Subhadeep Paul, and Paul De~Boeck.
\newblock Joint latent space model for social networks with multivariate attributes.
\newblock \emph{Psychometrika}, 88\penalty0 (4):\penalty0 1197--1227, 2023.

\bibitem[Wang et~al.(2019)Wang, Yin, and Zeng]{wang2019global}
Yu~Wang, Wotao Yin, and Jinshan Zeng.
\newblock Global convergence of admm in nonconvex nonsmooth optimization.
\newblock \emph{Journal of Scientific Computing}, 78\penalty0 (1):\penalty0 29--63, 2019.

\bibitem[Wang et~al.(2016)Wang, Sharpnack, Smola, and Tibshirani]{wang2016trend}
Yu-Xiang Wang, James Sharpnack, Alexander~J Smola, and Ryan~J Tibshirani.
\newblock Trend filtering on graphs.
\newblock \emph{Journal of Machine Learning Research}, 17\penalty0 (105):\penalty0 1--41, 2016.

\bibitem[Wilson et~al.(2017{\natexlab{a}})Wilson, Denny, Bhamidi, Cranmer, and Desmarais]{wilson2017stochastic}
James~D Wilson, Matthew~J Denny, Shankar Bhamidi, Skyler~J Cranmer, and Bruce~A Desmarais.
\newblock Stochastic weighted graphs: Flexible model specification and simulation.
\newblock \emph{Social Networks}, 49:\penalty0 37--47, 2017{\natexlab{a}}.

\bibitem[Wilson et~al.(2017{\natexlab{b}})Wilson, Palowitch, Bhamidi, and Nobel]{wilson2017community}
James~D Wilson, John Palowitch, Shankar Bhamidi, and Andrew~B Nobel.
\newblock Community extraction in multilayer networks with heterogeneous community structure.
\newblock \emph{Journal of Machine Learning Research}, 18\penalty0 (149):\penalty0 1--49, 2017{\natexlab{b}}.

\bibitem[Yan \& Sarkar(2021)Yan and Sarkar]{yan2021sdp}
Bowei Yan and Purnamrita Sarkar.
\newblock Covariate regularized community detection in sparse graphs.
\newblock \emph{Journal of the American Statistical Association}, 116\penalty0 (534):\penalty0 734--745, 2021.

\bibitem[Yang et~al.(2022)Yang, Wilson, Lu, and Cranmer]{yang2022functional}
Seo~Eun Yang, James~D Wilson, Zhong-Lin Lu, and Skyler Cranmer.
\newblock Functional connectivity signatures of political ideology.
\newblock \emph{PNAS Nexus}, 1\penalty0 (3):\penalty0 pgac066, 2022.

\bibitem[Yarotsky(2017)]{yarotsky2017error}
Dmitry Yarotsky.
\newblock Error bounds for approximations with deep {ReLU} networks.
\newblock \emph{Neural Networks}, 94:\penalty0 103--114, 2017.

\bibitem[Yu et~al.(2025)Yu, Yang, and Zhang]{yu2025graph}
Feng Yu, Archer~Yi Yang, and Teng Zhang.
\newblock A graph decomposition-based approach for the graph-fused {LASSO}.
\newblock \emph{Journal of Statistical Planning and Inference}, 235:\penalty0 106221, 2025.

\bibitem[Yuan \& Lin(2006)Yuan and Lin]{yuan2006model}
Ming Yuan and Yi~Lin.
\newblock Model selection and estimation in regression with grouped variables.
\newblock \emph{Journal of the Royal Statistical Society Series B: Statistical Methodology}, 68\penalty0 (1):\penalty0 49--67, 2006.

\bibitem[Zhu et~al.(2023)Zhu, Caliskan, Christenson, Spiliopoulos, Walker, and Kolaczyk]{zhu2023disentangling}
Xiaojing Zhu, Cantay Caliskan, Dino~P Christenson, Konstantinos Spiliopoulos, Dylan Walker, and Eric~D Kolaczyk.
\newblock Disentangling positive and negative partisanship in social media interactions using a coevolving latent space network with attractors model.
\newblock \emph{Journal of the Royal Statistical Society Series A: Statistics in Society}, 186\penalty0 (3):\penalty0 463--480, 2023.

\bibitem[Zhu et~al.(2025)Zhu, Xu, and Fan]{zhu2025simultaneous}
Xuening Zhu, Ganggang Xu, and Jianqing Fan.
\newblock Simultaneous estimation and group identification for network vector autoregressive model with heterogeneous nodes.
\newblock \emph{Journal of Econometrics}, 249:\penalty0 105564, 2025.

\end{thebibliography}

\newpage
\appendix

\section{Parameter Learning}
\label{appendix_par_learning}

\subsection{Updating \texorpdfstring{$\bm{\mu}$}{TEXT} and \texorpdfstring{$\bm{\phi}$}{TEXT}}
\label{appendix_mu_phi}

In this section, we derive the updates for prior parameter $\bm{\mu}$ and decoder parameter $\bm{\phi}$. Denote the collection of parameters $\{\bm{\phi},\bm{\mu}\}$ as $\bm{\theta}$. We first calculate the gradient of the log-likelihood $L(\bm{\theta})$ with respect to $\bm{\theta}$:
\begin{align*}
\nabla_{\bm{\theta}}\ L(\bm{\theta}) & = \nabla_{\bm{\theta}} \sum_{i \in \mathcal{V}} \ln\big(p(\bm{y}_i)\big)\\
& = \sum_{i \in \mathcal{V}} \frac{1}{p(\bm{y}_i)} \nabla_{\bm{\theta}}\, p(\bm{y}_i) \\
& = \sum_{i \in \mathcal{V}} \frac{1}{p(\bm{y}_i)} \nabla_{\bm{\theta}} \int p(\bm{y}_i,\bm{z}_i) d \bm{z}_i\\
& = \sum_{i \in \mathcal{V}} \frac{1}{p(\bm{y}_i)} \int \nabla_{\bm{\theta}}\, p(\bm{y}_i,\bm{z}_i) d \bm{z}_i\\
& = \sum_{i \in \mathcal{V}} \frac{1}{p(\bm{y}_i)} \int \left[p(\bm{y}_i,\bm{z}_i)\ \frac{1}{p(\bm{y}_i,\bm{z}_i)} \right] \Big[ \nabla_{\bm{\theta}}\, p(\bm{y}_i,\bm{z}_i) \Big] d \bm{z}_i\\
& = \sum_{i \in \mathcal{V}} \frac{1}{p(\bm{y}_i)} \int p(\bm{y}_i,\bm{z}_i) \Big[ \nabla_{\bm{\theta}} \ln\big(p(\bm{y}_i,\bm{z}_i)\big) \Big] d \bm{z}_i\\
& = \sum_{i \in \mathcal{V}} \int \frac{p(\bm{y}_i,\bm{z}_i)}{p(\bm{y}_i)} \Big[\nabla_{\bm{\theta}} \ln\big(p(\bm{y}_i,\bm{z}_i)\big)\Big] d \bm{z}_i\\
& = \sum_{i \in \mathcal{V}} \int p(\bm{z}_i|\bm{y}_i) \Big[\nabla_{\bm{\theta}} \ln\big(p(\bm{y}_i,\bm{z}_i)\big)\Big] d \bm{z}_i\\
& = \sum_{i \in \mathcal{V}} \mathbb{E} \Big( \nabla_{\bm{\theta}} \ln \big( p(\bm{Y}_i|\bm{Z}_i) p(\bm{Z}_i) \big) \Big| \bm{Y}_i \Big) \\
& = \sum_{i \in \mathcal{V}} \mathbb{E} \Big( \nabla_{\bm{\theta}} \ln\big(p(\bm{Y}_i|\bm{Z}_i)\big) \Big| \bm{Y}_i \Big) + \sum_{i \in \mathcal{V}} \mathbb{E} \Big( \nabla_{\bm{\theta}} \ln\big(p(\bm{Z}_i)\big) \Big| \bm{Y}_i \Big).
\end{align*}
Note that the expectation in the gradient is now with respect to the posterior density function $p(\bm{z}_i|\bm{y}_i) \propto p(\bm{y}_i|\bm{z}_i) \times p(\bm{z}_i)$. Furthermore, denote the objective function in (\ref{loss_mu}) as 
$$\mathcal{L}(\bm{\mu}_i) = -\ell(\bm{\phi},\bm{\mu}_i) + \frac{\gamma}{2} \sum_{j \in \mathcal{B}(i)} \lVert \bm{\mu}_i - \bm{\mu}_j - \bm{\nu}_{i,j} + \bm{w}_{i,j} \rVert_2^2.$$
The gradient of $\mathcal{L}(\bm{\mu}_i)$ with respect to the prior parameter $\bm{\mu}_i \in \mathbb{R}^d$ at a specific node $i$ is
\begin{align*}
\nabla_{\bm{\mu}_i}\ \mathcal{L}(\bm{\mu}_i) & = - \mathbb{E}\Big( \nabla_{\bm{\mu}_i} \ln \big(P(\bm{Z}_i)\big) \Big| \bm{Y}_i \Big) + \gamma \sum_{j \in \mathcal{B}(i)} (\bm{\mu}_i - \bm{\mu}_j - \bm{\nu}_{i,j} + \bm{w}_{i,j})\\
& = -\mathbb{E} ( \bm{Z}_i - \bm{\mu}_i | \bm{Y}_i ) + \gamma |\mathcal{B}(i)| \bm{\mu}_i + \gamma \sum_{j \in \mathcal{B}(i)} (- \bm{\mu}_j - \bm{\nu}_{i,j} + \bm{w}_{i,j})\\
& = -\mathbb{E} ( \bm{Z}_i | \bm{Y}_i ) + \big(1+\gamma |\mathcal{B}(i)|\big) \bm{\mu}_i + \gamma \sum_{j \in \mathcal{B}(i)} ( - \bm{\mu}_j - \bm{\nu}_{i,j} + \bm{w}_{i,j}).
\end{align*}
Setting $\nabla_{\bm{\mu}_i}\mathcal{L}(\bm{\mu}_i)=\bm{0}$ and rearranging yields
\begin{align*}
\bm{0} & = -\mathbb{E}( \bm{Z}_i |\bm{Y}_i ) + \big(1+\gamma |\mathcal{B}(i)|\big) \bm{\mu}_i + \gamma \sum_{j \in \mathcal{B}(i)} ( - \bm{\mu}_j - \bm{\nu}_{i,j} + \bm{w}_{i,j}) \\
\big(1+\gamma |\mathcal{B}(i)|\big) \bm{\mu}_i & = \mathbb{E}( \bm{Z}_i | \bm{Y}_i ) - \gamma \sum_{j \in \mathcal{B}(i)} ( - \bm{\mu}_j - \bm{\nu}_{i,j} + \bm{w}_{i,j})\\
\bm{\mu}_i & = \big(1+\gamma |\mathcal{B}(i)|\big)^{-1} \Big[ \mathbb{E}( \bm{Z}_i | \bm{Y}_i ) + \gamma \sum_{j \in \mathcal{B}(i)} ( \bm{\mu}_j + \bm{\nu}_{i,j} - \bm{w}_{i,j}) \Big].
\end{align*}
Similarly, the gradient of $\mathcal{L}(\bm{\phi}, \bm{\mu}) = -L(\bm{\phi}, \bm{\mu})$ with respect to the decoder parameter $\bm{\phi}$ is
$$\nabla_{\bm{\phi}}\ \mathcal{L}(\bm{\phi}, \bm{\mu}) = -\sum_{i \in \mathcal{V}} \mathbb{E} \Big( \nabla_{\bm{\phi}} \ln \big(p(\bm{Y}_i|\bm{Z}_i)\big) \Big| \bm{Y}_i \Big).$$

The posterior expectation in (\ref{sol_mu}) is approximated using Langevin samples generated under the parameter estimates from the preceding iteration. With this Monte Carlo estimate and the neighboring prior parameters held at their preceding values, the update in (\ref{sol_mu}) provides an explicit update obtained by minimizing the resulting node-wise surrogate objective. Thus, the subproblem in (\ref{loss_mu}) is approximately solved using a Monte Carlo EM-style update at each learning iteration.

\subsection{Updating \texorpdfstring{$\bm{\nu}$}{TEXT}}
\label{appendix_nu}

In this section, we present the derivation to update $\bm{\nu}_{i,j}$ for $(i,j) \in \mathcal{E}$, which is equivalent to solving a Group LASSO problem. Denote the objective function in (\ref{loss_nu}) as $\mathcal{L}(\bm{\nu}_{i,j})$. When $\bm{\nu}_{i,j} \neq \bm{0}$, the gradient of $\mathcal{L}(\bm{\nu}_{i,j})$ with respect to $\bm{\nu}_{i,j}$ is
\begin{align*}
\nabla_{\bm{\nu}_{i,j}}\mathcal{L}(\bm{\nu}_{i,j}) & = \lambda \frac{\bm{\nu}_{i,j}}{\|\bm{\nu}_{i,j}\|_2} - \gamma (\bm{\mu}_i - \bm{\mu}_j - \bm{\nu}_{i,j} + \bm{w}_{i,j}). 
\end{align*}
Setting the gradient equal to zero gives
\begin{align}
\bm{0} & = \frac{\lambda}{\|\bm{\nu}_{i,j}\|_2}\bm{\nu}_{i,j} + \gamma \cdot \bm{\nu}_{i,j} - \gamma (\bm{\mu}_i - \bm{\mu}_j + \bm{w}_{i,j}) \nonumber\\
\bm{\nu}_{i,j} & = \left(\frac{\lambda}{\|\bm{\nu}_{i,j}\|_2} + \gamma\right)^{-1} \Big[ \gamma (\bm{\mu}_i - \bm{\mu}_j + \bm{w}_{i,j}) \Big]
\label{eq_nu_temp}
\end{align}
which involves $\bm{\nu}_{i,j}$ on both sides. Calculating the Euclidean norm of (\ref{eq_nu_temp}) and rearranging the terms, we have
\begin{align*}
\|\bm{\nu}_{i,j}\|_2 & = \left(\frac{\lambda + \gamma \|\bm{\nu}_{i,j}\|_2}{\|\bm{\nu}_{i,j}\|_2}\right)^{-1} \left\| \gamma (\bm{\mu}_i - \bm{\mu}_j + \bm{w}_{i,j}) \right\|_2\\
\lambda + \gamma \|\bm{\nu}_{i,j}\|_2  & = \left\| \gamma (\bm{\mu}_i - \bm{\mu}_j + \bm{w}_{i,j}) \right\|_2\\
\|\bm{\nu}_{i,j}\|_2 & = \|\bm{\mu}_i - \bm{\mu}_j + \bm{w}_{i,j}\|_2 - \frac{\lambda}{\gamma}.    
\end{align*}
Plugging the result of $\|\bm{\nu}_{i,j}\|_2$ back into (\ref{eq_nu_temp}), the solution of $\bm{\nu}_{i,j}$ is
\begin{align*}
\bm{\nu}_{i,j} & = \left(\frac{\lambda + \gamma \|\bm{\mu}_i - \bm{\mu}_j + \bm{w}_{i,j}\|_2 - \lambda}{\|\bm{\mu}_i - \bm{\mu}_j + \bm{w}_{i,j}\|_2 - \frac{\lambda}{\gamma}}\right)^{-1} \big[ \gamma (\bm{\mu}_i - \bm{\mu}_j + \bm{w}_{i,j}) \big]\\
& = \left(\frac{\|\bm{\mu}_i - \bm{\mu}_j + \bm{w}_{i,j}\|_2 - \frac{\lambda}{\gamma}}{\gamma \|\bm{\mu}_i - \bm{\mu}_j + \bm{w}_{i,j}\|_2}\right) \big[ \gamma (\bm{\mu}_i - \bm{\mu}_j + \bm{w}_{i,j}) \big]\\
& = \left(1 - \frac{\lambda}{\gamma \|\bm{\mu}_i - \bm{\mu}_j + \bm{w}_{i,j}\|_2}\right) (\bm{\mu}_i - \bm{\mu}_j + \bm{w}_{i,j} ).    
\end{align*}
Furthermore, when $\bm{\nu}_{i,j} = \bm{0}$, a subgradient vector $\bm{b}$ of $\|\bm{\nu}_{i,j}\|_2$ needs to satisfy that $\|\bm{b}\|_2 \leq 1$. Since
\begin{align*}
\bm{0} & \in \lambda \bm{b} - \gamma (\bm{\mu}_i - \bm{\mu}_j + \bm{w}_{i,j}),\\
\lambda \bm{b} & \in \gamma (\bm{\mu}_i - \bm{\mu}_j + \bm{w}_{i,j}),
\end{align*}
we obtain the condition that $\bm{\nu}_{i,j}$ becomes zeros when $\gamma \|\bm{\mu}_i - \bm{\mu}_j + \bm{w}_{i,j}\|_2 \leq \lambda$. Therefore, we can update $\bm{\nu}_{i,j}$ for $(i,j) \in \mathcal{E}$ with the following equation:
$$\bm{\nu}_{i,j} = \left(1 - \frac{\lambda}{\gamma \|\bm{s}_{i,j}\|_2}\right)_+ \bm{s}_{i,j}$$
where $\bm{s}_{i,j} = \bm{\mu}_i - \bm{\mu}_j + \bm{w}_{i,j}$ and $(\cdot)_+ = \max(0,\cdot)$.

\subsection{Neural Decoder}

For flexibility of decoder $p_{\bm{\phi}}(\bm{y}_i|\bm{z}_i)$, the mean $\bm{h}_{\bm{\phi}}(\bm{\cdot})$ is parameterized by a three-layer ReLU neural network. Specifically, $\bm{h}_{\bm{\phi}}(\bm{\cdot})$ is taken as a three-layer rectified linear unit (ReLU) neural network:
\[
\bm{h}_{\bm{\phi}}(\bm{Z}_i) = \bm{W}_3 \text{ReLU} \big( \bm{W}_2 \text{ReLU} ( \bm{W}_1\bm{Z}_i + \bm{b}_1 ) + \bm{b}_2 \big) + \bm{b}_3,
\]
where $\bm{W}_1 \in \mathbb{R}^{h_1 \times d}, \bm{W}_2 \in \mathbb{R}^{h_2 \times h_1}$, and $\bm{W}_3 \in \mathbb{R}^{n \times h_2}$ are weights and $\bm{b}_1 \in \mathbb{R}^{h_1 \times 1}, \bm{b}_2 \in \mathbb{R}^{h_2 \times 1}$, and $\bm{b}_3 \in \mathbb{R}^{n \times 1}$ are bias parameters. Here, $\text{ReLU}(x) = \max(0, x)$ is the element-wise ReLU activation function at $x$, and model parameters are collected into $\bm{\phi}$.

Table~\ref{tab:params} summarizes the number of model parameters and data points, when the node size is $|\mathcal{V}| = 120$, the latent dimension is $d = 3$, and the number of features is $n = 100$. The neural network decoder consists of two hidden layers, each containing $32$ neurons.

\begin{table}[H]
\centering
\caption{Number of parameters and data points.}
\label{tab:params}
\resizebox{0.6\columnwidth}{!}{%
\begin{tabular}{l l c}
\toprule
Notation & Description & Quantity \\
\midrule
$\ \bm{\phi}\ $ & Neural networks & 
\makecell{$32 \times 3 + 32 \times 32 + 100 \times 32\ + $ \\ $32 \times 1 + 32 \times 1 + 100 \times 1 = 4{,}484$} \\
$\{\bm{\mu}_i\}_{i \in \mathcal{V}}$ & Prior means & $|\mathcal{V}| \times 3 = 360$ \\
\midrule
$\{\bm{Y}_i\}_{i \in \mathcal{V}}$ & Nodal features  & $|\mathcal{V}| \times 100 = 12{,}000$ \\
$\mathcal{G} = (\mathcal{V}, \mathcal{E})$ & Number of edges & $|\mathcal{E}|$ \\
\bottomrule
\end{tabular}
}
\end{table}

The negative log-likelihood that involves latent variable $\bm{Z}_i \in \mathbb{R}^d$ and observed data $\bm{Y}_i \in \mathbb{R}^n$ is lower bounded by
\begin{align}
- \ln \big( p(\bm{y}_i) \big) & = - \ln \left[ \int \frac{1}{(2\pi)^{n/2}} \exp\left( -\frac{1}{2} \| \bm{y}_i - \bm{h}_{\bm\phi}(\bm{z}_i)\|_2^2 \right) \cdot \frac{1}{(2\pi)^{d/2}} \exp \left( -\frac{1}{2} \|\bm{z}_i - \bm{\mu}_i\|_2^2 \right)d\bm{z}_i \right] \nonumber \\
& \geq - \ln \left[ \frac{1}{(2\pi)^{(n+d)/2} } \int \exp \left( -\frac{1}{2} \|\bm{z}_i - \bm{\mu}_i\|_2^2 \right) d\bm{z}_i \right] \label{exp_less_one}\\
& \geq - \ln \left[ \frac{1}{(2\pi)^{n/2} } \right] \nonumber\\
& \geq \frac{n}{2} \ln (2\pi) \nonumber
\end{align}
where (\ref{exp_less_one}) follows because $\exp(-\frac{1}{2} x^2) \leq 1$. While the likelihood function of some latent space models can be unbounded \citep{mattei2018leveraging}, the minimization problem in (\ref{eqn:original}) is valid and meaningful.

\subsection{Computation Complexity}
\label{ADMM_algo}

The algorithm to solve (\ref{eqn:original_sub_to}) via ADMM is presented in Algorithm \ref{alg:ADMM_latent}. The computation complexity is at least of order $O(A( |\mathcal{V}| skd + B |\mathcal{V}| sn + |\mathcal{V}| sd + |\mathcal{E}| d))$; additional gradient computation for neural networks is required in sub-routines. 

Elaborating, in each of $A$ iterations of ADMM, Langevin dynamics generate $s$ samples for each $i$, each requiring $k$ MCMC steps (and has dimension of $d$). Next, the decoder parameters $\bm{\phi}$ are updated via $B$ iterations of the Adam optimizer across all nodes using the $s$ generated samples. The output of neural networks has length $n$ for each node. Then, the node-specific prior parameters $\bm{\mu}_i$ are updated at each iteration using the MCMC samples, which incurs a computational cost that is linear in $|\mathcal{V}|sd$. Finally, the slack variables $\bm{\nu}_{i,j}$ are updated, resulting in a complexity of $O(|\mathcal{E}|d)$. The scaled dual variables $\bm{w}_{i,j}$ are updated in closed form with similar computation complexity. 

In summary, the ADMM procedure decomposes a complex optimization problem into smaller components, targeting the smaller components individually. While our methods are computationally demanding, substantial clustering performance is gained, as demonstrated in the simulation study.

\begin{algorithm}
\caption{Latent space graph-fused LASSO}
\label{alg:ADMM_latent} 

\begin{algorithmic}[1]

\State \textbf{Input}: learning iterations $A,B,D$, tuning parameter $\lambda$, penalty parameter $\gamma$, learning rate $\eta$, observed data $\{\bm{y}_i\}_{i \in \mathcal{V}}$, initialization $\{ \bm{\phi}^{(1)},\bm{\mu}^{(1)},\bm{\nu}^{(1)},\bm{w}^{(1)} \}$

\For{$a = 1,\cdots, A$}

    \For{$i \in \mathcal{V}$}
        \State draw samples $\bm{z}_{1,i}, \dots, \bm{z}_{s,i}$ according to (\ref{Langevin_sampling})
    \EndFor

    \For{$b = 1,\dots,B$}
        \State $\bm{\phi}_{(b+1)} = \mathrm{Adam}\big( \bm{\phi}_{(b)}, \nabla_{\bm{\phi}}\ \mathcal{L}(\bm{\phi},\bm{\mu}), \eta \big)$
    \EndFor

    \State $\bm{\mu}_i^{(a+1)} = \big(1+\gamma |\mathcal{B}(i)| \big)^{-1} \Big[ s^{-1}\sum_{u=1}^s \bm{z}_{u,i} + \gamma \sum_{j \in \mathcal{B}(i)} ( \bm{\mu}_j^{(a)} + \bm{\nu}_{i,j}^{(a)} - \bm{w}_{i,j}^{(a)}) \Big],\ \forall\ i \in \mathcal{V}$

    \State Let $\bm{\nu}_{i,j}^{(a+1)}$ be updated according to (\ref{update_nu}) for $(i,j) \in \mathcal{E}$

    \State $\bm{w}_{i,j}^{(a+1)} = \bm{\mu}_i^{(a+1)} - \bm{\mu}_j^{(a+1)} - \bm{\nu}_{i,j}^{(a+1)} + \bm{w}_{i,j}^{(a)},\ \forall\ (i,j) \in \mathcal{E}$

\EndFor

\State $\hat{\bm{\mu}} \leftarrow \bm{\mu}^{(a+1)}$

\State \textbf{Output}: learned parameters $\hat{\bm{\mu}}$

\end{algorithmic}
\end{algorithm}

\section{Excess Risk}
\label{appendix_excess_risk}

In this section, we provide the proof for Theorem \ref{thm1}.

Write $N\coloneqq |\mathcal{V}|$ and let $\Theta(C,U)$ denote the constraint set in (\ref{constraints}). For $\bm{\theta}=(\bm{\phi},\bm{\mu})\in\Theta(C,U)$, define
\[
\ell_{\bm{\theta},i}(\bm{y}_i)
\coloneqq -\ln [p_{\bm{\phi},\bm{\mu}_i}(\bm{y}_i)],
\qquad
f_{\bm{\theta},i}(\bm{y})
\coloneqq \ell_{\bm{\theta},i}(\bm{y}_i)-\ell_{\bm{\theta}^*,i}(\bm{y}_i),
\]
where $\bm{\theta}^*=(\bm{\phi}^*,\bm{\mu}^*)$. Introduce the population and empirical operators
\[
P f_{\bm{\theta}} \coloneqq \frac{1}{N}\sum_{i\in\mathcal{V}}\mathbb{E}\bigl[f_{\bm{\theta},i}(\bm{Y}_i)\bigr],
\qquad
P_N f_{\bm{\theta}} \coloneqq \frac{1}{N}\sum_{i\in\mathcal{V}}f_{\bm{\theta},i}(\bm{y}_i).
\]
When $\bm{\theta}$ is data-dependent, $P f_{\bm{\theta}}$ denotes the same integral with $\bm{\theta}$ held fixed at its realized value; equivalently, it is evaluated on an independent copy of the sample. Since the entropy term $\mathbb{E}[\ln (p_i^*(\bm{Y}_i))]$ does not depend on $\bm{\theta}$,
\begin{align}
P f_{\bm{\theta}} & = \frac{1}{N}\sum_{i\in\mathcal{V}}\mathbb{E}\bigl[\ell_{\bm{\theta},i}(\bm{Y}_i)-\ell_{\bm{\theta}^*,i}(\bm{Y}_i)\bigr] \nonumber\\
& = \frac{1}{N}\sum_{i\in\mathcal{V}}\Big\{ \mathbb{E}\bigl[\ell_{\bm{\theta},i}(\bm{Y}_i)\bigr] - \mathbb{E}\bigl[\ell_{\bm{\theta}^*,i}(\bm{Y}_i)\bigr] + \mathbb{E}\bigl[\ln (p_i^*(\bm{Y}_i))\bigr] - \mathbb{E}\bigl[\ln (p_i^*(\bm{Y}_i))\bigr] \Big\} \nonumber\\
& =\frac{1}{N}\sum_{i\in\mathcal{V}} \Big\{ \mathrm{KL}\bigl(p_i^*\,\big|\,p_{\bm{\phi},\bm{\mu}_i}(\bm{y}_i)\bigr) -\mathrm{KL}\bigl(p_i^*\,\big|\,p_{\bm{\phi}^*,\bm{\mu}_i^*}(\bm{y}_i)\bigr)
\Big\}.
\label{eq:population-kl-identity}
\end{align}

By empirical optimality 
$$\sum_{i \in \mathcal{V}} \ell_{\hat{\bm{\theta}},i}(\bm{y}_i) \leq \sum_{i \in \mathcal{V}} \ell_{\bm{\theta}^*,i}(\bm{y}_i),$$ 
we have 
$$P_N f_{\hat{\bm{\theta}}} = \frac{1}{N}\sum_{i\in\mathcal{V}}f_{\hat{\bm{\theta}},i}(\bm{y}_i) = \frac{1}{N}\sum_{i\in\mathcal{V}} \Bigl( \ell_{\hat{\bm{\theta}},i}(\bm{y}_i)-\ell_{\bm{\theta}^*,i}(\bm{y}_i) \Bigr) \leq 0.$$ 
Therefore,
\begin{equation}
\label{eq:basic-inequality-subgaussian}
P f_{\hat{\bm{\theta}}}
\leq (P-P_N)f_{\hat{\bm{\theta}}}
\leq \sup_{\bm{\theta}\in\Theta(C,U)}(P-P_N)f_{\bm{\theta}}.
\end{equation}
The vectors $\bm{Y}_i$ are independent across nodes, so the usual symmetrization inequality for independent, not necessarily identically distributed, observations gives
\begin{align}
\mathbb{E} \bigl[ P f_{\hat{\bm{\theta}}} \bigr] & \leq \mathbb{E}\left[ \sup_{\bm{\theta}\in\Theta(C,U)} (P-P_N)f_{\bm{\theta}}\right] \nonumber\\
& = \mathbb{E}\left[ \sup_{\bm{\theta}\in\Theta(C,U)} \left( \frac{1}{N}\sum_{i\in\mathcal{V}}\mathbb{E}\bigl[f_{\bm{\theta},i}(\bm{Y}_i)\bigr] - \frac{1}{N}\sum_{i\in\mathcal{V}}f_{\bm{\theta},i}(\bm{Y}_i)\right)\right] \nonumber\\
& \leq 2\,\mathbb{E}\left[ \mathbb{E}_{\bm{\xi}}\left( \sup_{\bm{\theta}\in\Theta(C,U)} \frac{1}{N}\sum_{i\in\mathcal{V}}\xi_i f_{\bm{\theta},i}(\bm{Y}_i) \right)\right]
\label{eq:symmetrization-subgaussian}
\end{align}
where $\{\xi_i\}_{i\in\mathcal{V}}$ are independent Rademacher variables, independent of the data.

We next establish a data-dependent Lipschitz bound for the log marginal density. The subscript for node $i$ is omitted for notation simplicity. Let
\[
q_{\bm{\mu}}(\bm{z})
\coloneqq (2\pi)^{-d/2}\exp\left(-\frac{1}{2}\|\bm{z}-\bm{\mu}\|_2^2\right)
\]
and write
\begin{align*}
p_{\bm{\phi},\bm{\mu}}(\bm{y})
& =\frac{1}{(2\pi)^{(n+d)/2}} \int \exp\left\{-\frac{1}{2}\|\bm{y}-\bm{h}_{\bm{\phi}}(\bm{z})\|_2^2 -\frac{1}{2}\|\bm{z}-\bm{\mu}\|_2^2\right\}d\bm{z}\\
& = \int (2\pi)^{-n/2} \exp\left\{-\frac{1}{2}\|\bm{y}\|_2^2 + \bm{y}^{\top}\bm{h}_{\bm{\phi}}(\bm{z}) -\frac{1}{2}\|\bm{h}_{\bm{\phi}}(\bm{z})\|_2^2\right\} q_{\bm{\mu}}(\bm{z})\,d\bm{z}\\
& = (2\pi)^{-n/2} \exp\left(-\frac{1}{2}\|\bm{y}\|_2^2\right)  \int \exp \left\{ \bm{y}^{\top}\bm{h}_{\bm{\phi}}(\bm{z}) -\frac{1}{2}\|\bm{h}_{\bm{\phi}}(\bm{z})\|_2^2\right\} q_{\bm{\mu}}(\bm{z})\,d\bm{z}\\
& = (2\pi)^{-n/2}\exp\left(-\frac{1}{2}\|\bm{y}\|_2^2\right) Q_{\bm{\phi},\bm{\mu}}(\bm{y})
\end{align*}
where the term containing model parameters is
$$Q_{\bm{\phi},\bm{\mu}}(\bm{y}) \coloneqq \int \exp\left\{\bm{y}^{\top}\bm{h}_{\bm{\phi}}(\bm{z}) -\frac{1}{2}\|\bm{h}_{\bm{\phi}}(\bm{z})\|_2^2\right\} q_{\bm{\mu}}(\bm{z})\,d\bm{z}.$$
Because $\|\bm{h}_{\bm{\phi}}(\bm{z})\|_\infty\leq C$, we have
$$\|\bm{h}_{\bm{\phi}}(\bm{z})\|_2^2 = \sum_{j=1}^n h_{\bm{\phi},j}(\bm{z})^2 \leq nC^2,$$
and
$$\bm{y}^{\top}\bm{h}_{\bm{\phi}}(\bm{z}) = \sum_{j=1}^n y_j \cdot h_{\bm{\phi},j}(\bm{z}) \geq  - \sum_{j=1}^n |y_j| \cdot |h_{\bm{\phi},j}(\bm{z})| \geq - C \|\bm{y}\|_1.$$
Hence,
\begin{align}
Q_{\bm{\phi},\bm{\mu}}(\bm{y}) & \geq \exp\left(-C\|\bm{y}\|_1-\frac{nC^2}{2}\right) \int q_{\bm{\mu}}(\bm{z})\,d\bm{z} \nonumber\\
& = \exp\left(-C\|\bm{y}\|_1-\frac{nC^2}{2}\right).
\label{eq:Q-lower}
\end{align}
Define
\[
\bar q(\bm{z})\coloneqq \sup_{\|\bm{\mu}\|_\infty\leq C} q_{\bm{\mu}}(\bm{z}),
\qquad
K_q\coloneqq\int\bar q(\bm{z})\,d\bm{z},
\qquad
\widetilde q(\bm{z})\coloneqq \frac{1}{K_q}\bar{q}(\bm{z}).
\]
For the $k$th coordinate of $\bm{\mu} \in \mathbb{R}^d$ such that $|\mu_k| \leq C$, then
\begin{align*}
|z_k - \mu_k| & \geq \big| |z_k| - |\mu_k| \big|,\\
|z_k - \mu_k| & \geq (|z_k| - C)_+,\\ 
\|\bm{z} - \bm{\mu}\|_2^2 & \geq \sum_{k=1}^d(|z_k| - C)_+^2, 
\end{align*}
and
\[
\bar q(\bm{z})
\leq (2\pi)^{-d/2}
\exp\left\{-\frac{1}{2}\sum_{k=1}^d (|z_k|-C)_+^2\right\}
\]
where $(\cdot)_+ = \max(\cdot,0)$. Also, the constant $K_q$ is finite. By factorizing the integral over coordinates,
\begin{align*}
K_q & \leq (2\pi)^{-d/2} \prod_{k=1}^d \int_{\mathbb R} \exp\left\{-\frac12(|z_k|-C)_+^2\right\}dz_k\\
& =\left(1+\frac{2C}{\sqrt{2\pi}}\right)^d<\infty.
\end{align*}
Hence, $\widetilde q(\bm z)=K_q^{-1}\bar q(\bm z)$ defines a probability density on $\mathbb R^d$.

Set $a_{\bm{\phi}}(\bm{y},\bm{z})
=\bm{y}^{\top}\bm{h}_{\bm{\phi}}(\bm{z})
- \frac{1}{2}\|\bm{h}_{\bm{\phi}}(\bm{z})\|_2^2$. Then
\begin{align*}
a_{\bm{\phi}} - a_{\widetilde{\bm{\phi}}}
=\bm{y}^{\top}\left( \bm{h}_{\bm{\phi}}(\bm{z}) -\bm{h}_{\widetilde{\bm{\phi}}}(\bm{z}) \right)
-\frac{1}{2} \Big( \|\bm{h}_{\bm{\phi}}(\bm{z})\|_2^2 - \|\bm{h}_{\widetilde{\bm{\phi}}}(\bm{z})\|_2^2 \Big).
\end{align*}
For the first term with H\"older's inequality,
\begin{align*}
\left| \bm{y}^{\top} \left( \bm{h}_{\bm{\phi}}(\bm{z}) -\bm{h}_{\widetilde{\bm{\phi}}}(\bm{z}) \right) \right| \leq \|\bm{y}\|_{\infty} \, \left\|\bm{h}_{\bm{\phi}}(\bm{z}) - \bm{h}_{\widetilde{\bm{\phi}}}(\bm{z})\right\|_1.
\end{align*}

For the second term with $\|\bm{h}_{\bm{\phi}}(\bm{z})\|_\infty\leq C$,
\begin{align*}
\frac{1}{2} \left| \|\bm{h}_{\bm{\phi}}(\bm{z})\|_2^2 - \|\bm{h}_{\widetilde{\bm{\phi}}}(\bm{z})\|_2^2 \right| & \leq \frac{1}{2} \sum_{j=1}^n \left| h_{\bm{\phi},j}(\bm{z})- h_{\widetilde{\bm{\phi}},j}(\bm{z}) \right| \cdot \left| h_{\bm{\phi},j}(\bm{z}) + h_{\widetilde{\bm{\phi}},j}(\bm{z}) \right|\\
& \leq C \, \|\bm{h}_{\bm{\phi}}(\bm{z})-\bm{h}_{\widetilde{\bm{\phi}}}(\bm{z})\|_1.
\end{align*}
Then,
\[
\left|a_{\bm{\phi}}-a_{\widetilde{\bm{\phi}}}\right| \leq \big(\|\bm{y}\|_\infty+C\big) \left\|\bm{h}_{\bm{\phi}}(\bm{z})-\bm{h}_{\widetilde{\bm{\phi}}}(\bm{z})\right\|_1.
\]
Since $a_{\bm{\phi}}
=\bm{y}^{\top}\bm{h}_{\bm{\phi}}(\bm{z})
- \frac{1}{2}\|\bm{h}_{\bm{\phi}}(\bm{z})\|_2^2 \leq \bm{y}^{\top}\bm{h}_{\bm{\phi}}(\bm{z})$, we have $a_{\bm{\phi}} \leq C\|\bm{y}\|_1$ and
$$\max\left\{a_{\bm{\phi}},\, a_{\widetilde{\bm{\phi}}}\right\} \leq C\|\bm{y}\|_1.$$

Also, with 
$$Q_{\bm{\phi},\bm{\mu}}(\bm{y})= \int \exp\big(a_{\bm{\phi}}(\bm{y},\bm{z})\big) q_{\bm{\mu}}(\bm{z})\,d\bm{z},$$ 
we have
\begin{align*}
\left|Q_{\bm{\phi},\bm{\mu}}(\bm{y}) - Q_{\widetilde{\bm{\phi}},\bm{\mu}}(\bm{y}) \right| & = \left| \int \left[\exp(a_{\bm{\phi}}) - \exp(a_{\widetilde{\bm{\phi}}})\right]\, q_{\bm{\mu}}(\bm{z})\,d\bm{z} \right|\\
& \leq \int \left| \exp(a_{\bm{\phi}}) - \exp(a_{\widetilde{\bm{\phi}}})\right|\, q_{\bm{\mu}}(\bm{z})\,d\bm{z}.
\end{align*}

Using $|e^u-e^v|\leq e^{\max(u,v)}|u-v|$,
\begin{align*}
\left|Q_{\bm{\phi},\bm{\mu}}(\bm{y})
-Q_{\widetilde{\bm{\phi}},\bm{\mu}}(\bm{y})\right|
& \leq \int \exp\left(\max\{a_{\bm{\phi}},\, a_{\widetilde{\bm{\phi}}}\}\right) \left| a_{\bm{\phi}} - a_{\widetilde{\bm{\phi}}}\right|\, q_{\bm{\mu}}(\bm{z})\,d\bm{z} \\
& \leq e^{C \|\bm{y}\|_1} \big(\|\bm{y}\|_\infty+C\big) \int \|\bm{h}_{\bm{\phi}}(\bm{z})-\bm{h}_{\widetilde{\bm{\phi}}}(\bm{z})\|_1\, q_{\bm{\mu}}(\bm{z})\,d\bm{z}
\end{align*}

For two admissible pairs $(\bm{\phi},\bm{\mu})$ and $(\widetilde{\bm{\phi}},\widetilde{\bm{\mu}})$, define
\[
D_h(\bm{\phi},\widetilde{\bm{\phi}})
\coloneqq
\sum_{j=1}^n
\|\bm{h}_{\bm{\phi},j}-\bm{h}_{\widetilde{\bm{\phi}},j}\|_{L_1(\widetilde{q})}
= \sum_{j=1}^n \int \left|h_{\bm{\phi},j}(\bm{z})-h_{\widetilde{\bm{\phi}},j}(\bm{z})\right|\, \widetilde{q}(\bm{z})\,d\bm{z}
\]
Since $q_{\bm{\mu}}(\bm{z}) \leq K_q \widetilde{q}(\bm{z})$,
\begin{align*}
\int \|\bm{h}_{\bm{\phi}}(\bm{z})-\bm{h}_{\widetilde{\bm{\phi}}}(\bm{z})\|_1\, q_{\bm{\mu}}(\bm{z})\,d\bm{z} & \leq K_q \int \|\bm{h}_{\bm{\phi}}(\bm{z})-\bm{h}_{\widetilde{\bm{\phi}}}(\bm{z})\|_1\, \widetilde{q}(\bm{z})\,d\bm{z}\\
& = K_q \sum_{j=1}^n \int \left|h_{\bm{\phi},j}(\bm{z})-h_{\widetilde{\bm{\phi}},j}(\bm{z})\right|\, \widetilde{q}(\bm{z})\,d\bm{z}\\
& = K_q \, D_h(\bm{\phi},\widetilde{\bm{\phi}})
\end{align*}
Therefore,
\begin{align}
\left|Q_{\bm{\phi},\bm{\mu}}(\bm{y})
-Q_{\widetilde{\bm{\phi}},\bm{\mu}}(\bm{y})\right| & \leq K_q e^{C\|\bm{y}\|_1}\big(\|\bm{y}\|_\infty+C\big) D_h(\bm{\phi},\widetilde{\bm{\phi}}).
\label{eq:Q-h-difference}
\end{align}

For the prior means, Pinsker's inequality and
$$\mathrm{KL}\big(\mathcal{N}(\bm{\mu},\bm{I}_d)\,|\,\mathcal{N}(\widetilde{\bm{\mu}},\bm{I}_d)\big) = \frac{1}{2}\|\bm{\mu}-\widetilde{\bm{\mu}}\|_2^2$$ imply
\[
\int \left|q_{\bm{\mu}}(\bm{z})-q_{\widetilde{\bm{\mu}}}(\bm{z})\right|\,d\bm{z}
\leq \|\bm{\mu}-\widetilde{\bm{\mu}}\|_2.
\]
Consequently,
\begin{align}
\left|Q_{\widetilde{\bm{\phi}},\bm{\mu}}(\bm{y})
-Q_{\widetilde{\bm{\phi}},\widetilde{\bm{\mu}}}(\bm{y})\right|
& = \left| \int \exp\left(a_{\widetilde{\bm{\phi}}}\right) \left[q_{\bm{\mu}}(\bm{z}) - q_{\widetilde{\bm{\mu}}}(\bm{z})\right] d\bm{z} \right| \nonumber\\
& \leq \int \exp\left(a_{\widetilde{\bm{\phi}}}\right) \left|q_{\bm{\mu}}(\bm{z}) - q_{\widetilde{\bm{\mu}}}(\bm{z})\right| d\bm{z}  \nonumber\\
& \leq e^{C\|\bm{y}\|_1} \|\bm{\mu}-\widetilde{\bm{\mu}}\|_2.
\label{eq:Q-mu-difference}
\end{align}
Combining (\ref{eq:Q-lower})-(\ref{eq:Q-mu-difference}) with
$|\ln (u)-\ln (v)|\leq |u-v|/\min(u,v)$, and using the triangle inequality, yields
\begin{align}
\big| \ell_{\bm{\phi},\bm{\mu}}(\bm{y}) - \ell_{\widetilde{\bm{\phi}},\widetilde{\bm{\mu}}}(\bm{y})\big| & = \big|\ln (Q_{\bm{\phi},\bm{\mu}}(\bm{y})) - \ln (Q_{\widetilde{\bm{\phi}},\widetilde{\bm{\mu}}}(\bm{y})) \big| \nonumber\\
& \leq \big|\ln (Q_{\bm{\phi},\bm{\mu}}(\bm{y})) -\ln (Q_{\widetilde{\bm{\phi}},\bm{\mu}}(\bm{y}))\big|   \,+\,  \big| \ln (Q_{\widetilde{\bm{\phi}},\bm{\mu}}(\bm{y})) - \ln  (Q_{\widetilde{\bm{\phi}},\widetilde{\bm{\mu}}}(\bm{y})) \big| \nonumber\\
& \leq e^{C \|\bm{y}\|_1 + \frac{nC^2}{2} } \left( \big| Q_{\bm{\phi},\bm{\mu}}(\bm{y}) -Q_{\widetilde{\bm{\phi}},\bm{\mu}}(\bm{y})\big| \,+\,  \big|  Q_{\widetilde{\bm{\phi}},\bm{\mu}}(\bm{y}) - Q_{\widetilde{\bm{\phi}},\widetilde{\bm{\mu}}}(\bm{y}) \big| \right) \nonumber\\
& \leq e^{C \|\bm{y}\|_1 + \frac{nC^2}{2} } \left( K_q e^{C\|\bm{y}\|_1}\big(\|\bm{y}\|_\infty+C\big) D_h(\bm{\phi},\widetilde{\bm{\phi}}) \,+\,  e^{C\|\bm{y}\|_1} \|\bm{\mu}-\widetilde{\bm{\mu}}\|_2 \right) \nonumber\\
& \leq e^{2C \|\bm{y}\|_1 + \frac{nC^2}{2} } \left( K_q \big(\|\bm{y}\|_\infty+C\big) D_h(\bm{\phi},\widetilde{\bm{\phi}}) \,+\, \|\bm{\mu}-\widetilde{\bm{\mu}}\|_2 \right) \nonumber\\
& \leq e^{2C \|\bm{y}\|_1 + \frac{nC^2}{2} } \max\big\{K_q(1 + \|\bm{y}\|_\infty)(1+C),\, 1 \big\} \left( D_h(\bm{\phi},\widetilde{\bm{\phi}}) \,+\, \|\bm{\mu}-\widetilde{\bm{\mu}}\|_2 \right) \nonumber\\
& \leq \displaystyle A(\bm{y}) \left( D_h(\bm{\phi},\widetilde{\bm{\phi}}) + \|\bm{\mu}-\widetilde{\bm{\mu}}\|_2\right)
\label{eq:log-likelihood-Lipschitz}
\end{align}
where
\[
A(\bm{y})
\coloneqq K_0(1+\|\bm{y}\|_\infty)\exp(2C\|\bm{y}\|_1)
\]
for a constant $K_0 = \exp(\frac{nC^2}{2})\max\{K_q(1+C),1\}$ depending only on $C$, $n$, and $d$. Moreover,
\[
A(\bm{y})^2 \leq K\sum_{j=1}^n(1+y_j^2) \exp\left(4C\sum_{k=1}^n|y_k|\right)
\]
for some constant $K>0$.

The sub-Gaussian condition makes this random Lipschitz factor square integrable uniformly over nodes. Note that
\begin{align*}
\left(\frac{|x|}{\sigma} - a\sigma\right)^2 & \geq 0,\\
\frac{x^2}{\sigma^2} - 2 a |x| +a^2\sigma^2 & \geq 0,\\
a |x| & \leq \frac{x^2}{2\sigma^2} + \frac{a^2\sigma^2}{2},\\
\exp(a |x|) & \leq \exp\left({\frac{x^2}{2\sigma^2}}\right) \, \exp\left({\frac{a^2\sigma^2}{2}}\right).
\end{align*}
For $\|X\|_{\psi_2}\leq\sigma$ and $a>0$, we have
\begin{align}
\mathbb{E}\left(e^{a|X|}\right) & \leq \exp\left({\frac{a^2\sigma^2}{2}}\right) \mathbb{E}\left[\exp\left({\frac{x^2}{2\sigma^2}}\right) \right] \nonumber\\
& \leq \exp\left({\frac{a^2\sigma^2}{2}}\right) \sqrt{\mathbb{E}\left[\exp\left({\frac{x^2}{\sigma^2}}\right) \right]} \nonumber\\
& \leq \sqrt{2}\,e^{a^2\sigma^2/2}.  
\label{exponential-moment bound}
\end{align}

Taking expectation and using generalized H\"older's inequality, we have
\begin{align*}
\mathbb E[A(\bm Y)^2] & \leq K\sum_{j=1}^n \mathbb E\left[ (1+Y_{j}^2) \prod_{k=1}^n e^{4C|Y_{k}|} \right]\\
& \leq K\sum_{j=1}^n \left\{\mathbb E\left[(1+Y_{j}^2)^{n+1}\right]\right\}^{\frac{1}{n+1}} \prod_{k=1}^n \left\{\mathbb E \left[e^{4C(n+1)|Y_{k}|}\right]\right\}^{\frac{1}{n+1}}\\
& \leq K\sum_{j=1}^n \left\{\mathbb E\left[(1+Y_{j}^2)^{n+1}\right]\right\}^{\frac{1}{n+1}} \prod_{k=1}^n \left\{\sqrt{2}\exp\left\{8C^2(n+1)^2\sigma^2 \right\}\right\}^{\frac{1}{n+1}}
\end{align*}
where the last inequality follows from (\ref{exponential-moment bound}). 

Using $x^2\leq c_a e^{a|x|}$ for some constant $c_a>0$, 
\begin{align*}
\mathbb E[A(\bm Y)^2] & \leq K\sum_{j=1}^n \left\{\mathbb E\left[\left(1+c_a e^{a|Y_j|}\right)^{n+1}\right]\right\}^{\frac{1}{n+1}} \prod_{k=1}^n \left\{\sqrt{2}\exp\left\{ 8C^2(n+1)^2\sigma^2\right\}\right\}^{\frac{1}{n+1}}\\
& \leq K\sum_{j=1}^n \left\{\mathbb E\left[\left(1+c_a\right)^{n+1} \left(e^{a(n+1)|Y_j|}\right) \right]\right\}^{\frac{1}{n+1}} \prod_{k=1}^n \left\{\sqrt{2}\exp\left\{ 8C^2(n+1)^2\sigma^2\right\}\right\}^{\frac{1}{n+1}}\\
& \leq K\sum_{j=1}^n \left\{\left(1+c_a\right)^{n+1} \sqrt{2} \exp\left(\frac{a^2(n+1)^2 \sigma^2}{2}\right)\right\}^{\frac{1}{n+1}} \prod_{k=1}^n \left\{\sqrt{2}\exp\left\{ 8C^2(n+1)^2\sigma^2\right\}\right\}^{\frac{1}{n+1}}
\end{align*}
where the last inequality also follows from (\ref{exponential-moment bound}).

The subscript for node $i$ is now restored in the following proof. Since $n=O(1)$, the random Lipschitz factors satisfy
\begin{equation}
\label{eq:A-second-moment}
\sup_{i\in\mathcal{V}}\mathbb{E}\bigl[A(\bm{Y}_i)^2\bigr] \leq K_A^2
\end{equation}
for a constant $K_A$ depending only on $C$, $\sigma$, $n$, and $d$.

Condition on the observations, let $A_i \coloneqq A(\bm{y}_i)$ and define
\[
S_N\coloneqq\left(\frac{1}{N}\sum_{i\in\mathcal{V}}A_i^2\right)^{1/2},
\qquad
p_i\coloneqq\frac{A_i^2}{\sum_{a\in\mathcal{V}}A_a^2},
\qquad
\|\bm{u}\|_{L_2(\bm{p})} \coloneqq\left(\sum_{i\in\mathcal{V}}p_i u_i^2\right)^{1/2}.
\]
Note that $\sum_{i\in\mathcal{V}}p_i = 1$ and $A_i^2 / N = S_N^2 p_i$.

For vectors indexed by the nodes, let $\bm{f}_{\bm{\theta}}=(f_{\bm{\theta},i}(\bm{y}_i))_{i\in\mathcal{V}}$ and
$$\|\bm{v}\|_{\mathcal{V}}=\sqrt{\frac{1}{|\mathcal{V}|}\sum_{i \in \mathcal{V}} v_i^2}.$$

Applying (\ref{eq:log-likelihood-Lipschitz}), followed by Minkowski's inequality, gives
\begin{align}
\|\bm{f}_{\bm{\theta}}-\bm{f}_{\widetilde{\bm{\theta}}}\|_{\mathcal{V}} & = \sqrt{ \frac{1}{N} \sum_{i \in \mathcal{V}} \Big|f_{\bm{\theta},i}(\bm{y}_i) - f_{\widetilde{\bm{\theta}},i}(\bm{y}_i)\Big|^2} \nonumber\\
& = \sqrt{ \frac{1}{N} \sum_{i \in \mathcal{V}} \Big|\ell_{\bm{\theta},i}(\bm{y}_i) - \ell_{\widetilde{\bm{\theta}},i}(\bm{y}_i)\Big|^2} \nonumber\\
& \leq \sqrt{ \sum_{i \in \mathcal{V}} \left( \frac{1}{N} A_i^2 \right) \left( D_h(\bm{\phi},\widetilde{\bm{\phi}}) + \|\bm{\mu}_i-\widetilde{\bm{\mu}}_i\|_2\right)^2}\nonumber\\
& = \sqrt{ \sum_{i \in \mathcal{V}} \big( S_N^2 p_i \big) \left( D_h(\bm{\phi},\widetilde{\bm{\phi}}) + \|\bm{\mu}_i-\widetilde{\bm{\mu}}_i\|_2\right)^2}\nonumber\\
&\leq S_N \left[ D_h(\bm{\phi},\widetilde{\bm{\phi}}) +\left\{\sum_{i\in\mathcal{V}}p_i \|\bm{\mu}_i-\widetilde{\bm{\mu}}_i\|_2^2\right\}^{1/2} \right]\nonumber\\
&\leq S_N \left[ \sum_{j=1}^n \|\bm{h}_{\bm{\phi},j}-\bm{h}_{\widetilde{\bm{\phi}},j}\|_{L_1(\widetilde q)} +\sum_{k=1}^d \|\bm{\mu}_{k,\cdot}-\widetilde{\bm{\mu}}_{k,\cdot}\|_{L_2(\bm{p})} \right]
\label{eq:weighted-embedding}
\end{align}
where $\bm{\mu}_{k,\cdot} = (\mu_{k,i})_{i\in\mathcal{V}}$ denote the vector of the $k$th prior-mean coordinates across all nodes.

We now bound the covering numbers in (\ref{eq:weighted-embedding}). The neural network governed by $\bm{h}_{ \bm{\phi}}(\bm{\cdot})$ is assumed to have $L$ layers, $B$ neurons across all layers, and $W$ total parameters. Let
\[
\mathcal{H}_j = \big\{\bm{h}_{\bm{\phi},j}:\bm{\phi}\in\mathcal{P},\ \|\bm{h}_{\bm{\phi}}\|_\infty\leq C \big\}.
\]
Theorem 7 of \cite{bartlett2019nearly}, together with the standard reduction from VC dimension to pseudodimension for real-valued neural networks, gives
\[
v_H\coloneqq\max_{1\leq j\leq n}\operatorname{Pdim}(\mathcal{H}_j) \leq K_1 LW\ln(eB),
\]
for some constant $K_1>0$.

The standard pseudodimension covering bound therefore implies, uniformly over probability measures $Q$,
\begin{equation}
\label{eq:network-cover}
\ln \left[ N\bigl(\varepsilon,\mathcal{H}_j,L_1(Q)\bigr)\right] \leq K_2v_H\ln\left(1+\frac{K_2C}{\varepsilon}\right),
\qquad 0<\varepsilon\leq 2C.
\end{equation}

Let $\pi(1),\ldots,\pi(N)$ be a DFS ordering of the connected graph. Lemma 1 of \cite{padilla2018dfs} gives, for every admissible $\bm{\mu}$,
\begin{equation}
\label{eq:dfs-correct-factor}
\sum_{r=1}^{N-1} \|\bm{\mu}_{\pi(r+1)}-\bm{\mu}_{\pi(r)}\|_2 \leq 2\sum_{(a,b)\in\mathcal{E}} \|\bm{\mu}_a-\bm{\mu}_b\|_2
\leq 2U.
\end{equation}
Thus each coordinate sequence belongs to
\[
\mathcal{M}_k \coloneqq \left\{\bm{u}\in[-C,C]^N: \sum_{r=1}^{N-1}|u_{\pi(r+1)}-u_{\pi(r)}|\leq 2U\right\}.
\]
For completeness, the entropy bound needed below is uniform over the probability vector $\bm{p}$. Associate with $\bm{u}\in\mathcal{M}_k$ a step function on $[0,1]$ whose successive interval lengths are $p_{\pi(1)},\ldots,p_{\pi(N)}$ and whose values are $u_{\pi(1)},\ldots,u_{\pi(N)}$. This map preserves the $L_2(\bm{p})$ distance and produces a function bounded by $C$ with total variation at most $2U$. By Page  3050 of \cite{sadhanala2019additive} it then follows that
\begin{equation}
\label{eq:tv-weighted-cover}
\ln \left[ N\bigl(\varepsilon,\mathcal{M}_k,L_2(\bm{p})\bigr) \right] \leq K_3\frac{C+U}{\varepsilon},
\qquad 
0<\varepsilon\leq 2C.
\end{equation}
The constant in (\ref{eq:tv-weighted-cover}) is independent of $N$ and of the weights $\bm{p}$.

Let
\[
\mathcal{F}_{\bm{Y}} \coloneqq \{\bm{f}_{\bm{\theta}}:\bm{\theta}\in\Theta(C,U)\}\subset\mathbb{R}^N.
\]
Using product covers in (\ref{eq:weighted-embedding}), with centers chosen in the bounded coordinate classes (an external cover can be converted to such an internal cover by doubling its radius), and then applying (\ref{eq:network-cover}) and (\ref{eq:tv-weighted-cover}), we obtain, for $0<\kappa\leq K_4S_N$,
\begin{equation}
\label{eq:loss-cover-subgaussian}
\ln \left[ N\bigl(\kappa,\mathcal{F}_{\bm{Y}},\|\cdot\|_{\mathcal{V}}\bigr) \right] \leq K_5 \left[ v_H \ln\left(1+\frac{K_5S_N}{\kappa}\right) +\frac{(1+U)S_N}{\kappa} \right],
\end{equation}
for some constant $K_5>0$. Moreover, (\ref{eq:weighted-embedding}) and the uniform bounds on $\bm{h}_{\bm{\phi}}$ and $\bm{\mu}$ imply $\operatorname{diam}(\mathcal{F}_{\bm{Y}},\|\cdot\|_{\mathcal{V}})\leq K_6S_N$ for some constant $K_6>0$.

Dudley's entropy inequality for the conditional Rademacher process, followed by $\sqrt{a+b}\leq\sqrt a+\sqrt b$, yields
\begin{align}
\mathbb{E}_{\bm{\xi}} \left[ \sup_{\bm{\theta}\in\Theta(C,U)} \frac{1}{N}\sum_{i\in\mathcal{V}}\xi_i f_{\bm{\theta},i}(\bm{Y}_i) \right]
& \lesssim \frac{1}{\sqrt{N}} \int_0^{K_6S_N} \sqrt{\ln N\bigl(\kappa,\mathcal{F}_{\bm{Y}},\|\cdot\|_{\mathcal{V}}\bigr)}\,d\kappa \nonumber\\
& \lesssim \frac{S_N}{\sqrt{N}} \left(\sqrt{v_H}+\sqrt{1+U}\right).
\label{eq:dudley-final-subgaussian}
\end{align}
After the change of variables $\kappa=S_N u$, the two relevant integrals are bounded by universal constants:
\[
\int_0^{K_6}\sqrt{\ln(1+K_5/u)}\,du<\infty,
\qquad
\int_0^{K_6}u^{-1/2}\,du<\infty.
\]
Finally, (\ref{eq:A-second-moment}) and Jensen's inequality imply
\[
\mathbb{E} (S_N) \leq\left(\frac{1}{N}\sum_{i\in\mathcal{V}}\mathbb{E}(A_i^2) \right)^{1/2} \leq K_A.
\]
Combining this with (\ref{eq:symmetrization-subgaussian}), (\ref{eq:dudley-final-subgaussian}), and $v_H\leq K_1LW\ln(eB)$ gives
\[
\mathbb{E}(P f_{\hat{\bm{\theta}}}) \lesssim  \frac{\sqrt{LW\ln(eB)}+\sqrt{1+U}}{\sqrt{N}}.
\]

From (\ref{eq:population-kl-identity}), we have
\begin{align*}
\mathbb{E}( P f_{\hat{\bm{\theta}}} ) & = \frac{1}{|\mathcal{V}|} \sum_{i\in\mathcal{V}} \mathbb{E} \left[ \mathrm{KL}\bigl(p_i^*\,\big|\,p_{\hat{\bm{\phi}},\hat{\bm{\mu}}_i}(\bm{y}_i)\bigr) \right] - \frac{1}{|\mathcal{V}|}\sum_{i\in\mathcal{V}} \mathrm{KL}\bigl(p_i^*\,\big|\,p_{\bm{\phi}^*,\bm{\mu}_i^*}(\bm{y}_i)\bigr), \\
\frac{1}{|\mathcal{V}|} \sum_{i\in\mathcal{V}} \mathbb{E} \left[ \mathrm{KL}\bigl(p_i^*\,\big|\,p_{\hat{\bm{\phi}},\hat{\bm{\mu}}_i}(\bm{y}_i)\bigr) \right] & = \mathbb{E}( P f_{\hat{\bm{\theta}}} ) + \frac{1}{|\mathcal{V}|}\sum_{i\in\mathcal{V}} \mathrm{KL}\bigl(p_i^*\,\big|\,p_{\bm{\phi}^*,\bm{\mu}_i^*}(\bm{y}_i)\bigr), \\
\frac{1}{|\mathcal{V}|}\sum_{i\in\mathcal{V}} \mathbb{E} \left[\mathrm{KL}\bigl(p_i^*\,\big|\,p_{\hat{\bm{\phi}},\hat{\bm{\mu}}_i}(\bm{y}_i)\bigr)\right] & \leq K\frac{\sqrt{LW\ln(eB)}+\sqrt{1+U}}{\sqrt{|\mathcal{V}|}} + \frac{1}{|\mathcal{V}|}\sum_{i\in\mathcal{V}} \mathrm{KL}\bigl(p_i^*\,\big|\,p_{\bm{\phi}^*,\bm{\mu}_i^*}(\bm{y}_i)\bigr),
\end{align*}
for some constant $K>0$ depending only on $C$, $\sigma$, $n$, and $d$. This completes the proof.

\section{Experiments}
\label{appendix_additional_result}

\subsection{Simulation Study}

Nine competitors, $k$-means \citep{hartigan1979algorithm}, dynamic time warping (DTW) \citep{sakoe2003dtw}, covariate assisted spectral clustering (CASC) \citep{binkiewicz2017casc}, semi-definite programming (SDP) \citep{yan2021sdp}, spectral clustering on network-adjusted covariates (NAC) \citep{hu2024nac}, spectral clustering on ratios-of-eigenvectors (SCORE) \citep{Jin2015SCORE}, variational graph auto-encoder (VGAE) \citep{kipf2016variational}, deep graph infomax (DGI) \citep{velickovic2018deep}, graph masked auto-encoder (GraphMAE) \citep{hou2022graphmae}, are provided for comparison. 

For particulars of competitor methods, $k$-means constructs clusters by minimizing within-cluster squared distances. DTW method measures similarities between time series by aligning them via a non-linear time warping. CASC method leverages the nodal covariate $\bm{X}$ via $\bm{L}_{\tau} \bm{L}_{\tau} + \alpha \bm{X}\bm{X}^\top$ to improve spectral clustering performance; here, $\bm{L}_{\tau}$ is the regularized graph Laplacian. SDP method converts clustering issues into a convex semi-definite procedure, enabling efficient approximation of the optimal partition. Specifically, SDP maximizes the inner product between $\bm{A} + \lambda \bm{K}$ and a cluster label matrix; here, $\bm{A}$ is an adjacency matrix and $\bm{K} \in \mathbb{R}_+^{N \times N}$ is a kernel matrix whose $(i,j)$th entry quantifies the similarity between covariates from nodes $i$ and $j$. NAC method incorporates node covariates into spectral clustering through the network-adjusted covariate matrix $\bm{A} \bm{X} + \bm{D}_{\alpha} \bm{X} \in \mathbb{R}^{N \times p}$ with diagonal weight matrix $\bm{D}_{\alpha}$ and covariate matrix $\bm{X}$. The SCORE method identifies communities through the ratios between the first leading eigenvector and the other leading eigenvectors. VGAE method learns probabilistic node embeddings by encoding graph structure and nodal covariates into latent distributions. DGI method obtains node representations by maximizing the mutual information between local node embeddings and a global graph summary. The GraphMAE method extracts node representations via a self-supervised masked reconstruction task, in which randomly masked nodal covariates are recovered from the graph and the remaining observed features. For these three graph neural network methods (VGAE, DGI, GraphMAE), $k$-means clustering is applied to the learned node representations to obtain cluster assignments.

For competitors, the true number of clusters from the simulated data and their default configurations are used during implementation. For GFL, we run our experiments with T4 GPU. The algorithm is implemented for a preset number of iterations, as an early stopping strategies to avoid over-fitting, which is commonly used.

\subsection{Visualization}

\begin{figure}[!ht]
\begin{center}
\includegraphics[width=0.6\columnwidth]{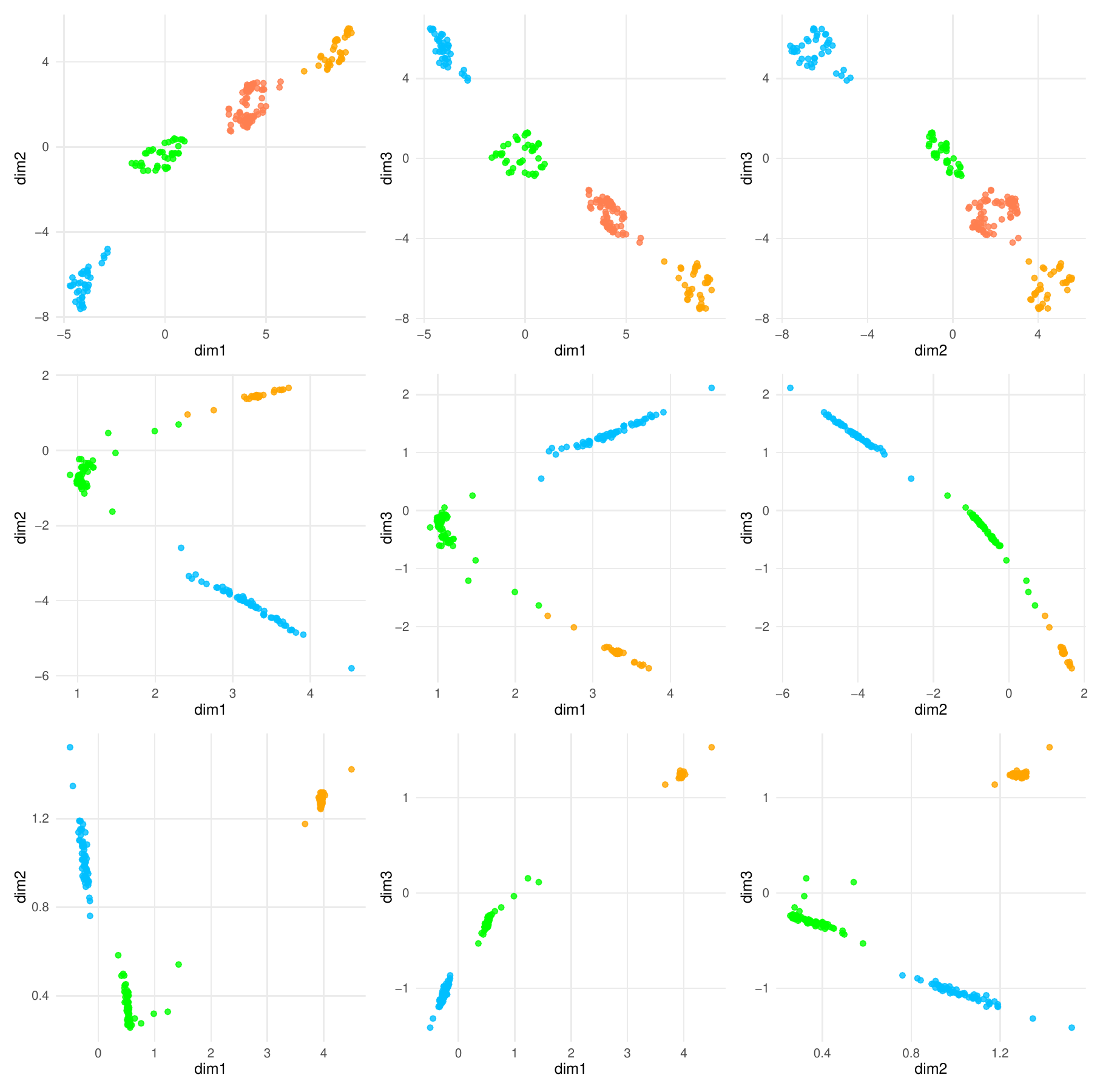}
\end{center}
\caption{Illustration of estimated $\hat{\bm{\mu}} \in \mathbb{R}^{N \times d}$ for a realization from Scenarios 1-3, where the latent dimension is $d=3$. Each row, from top to bottom, corresponds to a scenario. Each column, from left to right, displays a pairwise projection of the three latent dimensions. The node colors depict the ground truth labels.}
\label{fig_mu_sim}
\end{figure}

\begin{figure}[!ht]
\begin{center}
\includegraphics[width=0.75\columnwidth]{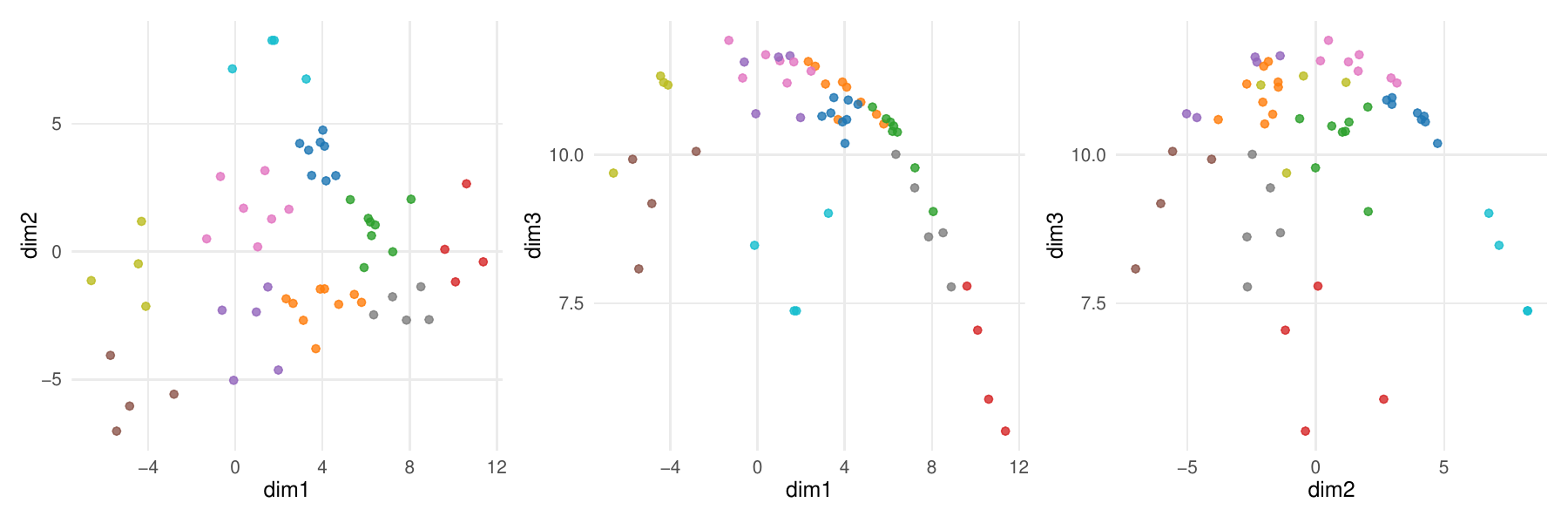}
\end{center}
\caption{Visualization of estimated $\hat{\bm{\mu}} \in \mathbb{R}^{58 \times 3}$. Each data point represents one of the $58$ counties in California. Node colors represent the clusters detected by our method.}
\label{fig_CA_mu}
\end{figure}

\begin{figure}[!ht]
\begin{center}
\includegraphics[width=0.75\columnwidth]{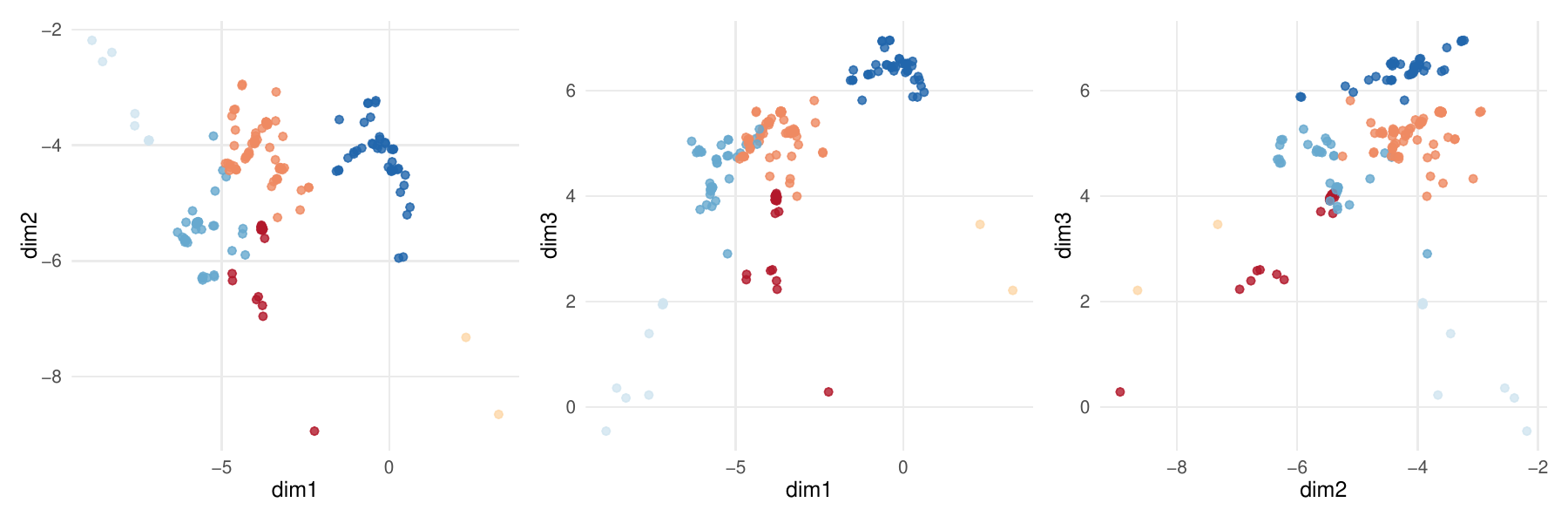}
\end{center}
\caption{Visualization of estimated $\hat{\bm{\mu}} \in \mathbb{R}^{207 \times 3}$. Each data point represents one of the $207$ traffic sensors in Los Angeles. Node colors represent the clusters detected by our method.}
\label{fig_Traffic_mu}
\end{figure}

For both simulation and real data experiments, the choice of latent dimension $d=3$ permits visualizations of learned prior parameters in latent space. Figure \ref{fig_mu_sim} illustrates the estimated $\hat{\bm{\mu}} \in \mathbb{R}^{N \times 3}$ for a realization from simulation study with $d=3$. We can see that the estimated prior parameters have captured meaningful boundary between clusters, enforced by graph-fused LASSO regularization. Figures \ref{fig_CA_mu} and \ref{fig_Traffic_mu} display the estimated $\hat{\bm{\mu}} \in \mathbb{R}^{N \times 3}$ from the California county temperature and Los Angeles traffic sensor data, respectively. From left to right, each panel displays a pairwise projection of the three latent dimensions. Taken together, the three-dimensional latent space exhibits clear boundaries between clusters.

\subsection{Sensitivity Analysis}
\label{sensitivity_analysis}

The three simulation studies are repeated for the proposed GFL method, with varying latent dimensions $d \in \{3,5,7,9\}$. Table \ref{tab:sim_sensitivity} reports the results, with the values for $d=3$ directly duplicated from Tables \ref{tab:sim1}, \ref{tab:sim2}, and \ref{tab:sim3} for comparison. The proposed method is robust when the latent dimension is relatively low. However, when $d=9$, the performance declines significantly, likely due to over-fitting. In this case, the latent representation, which contains both structure and attribute information, could include excessive noises that deteriorate the clustering results.

\begin{table}[!ht]
\caption{Sample means (one standard deviation) of our evaluation metrics for varied latent dimension $d \in \{3,5,7,9\}$ and Scenarios 1-3. The best metric's score is bolded.}
\label{tab:sim_sensitivity}
\centering
\resizebox{0.8\columnwidth}{!}{%
\begin{tabular}{c|c|cccccc}
\toprule
Scenario, $N$ & $d$ & NMI $\uparrow$ & ARI $\uparrow$ & ACC $\uparrow$ & HOM $\uparrow$ & COM $\uparrow$ & PUR $\uparrow$\\
\midrule
\multirow{4}{4em}{S1, $144$} 
& $3$ & $96.99\% (0.05)$ & $96.62\% (0.06)$ & $98.33\% (0.04)$ & $96.52\% (0.06)$ & $97.61\% (0.04)$ & $99.17\% (0.01)$ \\
& $5$ & $96.29\% (0.10)$ & $96.05\% (0.12)$ & $97.67\% (0.09)$ & $95.75\% (0.12)$ & $97.31\% (0.05)$ & $99.19\% (0.01)$ \\
& $7$ & $\bm{97.14}\% (0.04)$ & $\bm{97.15}\% (0.05)$ & $\bm{98.64}\% (0.02)$ & $\bm{97.43}\% (0.04)$ & $96.87\% (0.05)$ & $98.64\% (0.02)$ \\
& $9$ & $96.10\% (0.05)$ & $94.00\% (0.10)$ & $96.11\% (0.07)$ & $94.34\% (0.09)$ & $\bm{98.42}\% (0.02)$ & $\bm{99.44}\% (0.01)$ \\
\midrule
\multirow{4}{4em}{S1, $196$} 
& $3$ & $98.40\% (0.04)$ & $97.95\% (0.05)$ & $98.20\% (0.06)$ & $97.65\% (0.06)$ & $\bm{99.37}\% (0.01)$ & $\bm{99.84}\% (0.01)$ \\
& $5$ & $\bm{99.13}\% (0.01)$ & $\bm{99.34}\% (0.01)$ & $\bm{99.76}\% (0.00)$ & $\bm{99.15}\% (0.01)$ & $99.12\% (0.02)$ & $99.76\% (0.01)$ \\
& $7$ & $91.16\% (0.18)$ & $88.48\% (0.24)$ & $93.45\% (0.14)$ & $88.80\% (0.23)$ & $97.02\% (0.05)$ & $99.18\% (0.02)$ \\
& $9$ & $94.95\% (0.11)$ & $93.10\% (0.14)$ & $94.20\% (0.11)$ & $92.53\% (0.15)$ & $98.49\% (0.05)$ & $99.43\% (0.02)$ \\
\midrule\midrule
\multirow{4}{4em}{S2, $120$} 
& $3$ & $92.67\% (0.07)$ & $92.77\% (0.10)$ & $95.77\% (0.08)$ & $91.16\% (0.12)$ & $95.15\% (0.03)$ & $98.77\% (0.01)$ \\
& $5$ & $\bm{95.63}\% (0.03)$ & $\bm{96.89}\% (0.02)$ & $\bm{98.90}\% (0.01)$ & $\bm{95.52}\% (0.03)$ & $95.75\% (0.03)$ & $98.90\% (0.01)$ \\
& $7$ & $92.79\% (0.09)$ & $92.63\% (0.14)$ & $96.03\% (0.09)$ & $91.67\% (0.13)$ & $95.05\% (0.04)$ & $98.67\% (0.01)$ \\
& $9$ & $92.15\% (0.09)$ & $90.97\% (0.13)$ & $94.17\% (0.10)$ & $89.26\% (0.14)$ & $\bm{96.66}\% (0.04)$ & $\bm{99.17}\% (0.01)$ \\
\midrule
\multirow{4}{4em}{S2, $210$} 
& $3$ & $96.11\% (0.06)$ & $95.74\% (0.10)$ & $97.30\% (0.08)$ & $94.91\% (0.10)$ & $\bm{98.07}\% (0.02)$ & $\bm{99.58}\% (0.01)$ \\
& $5$ & $\bm{97.90}\% (0.02)$ & $\bm{98.63}\% (0.02)$ & $\bm{99.54}\% (0.01)$ & $\bm{97.89}\% (0.02)$ & $97.91\% (0.02)$ & $99.54\% (0.01)$ \\
& $7$ & $97.03\% (0.03)$ & $98.02\% (0.02)$ & $99.20\% (0.01)$ & $97.29\% (0.02)$ & $96.79\% (0.03)$ & $99.20\% (0.01)$ \\
& $9$ & $94.75\% (0.08)$ & $93.94\% (0.12)$ & $95.92\% (0.09)$ & $93.32\% (0.12)$ & $97.33\% (0.03)$ & $99.35\% (0.01)$ \\
\midrule\midrule
\multirow{4}{4em}{S3, $120$}
& $3$ & $94.41\% (0.08)$ & $93.44\% (0.13)$ & $95.83\% (0.09)$ & $92.89\% (0.13)$ & $97.24\% (0.04)$ & $99.17\% (0.01)$ \\
& $5$ & $\bm{98.54}\% (0.02)$ & $\bm{98.95}\% (0.02)$ & $\bm{99.57}\% (0.01)$ & $\bm{98.60}\% (0.03)$ & $\bm{98.48}\% (0.02)$ & $\bm{99.57}\% (0.01)$ \\
& $7$ & $97.82\% (0.02)$ & $98.56\% (0.02)$ & $99.40\% (0.01)$ & $98.01\% (0.02)$ & $97.65\% (0.02)$ & $99.40\% (0.01)$ \\
& $9$ & $96.04\% (0.06)$ & $95.85\% (0.09)$ & $96.87\% (0.07)$ & $95.51\% (0.10)$ & $97.36\% (0.05)$ & $99.00\% (0.03)$ \\
\midrule
\multirow{4}{4em}{S3, $210$} 
& $3$ & $92.42\% (0.11)$ & $88.99\% (0.17)$ & $91.83\% (0.13)$ & $88.28\% (0.17)$ & $\bm{99.23}\% (0.02)$ & $\bm{99.83}\% (0.01)$ \\
& $5$ & $\bm{99.11}\% (0.02)$ & $\bm{99.42}\% (0.01)$ & $\bm{99.79}\% (0.01)$ & $\bm{99.11}\% (0.02)$ & $99.12\% (0.02)$ & $99.79\% (0.01)$ \\
& $7$ & $97.02\% (0.07)$ & $96.30\% (0.10)$ & $96.21\% (0.10)$ & $96.07\% (0.11)$ & $98.74\% (0.03)$ & $99.68\% (0.01)$ \\
& $9$ & $93.32\% (0.10)$ & $90.42\% (0.16)$ & $92.93\% (0.12)$ & $89.81\% (0.16)$ & $99.11\% (0.02)$ & $99.79\% (0.01)$ \\
\bottomrule
\end{tabular}
}
\end{table}

\end{document}